\documentclass[twoside,12pt]{article}
\usepackage{epsfig}
\usepackage{amsmath}
\usepackage{amssymb}
\usepackage{times}
\usepackage{dsfont}
\usepackage{thophys}

\def\Journal#1#2#3#4{{#1} {#2} (#4) #3 }

\def\NPB{{\em Nucl. Phys.} B}

\def\PRL{\em Phys. Rev. Lett.}
\def\PREV{\em Phys. Rev.}
\def\PREP{\em Phys. Rep.}

\def\PRD{{\em Phys. Rev.} D}

\def\ANNP{\em Ann. Phys. (N.Y.)}

\newcommand{\be}{\begin{equation}}
\newcommand{\ee}{\end{equation}}
\newcommand{\bea}{\begin{eqnarray}}
\newcommand{\eea}{\end{eqnarray}}

\def\ti{\tilde}
\def\kon#1#2{\vbox{\halign{##&&##\cr\lower4pt
\hbox{$\scriptscriptstyle\vert$}\hrulefill &\hrulefill\lower4pt
\hbox{$\scriptscriptstyle\vert$}\cr $#1$&$#2$\cr}}}
\def\lra{\longleftrightarrow}
\def\fii{\varphi}
\def\al{\alpha}
\def\ga{\gamma}
\def\ro{\rho}
\def\si{\sigma}
\def\eh{{\scriptstyle{1\over 2}}}
\def\d{\partial}
\def\=d{\,{\buildrel\rm def\over =}\,}
\def\dh{\mathop{\vphantom{\odot}\hbox{$\partial$}}}
\def\dl{\dh^\leftrightarrow}
\def\sqr#1#2{{\vcenter{\vbox{\hrule height.#2pt\hbox{\vrule width.
#2pt height#1pt \kern#1pt \vrule width.#2pt}\hrule height.#2pt}}}}
 
\def\eps{\varepsilon}
\def\pdh{(2\pi )^{-3/2}}
\def\io{\int{d^3k\over\sqrt{2\omega}}\,}

\def\sq{\hbox{\rlap{$\sqcap$}$\sqcup$}}
\def\dh{\mathop{\vphantom{\odot}\hbox{$\partial$}}}
\def\dl{\dh^\leftrightarrow}

\def\te{\vartheta}
\def\reg{{\rm reg}}
\def\sgn{{\rm sgn}}
\def\supp{{\rm supp\,}}
\def\psm{\psi^{(-)}}
\def\psm{\psi^{(-)}}
\def\psp{\psi^{(+)}}
\def\pqp{\overline{\psi}^{(+)}}
\def\pqm{\overline{\psi}^{(-)}}
\def\ps{/ \! \! \! p}
\def\psq{{\overline{\psi}}}
\def\uq{\overline{u}}
\def\ds{/ \! \! \! \partial}
\def\i3p{(2\pi)^{-3/2}\int d^3p}
\def\tr{{\rm tr}\,}
\def\lo{{\cal L_+^\uparrow}}
\newcommand{\bd}{\begin{displaymath}}
\newcommand{\ed}{\end{displaymath}}

\begin{document}
\title{ \vspace{1cm} Regularization in Quantum Field Theory\\
from the Causal Point of View}
\author{A.\ Aste,$^{1}$ C. von Arx,$^{1}$ G.\ Scharf,$^{2}$\\
\\
$^1$Department of Physics, University of Basel\\
Klingelbergstrasse 82, CH-4056 Basel, Switzerland\\
$^2$Institute for Theoretical Physics, University of Z\"urich\\
Winterthurerstrasse 190, CH-8057 Z\"urich,
Switzerland}
\date{August 24, 2009}
\maketitle
\begin{abstract} The causal approach to perturbative quantum field
theory is presented in detail, which goes back to a seminal work by
Henri Epstein and Vladimir Jurko Glaser in 1973.
Causal perturbation theory is a mathematically
rigorous approach to renormalization theory, which makes it possible to put
the theoretical setup of perturbative quantum field theory on a sound mathematical
basis. Epstein and Glaser solved this problem for a special class of distributions,
the time-ordered products, that fulfill a causality condition, which itself is a
basic requirement in axiomatic quantum field theory. In their original work,
Epstein and Glaser studied only theories involving scalar particles. In this review,
the extension of the method to theories with higher spin, including gravity, is presented.
Furthermore, specific examples are presented in order to highlight the
technical differences between the causal method and other regularization
methods, like, e.g. dimensional regularization.
\end{abstract}
%\eject
\tableofcontents

\section{Introduction}
Quantum field theory (QFT) is more singular than quantum mechanics.
The basic mathematical objects of quantum mechanics are square integrable
functions, whereas the corresponding central objects in QFT
are generalized functions or distributions. A potential drawback of the theory of
distributions for physics is the fact that it is a purely linear theory,
in the sense that the product of two distributions cannot consistently be
defined in general, as has been proved by Laurent Schwartz \cite{Schwartz1966},
who was awarded the Fields medal for his work on distributions in 1950.

If one is careless about this point the well-known ultraviolet (UV)
divergences appear in perturbative quantum field theory (pQFT).
The occurrence of these divergences is
sometimes ascribed in a qualitative manner to problematic contributions of virtual
particles with "very high energy", or, equivalently, to physical phenomena at very short
spacetime distances, and put forward as an argument that the quantized version of
extended objects like strings which are less singular than point-like particles
should be used instead in QFT.
In view of the fact that UV divergences can be circumvented by a proper
treatment of distributions in pQFT, this argument for string theories is no longer
convincing. 

We illustrate the problem
mentioned above by a naive example of a "UV divergence"
by considering Heaviside-$\Theta$- and Dirac-$\delta$-distributions in 1-dimensional
"configuration space".
The product of these two distributions $\Theta(x) \delta(x)$ is
obviously ill-defined, however, the distributional Fourier transforms
\begin{equation}
\sqrt{2 \pi} \mathcal{F} \{ \delta \} (k)=
\sqrt{2 \pi} \hat{\delta} (k) = \int dx \, \delta(x) e^{-ikx} =1,
\end{equation}
\begin{equation}
\sqrt{2 \pi} \hat{\Theta} (k) = \lim_{\epsilon \searrow 0}
\int dx \, \Theta(x) e^{-ikx-\epsilon x} = \lim_{\epsilon \searrow 0}
\frac{ie^{-ikx-\epsilon x}}{k-i \epsilon} \Biggr|^{\infty}_{0}= -\frac{i}{k-i0},
\end{equation}
exist and one may attempt to calculate the ill-defined product in "momentum space",
which \emph{formally} goes over into a convolution
\begin{equation}
\mathcal{F} \{ \Theta \delta \} (k)=\frac{1}{\sqrt{2 \pi}}
\int dx \, e^{-ikx} \Theta(x)  \delta(x) =
\frac{1}{\sqrt{2 \pi}}
\int dx \, e^{-ikx} \int \frac{dk'}{\sqrt{2 \pi}} \hat{\Theta}(k') e^{+ik'x}
\int \frac{dk''}{\sqrt{2 \pi}} \hat{\delta}(k'') e^{+ik''x}.
\end{equation}
Throughout this paper, we use the symmetric definition of the (inverse) Fourier
transform according to Eq. (\ref{(1)}) and Eq. (\ref{(2)}).
Since $\int dx \, e^{i(k'+k''-k)x}=2 \pi \delta(k'+k''-k)$, we obtain
\begin{equation}
\mathcal{F} \{ \Theta \delta \} (k) = \frac{1}{\sqrt{2 \pi}} \int dk' \,
\hat{\Theta} (k') \hat{\delta} (k-k') = {
-\frac{i}{(2 \pi)^{3/2}} \int \frac{dk'}{k'-i0}}.
\end{equation}
The obvious problem in x-space leads to a "logarithmic UV divergence" in k-space.
It will become clear below that a concise description of the scaling properties
of distributions, related to the wide-spread notion of the superficial degree of divergence
of Feynman integrals, is crucial for the correct treatment of singular products
of distributions in pQFT.

In pQFT, the r\^{o}le of the {Heaviside $\Theta$-distribution} is taken over
by the time-ordering operator. The well-known textbook expression for the
perturbative scattering matrix given by
\begin{displaymath}
S  = \sum \limits_{n=0}^{\infty} \frac{(-i)^n}{n!}
\int \limits_{-\infty}^{+\infty} dt_1 \ldots
\int \limits_{-\infty}^{+\infty} dt_n \, {T}
[H_{int}(t_1) \ldots H_{int}(t_n)]
\end{displaymath}
\begin{equation}
 =  \sum \limits_{n=0}^{\infty} \frac{(-i)^n}{n!}
\int  d^4 x_1 \ldots \int  d^4 x_n \, {T}
[\mathcal{H}_{int}(x_1) \ldots
\mathcal{H}_{int}(x_n)], \label{Smatrix_textbook}
\end{equation}
where the interaction Hamiltonian {$H_{int}(t)$}
is given by the interaction Hamiltonian density $\mathcal{H}_{int}(x)$ via
$H_{int}(t)=\int d^3 x \,
\mathcal{H}_{int}(x)$, is problematic in the UV regime (and in the infrared
regime, when massless fields are involved).
A time-ordered expression \`a la
\begin{equation}
T [\mathcal{H}_{int}(x_1) \ldots \mathcal{H}_{int}(x_n)] = \! \!
\sum \limits_{Perm. \, \, \Pi} \Theta(x^0_{\Pi_1}-x^0_{\Pi_2}) \ldots
\Theta(x^0_{\Pi_{(n-1)}}-x^0_{\Pi_n}) \mathcal{H}_{int}(x_{\Pi_1})
\ldots \mathcal{H}_{int}(x_{\Pi_n})
\end{equation}
is formal (i.e., ill-defined), since the operator-valued distribution
products of the $\mathcal{H}_{int}$ are simply too singular to be multiplied by
$\Theta$-distributions.

In this review, the construction of pQFT is reviewed from a causal point of view
with a special focus on the regularization of distributions.
Typical examples are discussed in the causal framework and compared
to the corresponding treatment in the Pauli-Villars regularization or dimensional
regularization.
In the last section we describe a modern approach to quantum
gauge theories including gravity. This shows that the gauge principle
in a suitable formulation is a universal principle of nature because
it determines all interactions. Therefore, any regularization
method must be in accordance with it.

\section{Mathematical Preliminaries}
\subsection{\it Regularization of Distributions \label{sec:regdistr}}
Distributions are continuous linear functionals on certain function spaces.
There exist different spaces of distributions. For quantum field theory the
most important function space is the Schwartz space $\mathcal{S}(\mathds{R}^n)$. It consists of
infinitely differentiable complex-valued functions of rapid decrease, that
means the functions together with their derivatives fall off more quickly
than the inverse of any polynomial. The reason for the importance of $\mathcal{S}(\mathds{R}^n)$
is the fact that the Fourier transform
(the expression $px$ denotes a generalized $n-$dimensional Euclidean or
Minkowski scalar product depending on the respective situation)
\be
\mathcal{F} \{ f \}(p)= \hat{f} (p)=
(2\pi)^{-n/2}\int\limits_{\mathds{R}^n}f(x)e^{-ipx}d^nx , \label{(1)}
\ee
is a linear bi-continuous bijection from $\mathcal{S}(\mathds{R}^n)$ onto $\mathcal{S}(\mathds{R}^n)$. Indeed,
the inverse Fourier transform is given by
\be
\mathcal{F}^{-1} \{ g \}(x)= \check{g} (x)=
(2\pi)^{-n/2} \int\limits_{\mathds{R}^n}g(p)e^{ipx}d^np.
\label{(2)}
\ee

The dual space of $\mathcal{S}(\mathds{R}^n)$ denoted by $\mathcal{S}' (\mathds{R}^{n})$,
is the space of tempered
distributions. A tempered distribution $d(f)$ is a continuous linear
complex-valued functional on $\mathcal{S}(\mathds{R}^n)\ni f$; we also write $d(f)=\langle
d,f\rangle$. The functions $f\in \mathcal{S}(\mathds{R}^n)$ are called test functions. The 
Fourier transform of a tempered distribution $d$ is now simply defined
by its action on the test functions:
\be
\langle \mathcal{F} \{ d \} ,f\rangle \=d \langle d, \mathcal{F} \{f\}
\rangle,\quad f\in \mathcal{S}(\mathds{R}^n). \label{(3)}
\ee
In this way, by operating on test functions, various operations on
distributions like differentiation, convolution etc are defined.
Note that the definition Eq. (\ref{(3)}), $\hat{d}(f)=d(\hat{f})$,
is sometimes written in an intuitive manner by the help of formal integrals
\be
\int\limits_{\mathds{R}^n} \hat{d}(q) f(q) \, d^nq= 
\int\limits_{\mathds{R}^n} d(q) \hat{f}(q) \, d^nq \, ,
\ee
showing the close relation of the definition above to the Plancherel theorem.
Of course, the common physical distinction whether the integration variable $q$ is in
"real space" or "momentum space" is of no relevance here.
For mathematical details concerning the properties of distributions, we refer to
\cite{Streater,Constantinescu}.

The most important distributions for field theory are related to linear
partial differential equations, for example the Klein-Gordon equation
\be
(\sq +m^2)d(x)=\Bigl({\frac{\partial^2}{\partial x_0^2}}-
\sum_{j=1}^{n-1}{\frac{\partial^2}{\partial x_j^2}}
+m^2\Bigl)d(x)=0.\label{(4)}
\ee
An important \emph{distributional} solution of the 4-dimensional Klein-Gordon 
equation is the Jordan-Pauli distribution
\be
D_m (x)={\frac{i}{(2\pi)^3}}\int d^4p\,\delta(p^2-m^2){\rm sgn} (p_0)e^{-ipx}\label{(5)}
\ee
where the integral must be understood as a distributional Fourier transform;
the factor $i$ makes $D_m (x)$ real. If we decompose the sign-function, {\rm sgn}
$p_0=\Theta (p_0)-\Theta(-p_0)$, we obtain the decomposition of $D_m(x)=D^{(+)}_m(x)
+D^{(-)}_m(x)$ into positive and negative frequency parts, for example
\be
D^{(+)}_m(x)={\frac{i}{(2\pi)^3}}\int d^4p\delta(p^2-m^2)\Theta(-p_0)e^{ipx}=
{\frac{i}{(2\pi)^3}}\int d^4p\delta(p^2-m^2)\Theta(p_0)e^{-ipx}
\label{(6)}
\ee
In addition to these solutions of the homogeneous Klein-Gordon equation
we need weak solutions of the inhomogeneous equation
\be
(\sq +m^2)D_{ret}^m (x)=\delta (x).\label{(7)}
\ee
This retarded distribution which vanishes for negative time $ct=x_0<0$ 
($c$ is the velocity of light) is given by
\be
D_{ret}^m(x)=\Theta(x_0)D^m(x)\label{(8)}
\ee
and the corresponding advanced distribution by $D_{av}^m(x)=D_{ret}^m(-x)$.
Finally, the so-called Feynman propagator is defined as
\be
D_F^m(x)=D_{ret}^m(x)-D^{(-)}_m(x)=D_{av}^m(x)+D^{(+)}_m(x).\label{(9)}
\ee
Its Fourier transform is equal to
\be
D_F^m(x)=-(2\pi)^{-4}\int d^4p {\frac{e^{-ipx}}{p^2-m^2+i0}},
\label{(10)}
\ee
where the symbol $i0$ stands for $i\eps$ and the limit $\eps\to 0$ in the
distributional sense, i.e. in $\mathcal{S}'(\mathds{R}^n)$.

In the appendix, a concise list of the fundamental free field commutators and
propagators is given, where we also explicitly account for the most common
conventions concerning the signs and normalizations of the distributions. 

In standard QFT the Feynman propagator $D_F^m$ is associated with the inner 
lines of a Feynman
graph in the simplest case of scalar particles (spin 0). In a lowest-order loop graph
there arises the problem of multiplying two Feynman propagators $D_F^{m_1}(x)\cdot
D_F^{m_2}(x)$, a product which is ill-defined. In fact, in momentum space this product
corresponds to a formal convolution of the form
\be
\Sigma(p)=C\int d^4k\, D_{m_1}(k)D_{m_2}(p-k),\label{(11)}
\ee
where $C$ is a numerical constant; we shall always use the symbol $C$
for constants which we do not compute explicitly because they are not
interesting for our purpose. To simplify the notation the various $C$'s 
stand for different constants. By counting powers of $\vert k\vert$ we see
that the integral Eq. (\ref{(11)}) is logarithmically divergent in the ultraviolet
regime $\vert k\vert\to\infty$. To make it well defined we use a regularization
of the Feynman propagator $D_F^{m}(x)$
\be
D_F^\reg (k)=C\Bigl({\frac{1}{k^2-m^2+i0}}-{\frac{1}{k^2-M^2+i0}}\Bigl)=
C{\frac{m^2-M^2}{(k^2-m^2+i0)(k^2-M^2+i0)}}, \label{(12)}
\ee
where $C$ denotes a real normalization constant which depends on specifically
chosen conventions. Modifying the Feynman propagator according to Eq. (\ref{(12)})
at a high mass or energy scale given $M$ is the basic essence
of the so-called Pauli-Villars regularization. Note that the propagator term
containing $M$ has the "wrong sign" and does not correspond to the contribution
of a heavy physical particle. However, for $M\to\infty$,
$D_F^\reg (k)$ converges to $D_F^m(k)$ in the sense of tempered distributions.
We present here one possible approach to calculate the scalar self-energy diagram.
Using the Fourier transform
\be
{\frac{1}{k^2-m^2+i0}}={\frac{1}{i}}\int\limits_0^\infty e^{is(k^2-m^2+i0)}ds
\ee
the regularized propagator is equal to
\be
D_F^\reg (k)=C\int\limits_0^\infty ds\,e^{isk^2-s0}\Bigl(e^{-ism^2}-
e^{-isM^2}\Bigl).
\ee
Substituting the Feynman propagators in the self-energy integral Eq. (\ref{(11)}) by
regularized ones, we obtain a finite integral
\be
\Sigma^\reg(p)=C\int d^4k\int\limits_0^\infty ds_1\int\limits_0^\infty ds_2
e^{is_1k^2-s_10}\Bigl((e^{-is_1m_1^2}-e^{-is_1M^2}\Bigl)\times
\ee
\be
\times e^{is_2(p-k)^2-s_20}\Bigl(e^{-is_2m_2^2}-e^{-is_2M^2}\Bigl).\label{(13)}
\ee
Here the 4-dimensional $k$-integral can be carried out by means of the
Gauss-Fresnel integral
\be
\int e^{i(ak^2+bk)}d^4k={\frac{\pi^2}{ia^2}}\exp\Bigl(-{\frac{ib^2}{4a}}
\Bigl),\quad a>0.
\ee
The result is
\be
\Sigma^\reg(p)=C\int\limits_0^\infty ds_1\int\limits_0^\infty ds_2
{\frac{e^{-(s_1+s_2)0}}{(s_1+s_2)^2}}\exp\Bigl(\frac{i{s_1s_2}{ s_1+s_2}}p^2\Bigl)\Bigl(e^{-is_1m_1^2}-e^{-is_1M^2}\Bigl)
\ee
\be
\times\Bigl(e^{-is_2m_2^2}-e^{-is_2M^2}\Bigl).\label{(14)}
\ee
Now we introduce the new integration variables $t_1=s_1/(s_1+s_2)$ and
$t_2=s_1+s_2$, then we have
\be
\Sigma_\eps^\reg(p)=C\int\limits_0^1 dt_1\int\limits_0^\infty dt_2
{\frac{1}{t_2^2}}e^{-\eps t_2+it_1t_2(1-t_1)p^2}
\ee
\be
\Bigl(e^{-it_1t_2m_1^2}-e^{-it_1t_2M^2}\Bigl)
\Bigl(e^{-i(1-t_1)t_2m_2^2}-e^{-i(1-t_1)t_2M^2}\Bigl).\label{(15)}
\ee
We have written $i\eps$ for $i0$ and take the limit $\eps\to 0$ later on.
To perform the $t_2$-integration we need the integral
\be
\int\limits_a^\infty{dx\over x^2}e^{-\eps x+izx}\=d J_a(z)
\ee
where we the lower limit of integration is $a>0$ in order to avoid the
singularity at $x=0$; note that Eq. (\ref{(15)}) is integrable at $t_2=0$. By differentiating
twice with respect to $z$ the denominator $x^2$ is removed and the integral
can be easily evaluated
\be
J_a''(z)={e^{iza-\eps a}\over iz-\eps}.
\ee
Now the limit $a\to 0$ is possible and two integrations in $z$ yield
\be
J_0(z)=-iz[\log(iz-\eps)-1].
\ee
Using this result in Eq. (\ref{(15)}) the regularized self-energy integral becomes
\be
\Sigma_\eps^\reg(p)=C\int\limits_0^1 dt_1\Bigl[z_1\log(iz_1-\eps)
-z_2\log(iz_2-\eps)-
-z_3\log(iz_3-\eps)+z_4\log(iz_4-\eps)\Bigl],\label{(16)}
\ee
where
\be
z_1=t_1(1-t_1)p^2-t_1m_1^2-(1-t_1)m_2^2
\ee
\be
z_2=t_1(1-t_1)p^2-t_1m_1^2-(1-t_1)M^2
\ee
\be
z_3=t_1(1-t_1)p^2-t_1M^2-(1-t_1)m_2^2\label{(17)}
\ee
\be
z_4=t_1(1-t_1)p^2-t_1M^2-(1-t_1)M^2.
\ee

The integral Eq. (\ref{(16)}) still diverges for $M\to\infty$. We have to split off the
divergent part. This process, called renormalization, must always be combined
with regularization. In order to obtain a unique finite result we proceed as 
follows. We compute the special value $\Sigma_\eps^\reg(0)$ and subtract it
from (16). Then the limit
\be
\lim_{\eps\to 0}\lim_{M\to\infty}(\Sigma_\eps^\reg(p)-\Sigma_\eps^\reg(0))\=d
\Sigma'(p)\label{(18)}
\ee
is finite. It satisfies the normalization condition
\be
\Sigma'(0)=0.\label{(19)}
\ee
We will not calculate the finite self-energy $\Sigma'(p)$ explicitly because 
later on we shall discuss methods which give the result in a more elegant way.
The subtraction of a constant in Eq. (\ref{(18)}) is equivalent to the subtraction of a
local term $\sim\delta(x)$ in $x$-space. If we considered a more singular
distribution, then a certain polynomial in $p$ must be subtracted which corresponds
to a sum of derivatives of the $\delta$-distribution in $x$-space.

\subsection{\it Scaling Properties of Distributions \label{subsec:scaling}}
From the last section it is clear that the singular behavior of a tempered
distribution $\hat d(p)$ at infinity or of its (inverse) Fourier transform
$d(x)$ at $x=0$ is of central importance in QFT. To study these properties
the so-called quasi-asymptotics of a tempered distribution is very useful. The
definition is the following:

{\bf Definition 2.1.} The distribution $d(x)\in \mathcal{S}'(\mathds{R}^m)$ 
has a quasi-asymptotics $d_0(x)$ at $x=0$ with respect to a positive 
continuous function $\rho (\delta)$, $\delta>0$, if the limit
\be
\lim_{\delta\to 0}\rho (\delta)\delta^m d(\delta x)=d_0(x)
\not\equiv 0 \label{(20)}
\ee
exists in $\mathcal{S}'(\mathds{R}^m)$.

In the smeared out form of Eq. (\ref{(20)}) with a test function
$\fii\in \mathcal{S}'(\mathds{R}^m)$
\be
\lim_{\delta\to 0}\rho(\delta)\Bigl\langle d(x),\,\fii\Bigl(
{\frac{x}{\delta}}\Bigl)\Bigl\rangle =\langle d_0,\,\fii\rangle.\label{(21)}
\ee
we go over to momentum space to find an equivalent condition for the
Fourier transform $\hat d(p)$. Since
\be
\Bigl\langle d(x),\,\fii\Bigl({\frac{x}{\delta}}\Bigl)\Bigl\rangle =
\Bigl\langle\hat d(p),\,\Bigl(\fii\bigr({\frac{x}{\delta}}\bigr)\Bigl) 
\check{\phantom{I}}
(p)\Bigl\rangle =\delta^m\langle\hat d(p),\,\check\fii (\delta p)\rangle
=\Bigl\langle\hat d\Bigl({\frac{p}{\delta}}\Bigl),\,\check\fii (p)\Bigl
\rangle,\label{(22)}
\ee
where $\check\fii$ denotes the inverse Fourier transform, we get the
following equivalent definition:

{\bf Definition 2.2.} The distribution $\hat d(p)\in{\cal S}'(R^m)$
has quasi-asymptotics $\hat d_0(p)$ at $p=\infty$ if
\be
\lim_{\delta\to 0}\rho(\delta)\Bigl\langle\hat d\Bigl({\frac{p}{\delta}}
\Bigl),\,\check\fii(p)\Bigl\rangle =\langle\hat d_0,\,\check\fii
\rangle\label{(23)}
\ee 
exists for all $\check\fii\in {\cal S}(R^m)$.

In momentum space the quasi-asymptotics controls the ultraviolet
behavior of the distribution. Let us consider a scaling transformation
\bea
\lim_{\delta\to 0}\rho(\delta)\langle\hat d({\frac{p}{\delta}}),\,\check 
\fii(ap)\rangle =\langle\hat d_0(p),\,\check\fii(ap)\rangle
\eea
\be
=a^{-m}\lim_{\delta\to 0}\rho (\delta)\Bigl\langle\hat d\Bigl({\frac{p}{a \delta}
a\delta}\Bigl),\,\check\fii(p)\Bigl\rangle =a^{-m}\lim_{\delta\to 0}
{\rho(\delta)\over\rho(a\delta)}\rho(a\delta)\Bigl\langle\hat d\Bigl(
{\frac{p}{a \delta}}\Bigl),\,\check\fii(p)\Bigl\rangle.\label{(24)}
\ee
Since
\be
\lim_{\delta\to 0}\rho(a\delta)\langle\hat d({p\over a\delta}),
\,\check\fii(p)\rangle =\langle\hat d_0(p),\,\check\fii(p)\rangle
\ee
exists, we may conclude that the limit
\be
\lim_{\delta\to 0}{\rho(a\delta)\over\rho(\delta)}=a^{-m}{\langle
\hat d_0(p),\,\check\fii(p)\rangle\over\langle\hat d_0(p),\,\check
\fii(ap)\rangle}\=d\rho_0(a)\label{(25)}
\ee
exists, too, assuming that the denominator is different from 0. By
another scaling transformation it follows
\be
\rho_0(ab)=\rho_0(a)\rho_0(b),\label{(26)}
\ee
which implies $\rho_0(a)=a^\omega$ with some real $\omega$. We therefore
call $\rho(\delta)$ the power-counting function.

With help of the power-counting function we can now define the singular
order of a distribution.

{\bf Definition 2.3.} The distribution $d\in {\cal S}'(R^m)$ is
called singular of order $\omega$, if it has a quasi-asymptotics
$d_0(x)$ at $x=0$, or its Fourier transform has quasi-asymptotics $\hat
d_0(p)$ at $p=\infty$, respectively, with power-counting function 
$\rho(\delta)$ satisfying
\be
\lim_{\delta\to 0}{\rho(a\delta)\over\rho(\delta)}=a^\omega,\label{(27)}
\ee
for each $a>0$.

Eq. (\ref{(25)}) implies
\bea
a^m\langle\hat d_0(p),\,\check\fii(ap)\rangle =\langle\hat d_0
({p\over a}),\,\check\fii(p)\rangle =a^{-\omega}\langle\hat d_0(p),
\,\check\fii(p)\rangle
\eea
\be
=\langle d_0(x),\,\fii({x\over a})\rangle =a^m\langle d_0(ax),\,
\fii(x)\rangle =a^{-\omega}\langle d_0(x),\,\fii(x)\rangle,\label{(28)}
\ee
i.e. $\hat d_0$ is homogeneous of degree $\omega$:
\be
\hat d_0({p\over a})=a^{-\omega}\hat d_0(p), \label{(29)}
\ee
\be
d_0(ax)=a^{-(m+\omega)}d_0(x).\label{(30)}
\ee
This implies that $d_0$ has quasi-asymptotics $\rho(\delta)=\delta^ 
\omega$ and the singular order $\omega$, too. A positive measurable
function $\rho(\delta)$, satisfying Eq. (\ref{(27)}), is called regularly
varying at zero by mathematicians \cite{Senata}.
The power-counting function satisfies the following estimates: 
If $\eps>0$ is an
arbitrarily small number, then there exist constants $C, C'$ and
$\delta_0$, such that
\be
C\delta^{\omega+\eps}\ge \rho(\delta)\ge C'\delta^{
\omega-\eps},\label{(31)}
\ee
for $\delta<\delta_0$.
 
We want to apply the definitions to the following examples:

1) $d=1$: From Eq. (\ref{(20)}) we get $\rho(\delta)=\delta^{-m}$ and $\omega=-m$.

2) $d(x)=D^a\delta(x)$ where
\be
D^a\=d {\d^{a_1+\ldots +a_m}\over\d x_1^{a_1}\ldots\d x_m^{a_m}}\quad
 ,\quad |a|=a_1+\ldots +a_m.
\ee
Since
\be
\hat d(p)=(2\pi)^{-m/2}(ip)^a,
\ee
we obtain $\rho(\delta)=\delta^{|a|}$ from Eq. (\ref{(23)}) and $\omega=|a|$.

3) Let us consider the Jordan-Pauli distribution Eq. (\ref{(5)}) which has the following
form in $x$-space
\be
D_m (x)={\sgn x^0\over 2\pi}\Bigl[\delta (x^2)-{\frac{m}{2}}{\Theta (x^2)\over
\sqrt{x^2}}J_1(m\sqrt{x^2})\Bigl],\label{(32)}
\ee
where $J_1$ is the Bessel function. We shall write the $\delta$-distribution
with argument always in contrast to the positive scaling factor $\delta$.
The one-dimensional $\delta$-distribution satisfies
\be
\delta (\delta^2x^2)={\frac{\delta (x^2)}{\delta^2}},
\ee
whereas the term with the Bessel function stays bounded for
$\delta\sqrt{x^2}\to 0$. Hence
\be
\lim_{\delta\to 0}\delta^2 D_m (\delta x)={\sgn x^0\over 2\pi}\delta
(x^2)=D_0(x)\label{(33)}
\ee
which is just the mass zero Jordan-Pauli distribution. This illustrates
the general fact that the quasi-asymptotics $d_0$ is given by the 
corresponding mass zero distribution. Since the
Jordan-Pauli distribution is typically considered in $\mathds{R}^{4}$ $(m=4)$,
we find $\rho(\delta)=\delta^{-2}$ and $\omega (D_m)=-2$.

4) The positive frequency part Eq. (\ref{(6)})
\be
\hat D^{(+)}_m (p)={i\over 2\pi}\Theta (p^0)\delta (p^2-m^2)\label{(34)}
\ee
is best considered in momentum space. Since
\bea
\int\Theta\Bigl({p_0\over\delta}\Bigl)\delta\Bigl({p^2\over\delta^2}- 
m^2\Bigl)\fii(p)\,d^4p=\delta^2\int\Theta(p_0)
\delta(p^2-\delta^2m^2)\fii(p)\,d^4p
\eea
\be
=\delta^2\int{d^3p\over 2\sqrt{\vec p^2+\delta^2m^2}}\fii (\sqrt{
\vec p^2+\delta^2m^2},\vec p),\label{(35)}
\ee
we find
\be
\lim_{\delta\to 0}\delta^{-2}\hat D^{(+)}_m \Bigl({\frac{p}{\delta}}\Bigl)
=\hat D_0^{(+)}(p)\label{(36)}
\ee
which implies $\omega(D^{(+)}_m)=-2$, in agreement with the foregoing
example. We obviously have $\omega(D^{(-)}_m)=-2$, too.

We notice from example 2 that the degree of singularity at $x=0$ increases
with $\omega>0$. The distributions with negative $\omega$ have only mild
singularities. This difference will be important in the next section.

\subsection{\it Splitting of Distributions}

In QFT the problem arises of multiplying certain distributions which are
singular at $x=0$ by the discontinuous step function $\Theta (x_0)$, where
$x_0$ is time. We will
consider this problem only for distributions with a causal support: Let
$d(x)=d(x_1,\ldots x_n)\in \mathcal{S}' (\mathds{R}^{m})$ where $x_j\in \mathds{R}^{4}$,
$m=4n$, be a tempered distribution depending on $n$ space-time arguments. By
\be
\overline{V^+}(0)=\{ x\, |\, x^2=x_0^2-\vec x^2\ge 0\> ,\> x_0\ge 0\}
\label{(37)}
\ee
we denote the closed forward cone, and by
\be
\overline{V^-}(0)=\{ x\, |\, x^2\ge 0\> ,\> x_0\le 0\ \}
\ee
the closed backward cone. The $n$-dimensional generalizations are 
\be
\Gamma_n^{\pm}(0)=\{ (x_1,\ldots x_n)\, |\, x_j\in\overline{V^\pm}(0)\,
 ,\,\forall j=1,\ldots n\} .\label{(38)}
\ee
The distribution $d(x)$ has causal support if
\be
\supp d\subseteq\Gamma _n^+(x)\cup\Gamma _n^-(0).\label{(39)}
\ee
This means that all $n$ space-time points are either in the forward
light-cone or in the backward cone. The splitting problem now is to decompose
such a distribution into a retarded minus advanced part
\be
d(x)=r(x)-a(x),\label{(40)}
\ee
where $\supp r\subseteq\Gamma_n^+$ and $\supp a\subseteq\Gamma_n^-$.

The simplest example of a causal distribution is the Jordan-Pauli distribution given by
Eq. (\ref{(32)}) where the splitting into retarded minus advanced distributions is trivially
possible, see Eq. (\ref{(8)}). It is misleading that the Feynman propagator $D_F^m(x)$ is also
called "causal" sometimes, because it does not have a causal support due
to the presence of $D^{(-)}_m$ in Eq. (\ref{(9)}). In the general case we have to 
distinguish two cases:

a) Singular order $\omega <0$: In this case, the power-counting function goes to
infinity by Eq. (\ref{(31)})
\be
\rho(\delta)\to\infty\quad {\rm for}\quad\delta\to 0.
\ee
This implies
\be
\Bigl\langle d(x),\,\fii\Bigl({\frac{x}{\delta}}\Bigl)\Bigl\rangle\to
{\langle d_0,\,\fii\rangle\over\rho(\delta)}\to 0.\label{(42)}
\ee
We choose a monotonous $C^\infty$-function $\chi_0$ over $\mathds{R}^1$
with
\be
\chi_0(t)=
\begin{cases}
0 & \, \mbox{for} \quad t\le 0\\ <1 &\, \mbox{for} \quad 0<t<1\\ 1 & \, \mbox{for} \quad t\ge 1.
\end{cases}
\label{(43)}
\ee
In addition we choose a vector $v=(v_1,\ldots v_{n-1})\in\Gamma^+$, which
means that all four-vectors $v_j$ are inside the forward cone $V^+$.
Then
\be
v\cdot x=\sum_{j=1}^{n-1}v_j\cdot x_j=0\label{(44)}
\ee
is a space-like hyperplane that separates the causal support: All
products $v_j\cdot x_j$ are either $\ge 0$ for $x\in\Gamma^+$ or $\le 0$ for
$x\in\Gamma^-$. Then as a consequence of Eq. (\ref{(42)}) the limit
\be
\lim_{\delta\to 0}\chi_0\Bigl({v\cdot x\over\delta}\Bigl)d(x)\=d
\Theta(v\cdot x)d(x)=r(x)\label{(45)}
\ee
exists. This is the case of trivial splitting where the multiplication by
step function is possible. The result is independent of $v$.

b) $\omega \ge 0$: Now the power-counting function satisfies
\be
{\rho(\delta)\over\delta^{\omega+1}}\to\infty\quad{\rm for}
\quad\delta\to 0.\label{(46)}
\ee
To get a vanishing scaling limit as in Eq. (\ref{(42)}) we choose a multi-index
$b$ with $|b|=\omega+1$ and consider
\be
\langle d(x)x^b,\,\psi({\frac{x}{\delta}})\rangle =\langle d(\delta y)
y^b,\psi(y)\rangle\delta^{m+\omega+1}
\to\langle d_0(y),\,y^b\psi\rangle{\delta^{\omega+1}\over\rho(\delta
)}\to 0.\label{(47)}
\ee
It follows that the splitting as in case a) is possible if the test function
$\fii$ satisfies
\be
D^a\fii (0)=0\quad {\rm for}\quad |a|\le\omega .\label{(48)}
\ee
To achieve that, we introduce an auxiliary function $w(x)\in {\cal S}  
(\mathds{R}^m)$ with
\be
w(0)=1\> ,\> D^aw(0)=0\quad {\rm for}\quad 1\le |a|\le\omega ,
\label{(49)}
\ee
and define
\be
(W\fii )(x)\=d\fii (x)-w(x)\sum_{|a|=0}^\omega {x^a\over a!}
(D^a\fii )(0)\label{(50)}
=\sum_{|b|=\omega+1}x^b\psi_b(x).
\ee
The function $w(x)$ serves for the purpose of getting rapid decrease for
$|x|\to\infty$. Now the decomposition according to a) Eq. (\ref{(45)}) is possible
\be
\langle r(x),\,\fii\rangle\=d\langle \Theta(v\cdot x)d,\,W\fii\rangle ,
\label{(51)}
\ee
\be
a(x)=r-d.
\ee
After construction $r(x)$
defines a tempered distribution with $\supp r\subseteq\Gamma ^+(0)$. 
It agrees with $d(x)$ on $\Gamma^+(0)\setminus\{ 0\}$ in the
sense of distributions, because
a test function $\fii\in \mathcal{S}$ with $\supp\fii\subset\Gamma^+(0)
\setminus\{ 0\}$ vanishes at $x=0$, together with all its derivatives,
so that the additional subtracted terms in Eq. (\ref{(50)}) are 0. But without
these terms, there is no splitting of $d(x)$ which makes sense for
arbitrary $\fii\in \mathcal{S}$, because the limit Eq. (\ref{(45)}) exists 
on subtracted test functions only. 
{\it If one does the splitting incorrectly
by simple multiplication with $\Theta (v\cdot x)$ as in a), 
one is punished by the well-known ultraviolet divergences in quantum field theory.} 
As we will discuss in detail later, these divergences
appear in loop graphs which have $\omega\ge 0$. For those graphs the
naive splitting with $\Theta(v\cdot x)$ is impossible and, as a
consequence, the Feynman rules do not hold.

Again we have 
\be
\omega(r)=\omega(d)=\omega(a),\label{(52)}
\ee
This is a direct consequence of the definitions Eq. (\ref{(50)}) and Eq. (\ref{(51)}), because the
limit
\bea
\lim_{\delta\to 0}\rho (\delta)\Bigl\langle r(x),\,\fii\Bigl(
{\frac{x}{\delta}}\Bigl)\Bigl\rangle =\lim_{\delta\to 0}\rho (\delta)
\Bigl\langle d(x),\,\Theta W\Bigl(\fii\Bigl({\frac{x}{\delta}}\Bigl)\Bigl)
\Bigl\rangle
\eea
\be
=\lim_{\delta\to 0}\rho (\delta)\langle d(x),\,(\Theta W\fii)
({\frac{x}{\delta}})\rangle =\langle d_0(x),\,(\Theta W\fii)(x)\rangle
\ee
exists with the same power counting function as $d(x)$.
But in sharp contrast to case a), the splitting b) is not unique. 
If $\tilde r(x)$ is the retarded part of another decomposition, then the
difference 
\be
\tilde r-r=\sum_{|a|=0}^\omega\tilde C_a D^a\delta
(x)\label{(53)}
\ee
is again a distribution with point support. Since $\omega>0$, this time
the splitting is only determined up to a finite sum of local terms according to
Eq. (\ref{(53)}).
These undetermined local terms are not fixed by causality, additional
physical normalization conditions are necessary to fix them.

Before we proceed, it might help to provide some intuitive understanding of
the distribution splitting process. One should remember the fact that the distributions
appearing in local quantum field theory are more singular than ordinary
functions, such that the products of the distributions are not necessarily
well defined \emph{ab initio}. E.g., the Feymnan propagator can be calculated
in configuration space \cite{Bogol}
\be
D_F^m(x)=\frac{1}{4 \pi} \delta(x^2) - \frac{m}{8 \pi \sqrt{x^2}} \Theta(x^2)
[J_1(m \sqrt{x^2})- iN_1(m \sqrt{x^2})] + \frac{im}{(2 \pi)^2 \sqrt{-x^2}}
\Theta(-x^2) K_1(m \sqrt{-x^2}), \label{feyn_config}
\ee
where $J_1$, $N_1$, and $K_1$ are Bessel functions. For $x^2 \sim 0$,
Eq. (\ref{feyn_config}) can be decomposed according to
\be
D_F^m(x)=\frac{1}{(2 \pi)^2} \frac{1}{x^2-i0} + O(m^2 x^2)= D_F^0(x) + O(m^2 x^2).
\ee
A formal product like $D_F^m(x)D_F^m(x)$ contains the highly singular (formal) expression
$1/(x^2-i0)^2 \sim 1/x^4$, and it is not trivial to understand the precise meaning of such a
singular expression in the vicinity of the light-cone where $x^2 \sim 0$.
In general, the definition of distributional products works better in momentum space,
where the analytic behavior of distributions appearing in quantum field theory is smoother
and where the product goes over into a convolution.
Still, the true difficulty is located in the point $x = 0$, where the distributional
behavior of the product of two Feynman propagators is no longer mathematically meaningful.
In momentum space, this problem leads to a logarithmically divergent integral.
This is the point where the causal method provides the well-defined tools to isolate
this ill-defined part of the product from the regular part on $\mathds{R}^4 \backslash \{0\}$.
Generally, perturbation theory itself is unable to describe local, "zero-distance" interactions
without further input. At least, if the mathematics is done correctly and distributions are
treated correctly, then all results remain finite at every calculational step, i.e. well-defined.
The point $x=0$ is essential because the distributions are most singular there. The subtracted
terms have no direct physical meaning because they remain with free parameters,
this is the freedom of (finite) renormalization, which is discussed in further detail in
the sequel.

For practical reasons, explicit calculations in quantum field theory are usually done 
in momentum space. As a natural consequence, we must investigate the splitting procedure in $p$-space. 
We need the distributional Fourier transforms
\be
\mathcal{F}^{-1}[\Theta(v\cdot x)]\=d\check\chi (k)\label{(54)}
\ee
\be
\mathcal{F}^{-1}[x^aw](p)=(iD_p)^a\check w(p).\label{(55)}
\ee
Since
\bea
(D^a\fii )(0)=\langle (-)^aD^a\delta ,\,\fii\rangle =(-)^a\langle
\widehat{D^a\delta},\,\check\fii\rangle
\eea
\be
=(-)^a(2\pi )^{-m/2}\langle (-ip)^a,\,\check\fii\rangle =(2\pi )
^{-m/2}\langle (ip)^a,\,\check\fii\rangle ,\label{(56)}
\ee
we conclude from Eq. (\ref{(51)}) that 
\bea
\langle \hat r,\,\check\fii\rangle =\langle\hat d,\,(\Theta W\fii)
\check{\phantom I}\rangle =(2\pi )^{-m/2}\biggl< \hat d,\,
\check\chi\ast\Bigl[\check\fii-
-\sum_{|a|=0}^\omega {1\over a!}(iD_p)^a\check w(p)
(2\pi )^{-m/2}\langle (ip ' )^a,\,\check\fii\rangle\Bigl]\biggl>
_p\label{(57)}
\eea
\be
=(2\pi)^{-m/2}\Bigl\langle\hat\chi\ast\hat d,\,\check\fii-\sum_{|a|=0}^\omega 
\ldots\>\Bigl\rangle ,\label{(58)}
\ee
where the asterisk means convolution. We stress the fact that the
convolution $\hat\chi\ast\hat d$ is only defined on subtracted test
functions, not on $\check\fii$ alone. Interchanging $p'$ and $p$ in the
subtraction terms, we may write
\be
\langle\hat r,\,\check\fii\rangle =(2\pi)^{-m/2}\int dk\,\hat\chi (k)\Bigl
\langle\hat d(p-k)
-(2\pi)^{-m/2}\sum_a{(-)^a\over a!}p^a\int dp'\,\hat d(p'-k)D^a_{p'}
\check w(p'),\,\check\fii\Bigl\rangle_p.\label{(59)}
\ee
After partial integration in the $p'$-integral this is equivalent to the
following result for the retarded distribution
\be
\hat r(p)=(2\pi )^{-m/2}\int dk\,\hat\chi (k)\Bigl[\hat d(p-k)
-(2\pi )^{-m/2}\sum_{|a|=0}^\omega{p^a\over a!}\int dp '\,\bigr(D^a
_{p'}\hat d(p ' -k)\bigr)\check w(p ' )
\Bigl].\label{(60)}
\ee
Here the $k$-integral is understood in the sense of distributions as in
(59). 

By considering the Fourier transform of Eq. (\ref{(53)}) we see that $\hat r(p)$ is only
determined up to a polynomial in $p$ of degree $\omega$. Consequently the
general result for the retarded distribution reads
\be
\hat{\tilde r}(p)=\hat r(p)+\sum_{|a|=0}^\omega C_ap^a\label{(61)}
\ee
with $\hat r(p)$ given by Eq. (\ref{(60)}). We now assume that there exists a
point $q\in \mathds{R}^m$ where the derivatives $D^b\hat r(q)$ exist in
the usual sense of functions for all $|b|\le\omega$. Let us define
\be
\hat r_q(p)=\hat r(p)-\sum_{|b|=0}^\omega {(p-q)^b\over b!}D^b
\hat r(q).\label{(62)}
\ee
This is another retarded distribution because we have only added a
polynomial in $p$ of degree $\omega$. Furthermore, this solution of the
splitting problem is {\it uniquely} specified by the normalization
condition
\be
D^b\hat r_q(q)=0,\quad |b|\le\omega.\label{(63)}
\ee
We compute
\bea
D^b\hat r(q)=(2\pi)^{-m/2}\int dk\,\hat\chi (k)\Bigl[\bigr(D^b\hat d
\bigr)(q-k)
\eea
\be
-(2\pi)^{-m/2}\sum_{b\le a}{a!q^{a-b}\over (a-b)!a!}\int dp'\,
\check w(p')D^a_{p'}\hat d(p'-k)\Bigl]\label{(64)}
\ee
from Eq. (\ref{(60)}) and substitute this into Eq. (\ref{(62)}). Since
\be
\sum_{b\le a}{(p-q)^b\over b!}{q^{a-b}\over (a-b)!}={1\over a!}
\sum_{b\le a}{a\choose b}(p-q)^bq^{a-b}={p^a\over a!},
\ee
the subtracted terms in Eq. (\ref{(60)}) drop out
\be
\hat r_q(p)=(2\pi)^{-m/2}\int dk\,\hat\chi (k)\Bigl[\hat d(p-k)-
\sum_{|b|=0}^\omega {(p-q)^b\over b!}\bigr(D^b\hat d\bigr)(q-k)
\Bigl].\label{(65)}
\ee
This is the splitting solution with normalization point $q$. It is
uniquely specified by Eq. (\ref{(63)}), that means it does not depend on the
time-like vector $v$ in Eq. (\ref{(54)}). The subtracted terms are the beginning
of the Taylor series at $p=q$. This is an ultraviolet "regularization"
in the usual terminology. It should be stressed, however, that here this
is a consequence of the causal distribution splitting and not an ad hoc
recipe.

It is well-known that causality can be expressed in
momentum space by dispersion relations. Therefore we look for a
connection of the result Eq. (\ref{(65)}) with dispersion relations. We take
$q=0$ in Eq. (\ref{(65)}), which is possible if all fields are massive, for 
example, and
consider time-like $p\in\Gamma^+$. We choose a special coordinate system
such that $p=(p_1^0,\vec 0, 0,\ldots)$. Note that this coordinate system
is not obtained by a Lorentz transformation from the original one, but
by an orthogonal transformation in $\mathds{R}^m$.
Furthermore we take $v$ parallel
to $p$, i.e. $v=(1,\vec 0, 0,\ldots)$. Then $v$ varies with $p$, but
this is admissible because Eq. (\ref{(65)}) is actually independent of $v$.
We now have $\Theta (v\cdot x)=\Theta (x_1^0)$ and the Fourier transform
(54) is given by
\be
\hat\chi(k)=(2\pi)^{m/2-1}\delta (\vec k_1,k_2,\ldots k_{n-1}) 
{i\over k_1^0+i0}.
\ee
We always use the mathematical notation $i0$ for $i\eps$ with the
subsequent distributional limit $\eps\to 0$.
Using this result in Eq. (\ref{(65)}) we shall obtain
\bea
\hat r_0(p_1^0)={i\over 2\pi}\int\limits_{-\infty}^{+\infty} 
dk_1^0\,{1\over k_1^0+i0}\biggl[\hat d(p_1^0-k_1^0,0,\ldots )
\eea
\be
-\sum_{a=0}^\omega {(p_1^0)^a\over a!}(-)^aD_{k_1^0}^a\hat
d(q_1^0-k_1^0,0,\ldots) \vtop{
\hbox{$\Bigl\vert$}\hbox{$\scriptstyle q_1^0=0$}}\biggl].\label{(66)}
\ee

The transformation of this result to the usual form of a dispersion
integral leads to the following result:
\be
r_0(p_1^0)={i\over 2\pi}(p_1^0)^{\omega +1}\int\limits_{-\infty}^{+\infty} 
dk_0\,{\hat d(k_0)\over (k_0-i0)^{\omega +1}(p_1^0-k_0+i0)}.\label{(67)}
\ee
The proof is given in \cite{Scharf2}, proposition 2.4.1.
This expression is a subtracted dispersion relation. 
To write down the result for arbitrary $p\in\Gamma^+$, we use
the variable of integration $t=k_0/p_1^0$ and arrive at
\be
\hat r_0(p)={i\over 2\pi}\int\limits_{-\infty}^{+\infty} 
dt\,{\hat d(tp)\over (t-i0)^{\omega +1}(1-t+i0)}. \label{(68)}
\ee
For later reference we call this the central splitting solution, because
it is normalized at the origin ($q=0$ in Eq. (\ref{(63)})). The latter fact
has two important consequences. (i) The central splitting 
solution does not introduce a new mass scale into the theory. If $q\ne 0$,
then $|q^2|=M^2$
defines such a scale. (ii) Most symmetry properties of $\hat d(p)$ are
preserved under central splitting, as we will see later, because the
origin $q=0$ is a very symmetrical point. In the self-energy computation
in sect. 2.1 by means of Pauli-Villars regularization we have also calculated
this central solution Eq. (\ref{(19)}).

It is easy to verify that the dispersion integral 
Eq. (68) is convergent for $|t|\to\infty$. But it would be
ultraviolet divergent, if $\omega$ in Eq. (\ref{(68)}) is chosen too small. 
Consequently, {\it the correct distribution splitting with the right singular
order $\omega$ is terribly important. Incorrect distribution splitting
leads to ultraviolet divergences.} This is the origin of the
ultraviolet problem in quantum field theory. 

\section{Perturbative S-Matrix Theory}

In perturbation theory all quantities are expanded in terms of free fields.
To decide which free fields are relevant we notice that all interactions
in nature can be described by quantum gauge theories, {\it gravity
included}. Therefore, it is important to discuss quantized free gauge fields
and their gauge structure. The latter is defined by means of ghost fields.
In contrast to the functional integral approach to QFT where the Faddeev-
Popov ghosts play indeed a somewhat ghost-like r\^{o}le, these are genuine dynamical 
fields in the causal approach. Regarding regularization it is a subtle
problem to perform it in a way such that gauge invariance is conserved.
In this respect dimensional regularization is technically advantageous.

\subsection{\it Free Fields \label{sec:free}}
\subsubsection{\it Scalar Fields \label{sec:scalarfield}}
First we consider a neutral scalar field with mass $m$ which is a solution
of the Klein-Gordon equation
\be
(\sq +m^2)\fii(x)=0.\label{(1.1.1)}
\ee
A real $classical$ solution of this equation is given by
\be
\fii(x)=\pdh\int{d^3p\over\sqrt{2E}}\Bigl(a(\vec p)e^{-ipx}+a^*(\vec p)
e^{ipx}\Bigl),\label{(1.1.2)}
\ee
where
\be
px=p^0x^0-\vec p\cdot\vec x=p_\mu x^\mu,\quad E=+\sqrt{\vec p^2+m^2}
=p^0.\label{(1.1.3)}
\ee
In quantum field theory $a(\vec p)$ and $a^*(\vec p)$ become
operator-valued distributions satisfying the commutation relations
\be
[a(\vec p),a^\dagger(\vec q)]=\delta^{(3)}(\vec p-\vec q).\label{(1.1.7)} 
\ee
where $\delta^{(3)}$ denotes the Dirac's $\delta$-distribution,
all other commutators vanish. The quantized Bose field is now given by
\be
\fii(x)=\pdh\int{d^3p\over\sqrt{2E}}\Bigl(a(\vec p)e^{-ipx}+a^\dagger(\vec p)
e^{ipx}\Bigl).\label{(1.1.8)}
\ee
The cross denotes the Hermitian conjugate. In fact, there exists a Fock-Hilbert
space representation of the Bose field which proves the consistency of the
quantization. $\fii$ is obviously Hermitian $\fii^\dagger(x)=\fii(x).$

Let us call the second term in Eq. (\ref{(1.1.8)}) involving $a^\dagger$ the creation part
$\fii^{(+)}$ and the first term with $a(\vec p)$ the absorption part
$\fii^{(-)}$. Then by Eq. (\ref{(1.1.7)}) their commutator is equal to
\be
[\fii^{(-)}(x),\fii^{(+)}(y)]=(2\pi)^{-3/2}\int{d^3p\over 2E} e^{-ip(x-y)}
=-iD_m^{(+)}(x-y).\label{(1.1.10)}
\ee
In the same way we get
\be
[\fii^{(+)}(x),\fii^{(-)}(y)]=-iD_m^{(-)}(x-y).
\ee
Then the commutation relation for the total scalar field is given by
the Jordan-Pauli distribution
\be
[\fii(x),\fii(y)]=-iD_m(x-y).\label{(1.1.13)}
\ee

The charged scalar field involves a slight generalization of the neutral one:
\be 
\fii(x)=\pdh\int{d^3p\over\sqrt{2E}}\Bigl(a(\vec p)e^{-ipx}+b^\dagger(\vec p)
e^{ipx}\Bigl).\label{(1.1.47)}
\ee
It contains two different kinds of particles whose absorption and
emission operators satisfy
\be 
[a(\vec p),a^\dagger(\vec q)]=\delta(\vec p-\vec q)=[b(\vec p),b^\dagger(\vec q)
]\label{(1.1.48)}
\ee
and all other commutators vanish. Then it follows
\be 
[\fii(x),\fii(y)^\dagger]=-iD_m(x-y)\label{(1.1.49)}
\ee
but $[\fii(x),\fii(y)]=0$.

\subsubsection{\it Spin-1/2 Fields \label{sec:fermifield}}
Spin-1/2 fields are needed to describe leptons and quarks.
Spinor fields are solution of the Dirac equation
\be 
i\gamma^\mu\d_\mu\psi(x)=m\psi(x).\label{(1.8.1)}\ee 
The $\gamma$-matrices obey the anticommutation relation
\be \gamma^\mu\gamma^\nu+\gamma^\nu\gamma^\mu=2g^{\mu\nu}.\label{(1.8.2)}\ee
To define the quantized Dirac field we consider a solution of Eq. (\ref{(1.8.1)}) of
the following form
\be \psi(x)=\i3p\sum_{s=\pm 1}[u_s(\vec p)e^{-ipx}b_s(\vec p)+
v_s(\vec p)e^{ipx}d_s^\dagger(\vec p)],\label{(1.8.3)}\ee
\be \=d \psi^{(-)}+\psi^{(+)}.\ee
The $u$- and $v$-spinors herein are obtained from the Fourier
transformed equations
\be (p_\mu\gamma^\mu-m)u_s(\vec p)=0\ee
\be (p_\mu\gamma^\mu+m)v_s(-\vec p)=0,\label{(1.8.4)}\ee
with the normalization
\be u_s^\dagger(\vec p)u_{s'}(\vec p)=\delta_{ss'}=v_s^\dagger(\vec p)v_{s'}(\vec p)
\label{(1.8.5)}\ee
\be u_s^\dagger(\vec p)v_{s'}(-\vec p)=0=v_s^\dagger(-\vec p)u_{s'}(\vec p)\ee
\be u_s^\dagger(\vec p)\gamma^0 u_{s'}(\vec p)={m\over E}\delta_{ss'}=
-v_s(\vec p)^\dagger\gamma^0 v_{s'}(\vec p).\label{(1.8.6)}\ee
The $u$- and $v$-spinors span the positive and negative spectral subspaces
of the Dirac operator, respectively, which are defined by the projection
operators 
\be P_+(\vec p)=\sum_s u_s(\vec p)u_s^\dagger(\vec p)=\Bigl({\ps +m\over 2E}
\Bigl)\gamma^0\label{(1.8.7)}\ee
\be P_-(\vec p)=\sum_s v_s(-\vec p)v_s^\dagger(-\vec p)=\Bigl({\ps -m\over 2E}
\Bigl)\gamma^0,\label{(1.8.8)}\ee
where
\be \ps=p_\mu\gamma^\mu,\quad p_0=E=\sqrt{\vec p^2+m^2}.\label{(1.8.9)}\ee
The projections are orthogonal
\be P_\pm(\vec p)^2=P_\pm(\vec p),\quad P_+(\vec p)+P_-(\vec p)={\bf 1}.
\label{(1.8.10)}\ee

The quantization of the Dirac field is easily achieved by
considering the $b$'s and $d$'s as operator-valued distributions
satisfying the anticommutation relations
\be 
\{b_s(\vec p),b_{s'}^\dagger(\vec q)\}=\delta_{ss'}\delta^{(3)}(\vec p-
\vec q)=\{d_s(\vec p),d_{s'}^\dagger(\vec q)\},\label{(1.8.11)}\ee
and all other anticommutators vanish. We do not treat the Majorana case
for neutral spin-1/2 fermions here. Then, the operators $b$ and $d$ 
can be interpreted as annihilation and their adjoints $b^\dagger, d^\dagger$ as
creation operators and the Fock space can be constructed from a unique
vacuum in the usual way (\cite{Scharf}, sect. 2.2). To get the anticommutation
relations for the whole Dirac field we need the adjoint Dirac field
\be \psi ^\dagger(x)=\i3p\, [b^\dagger_s(\vec p)u_s(\vec p)^\dagger e^{ipx}+d_s(\vec p)v_s
(\vec p)^\dagger e^{-ipx}].\ee
Multiplying by $\gamma ^0$, we get the so-called Dirac adjoint
\be \psq (x)=\psi ^\dagger(x)\gamma ^0=\pqp +\pqm\ee
\be \pqp =\i3p\, b_s^\dagger(\vec p)\overline u_s(\vec p)e^{ipx}\ee
\be \pqm (x)=\i3p\,d_s(\vec p)\overline v_s(\vec p)e^{-ipx}.\label{(1.8.12)}\ee
With the aid of Eq. (\ref{(1.8.11)}) we find
\be \{\psm _a(x),\, \pqp _b(y)\} =(2\pi )^{-3}\int d^3p\, u_{sa}(\vec p)
\uq _{sb}(\vec p)e^{-ip(x-y)}.\label{(1.8.13)}\ee
In the result Eq. (\ref{(1.8.13)}), the covariant positive spectral projection
operator Eq. (\ref{(1.8.7)}) appears  
\be \{\psm (x),\, \pqp (y)\} =(2\pi )^{-3}\int {d^3p\over 2E}(\ps +m)
e^{-ip(x-y)}\=d {1\over i}S^{(+)}_m(x-y)\label{(1.8.14)}\ee
\be =-i(i\ds +m)D_m^{(+)}(x-y).\ee
In the same way, one obtains the other non-vanishing anticommutator
\be \{ \psp (x),\,\pqm (y)\}\=d {1\over i}S^{(-)}_m(x-y)=(2\pi )^{-3}\int
 {d^3p\over 2E}(\ps -m)e^{ip(x-y)}\label{(1.8.15)}\ee
\be =-i(i\ds +m)D_m^{(-)}(x-y).\ee
This gives the anticommutation relation for the total Dirac field
\be \{\psi (x),\,\psq (y)\}=\frac{1}{i} S_m(x-y),\label{(1.8.16)}\ee
with
\be S_m(x)=S^{(-)}_m(x)+S^{(+)}_m(x)=(i\ds +m)D_m(x-y).\label{(1.8.17)}\ee
The anticommutators between two $\psi$'s and two $\psq$'s vanish.

\subsubsection{\it Vector Fields \label{sec:vectorfield}}
Next we consider the massless vector field which obeys the wave equation
$\sq A^\mu(x)=0$. Its Fourier decomposition reads
\be 
A^\mu (x)=\pdh\io\Bigl(a^\mu (\vec k)e^{-ikx}+a^\mu (\vec k)^\dagger
e^{ikx}\Bigl),\label{(1.3.7)}
\ee
and it is quantized in Lorentz-invariant form according to
\be 
[A^\mu (x),\, A^\nu (y)]=g^{\mu\nu}iD_0(x-y).\label{(1.3.14)}
\ee
We need also the commutators of the absorption and emission parts alone
\be
[A^\mu _-(x),\, A^\nu _+(y)]=g^{\mu\nu}iD^{(+)}_0(x-y),
\ee
\be 
[A^\mu _+(x),\, A^\nu _-(y)]=g^{\mu\nu}iD^{(-)}_0(x-y).\label{(1.3.18)}
\ee
We are working in the Feyman gauge for the sake of convenience and covariance.
However, asymptotic massless spin 1 particles only have two polarization degrees of freedom.
Consequently, the four polarization types of emission operators introduced
above create unphysical particle states, and the space of physical states is
a subspace of the full Fock-Hilbert space.

This observation is closely related to the issue of gauge transformations, therefore
we also comment here on gauge transformations in perturbative quantum field theory.
The massless vector fields describe non-interacting photons and gluons in the standard model.
In classical electrodynamics the vector potential can be changed by a gauge transformation
\be
A'^\mu(x)=A^\mu(x)+\lambda\d^\mu u(x),\label{(1.4.1)}
\ee
where $u(x)$ is assumed to fulfill the wave equation
$\sq u(x)=0$ because we want the transformed field $A'^\mu(x)$ to satisfy the wave
equation also. In QFT, the quantized $A'^\mu(x)$ should fulfill the same
commutation relations Eq. (\ref{(1.3.14)}) as $A^\mu(x)$. This is true if the gauge
transformation Eq. (\ref{(1.4.1)}) is of the following form
\be 
A'^\mu(x)=e^{-i\lambda Q}A^\mu(x)e^{i\lambda Q},\label{(1.4.3)}
\ee 
where $Q$ is some operator in the Fock-Hilbert space. Expanding this by means 
of the Lie series
\be 
=A^\mu(x)-i\lambda [Q,A^\mu(x)]+O(\lambda^2).\ee 
and comparing with Eq. (\ref{(1.4.1)}) we conclude
\be 
[Q,A^\mu(x)]=i\d^\mu u(x).\label{(1.4.5)}
\ee

The operator $Q$ will be called gauge charge because it is the
infinitesimal generator of the gauge transformation defined by Eq. (\ref{(1.4.1)}). For the
following it is important to have $Q$ nilpotent
\be
Q^2=0.\label{(1.4.9)}
\ee 
The important consequence of this property is the fact that the factor space
$F_{\rm ph}={\rm Ker}\, Q/\overline{{\rm Ran}\, Q}$ is isomorphic to the subspace of physical
states. Here, Ran is the range and Ker the kernel of the operator $Q$. The overline denotes
the closure; note that Ran $Q$ is not closed because $0$ is in the essential spectrum of $Q$.
We will not discuss this in detail but refer to the literature \cite{Scharf2}.
Such a nilpotency according to Eq. (\ref{(1.4.9)}) is characteristic for Fermi operators.
Therefore, we assume $u(x)$ to be a fermionic scalar field with mass zero, a
so-called ghost field. This field has the following Fourier decomposition
\be 
u(x)=\pdh\int{d^3p\over\sqrt{2E}}\Bigl(c_2(\vec p)e^{-ipx}+c_1(\vec p)
^\dagger e^{ipx}\Bigl).\label{(1.2.1)}\ee
In addition, we introduce a second scalar field
\be 
\tilde u(x)=\pdh\int{d^3p\over\sqrt{2E}}\Bigl(-c_1(\vec p)e^{-ipx}+ 
c_2(\vec p)^\dagger e^{ipx}\Bigl).\label{(1.2.2)}\ee
The absorption and emission operators
$c_j$, $c_k^\dagger$ obey the anticommutation relations
\be 
\{c_j(\vec p),c_k(\vec q)^\dagger\}=\delta_{jk}\delta^{(3)}(\vec p-\vec q).
\label{(1.2.3)}\ee
Some remarks are in order here. Firstly, also bosonic fields would do the job in
the case of an abelian theory like quantum electrodynamics (QED). However, non-abelian
gauge theories like quantum chromodynamics (QCD) require fermionic ghosts.
In order to avoid any conflict with the spin-statistics theorem, states containing
ghosts necessarily do not belong to the physical sector of the Fock-Hilbert space
of the theory under consideration. Secondly, when the spin-1 fields become massive,
as it is the case in the standard model for the $W^\pm$- and $Z$-boson fields, the corresponding
ghost fields also become massive. The ghost mass then depends on the chosen formalism
(i.e., the gauge fixing, \cite{ADS}). In the following, we allow the ghost fields to be massive,
but $m=0$ holds true whenever the related vector fields are massless. The formalism
used below is chosen so that the ghost mass is equal to the vector boson mass. 

Again, the absorption and emission parts (with the adjoint operators) are
denoted by (-) and (+). They satisfy the following anticommutation relations
\be 
\{u^{(-)}(x),\tilde u^{(+)}(y)\}=(2\pi)^{-3}\int {d^3p\over 2E}\,e^{-ip(x-y)}=
-iD_m^{(+)}(x-y)\ee
\be 
\{u^{(+)}(x),\tilde u^{(-)}(y)\}=-(2\pi)^{-3}\int {d^3p\over 2E}\,e^{ip(x-y)}=
-iD_m^{(-)}(x-y).\label{(1.2.5)}\ee
All other anticommutators vanish. This implies
\be 
\{u(x),\tilde u(y)\}=-iD_m(x-y)\label{(1.2.6)}\ee
and $\{u(x), u(y)\}=0$.
Then it is not hard to verify that the nilpotent gauge charge $Q$
satisfying Eq. (\ref{(1.4.5)}) is given by
\be 
Q=\int d^3x\,[\d_\nu A^\nu\d_0u-(\d_0\d_\nu A^\nu)u]
\=d\int d^3x\,\d_\nu A^\nu{\dl}_0u\label{(1.4.6)}
\ee
where the integrals are taken over any plane $x^0={\rm const.}$

Now we return to the defining property of $Q$ as being the infinitesimal
generator of gauge transformations given by Eq. (\ref{(1.4.3)}) and Eq. (\ref{(1.4.5)}). 
We introduce the notation
$$d_Q F=[Q,F],$$
if $F$ contains only Bose fields and an even number of ghost fields,
and
\be 
d_Q F=\{Q,f\}=QF+FQ,\label{(1.4.27)}\ee
if $F$ contain an odd number of ghost fields. Then $d_Q$ has all
properties of an anti-derivation, in particular the identity
\be 
\{AB,C\}=A\{B,C\}-[A,C]B\label{(1.4.10)}\ee 
implies the product rule
\be d_Q(F(x)G(y))=(d_QF(x))G(y)+(-1)^{n_F}F(x)d_QG(y),\label{(1.4.28)}\ee
where $n_F$ is the ghost number of $F$, i.e. the number of $u$'s minus
the number of $\tilde u$'s. The gauge variations $d_Q$ of our free fields 
now are
\be d_Q A^\mu=i\d^\mu u,\quad d_QA_\pm^\mu=i\d^\mu u_\pm \label{(1.4.29)}\ee
\be d_Q u=0,\quad d_Q\tilde u=\{Q,\tilde u\}=-i\d_\mu A^\mu,\quad 
d_Q\tilde u_\pm=-i\d_\mu A_\pm^\mu.\label{(1.4.30)}\ee
The latter follows from the anticommutation relation Eq. (\ref{(1.2.6)}).
$d_Q$ changes the ghost number by one, i.e. a Bose field
goes over into a Fermi field and vice verse. Then the nilpotency $Q^2=0$
implies for a Bose field $F_B$
$$d_Q^2F_B=\{Q,[Q,F_B]\}=Q(QF_B-F_BQ)+(QF_B-F_BQ)Q=0,$$ 
and for a Fermi field $F$
$$d_Q^2F=[Q,\{Q,F\}]=Q(QF_B+F_BQ)-(QF_B-F_BQ)Q=0,$$
hence
$$d_Q^2=0\eqno(1.4.31)$$
is also nilpotent. The gauge variation $d_Q$ has some similarity with the
Becchi-Rouet-Stora-Tyutin (BRST) transformation \cite{BRS,Tyutin}
in the functional approach. However, the BRST transformation
operates on interacting fields (mainly classical) and the quantum gauge
invariance which we are going to define is completely different from
BRST invariance.

Now we consider massive vector fields.
These fields will be used to represent the $W^\pm$- and $Z$-bosons
of the electroweak theory, for example. They obey the Klein-Gordon
equation
\be 
(\sq+m^2)A^\mu(x)=0.\label{(1.5.1)}\ee
Since a spin-1 field has three physical degrees of freedom, we need 
one subsidiary condition to define unphysical states. As this we can
choose the Lorentz condition
\be 
\d_\mu A^\mu(x)=0.\label{(1.5.2)}\ee
The commutation relations are similar to the massless case,
for example (see Eq. (\ref{(1.3.14)}))
\be 
[A^\mu (x),\, A^\nu (y)]=g^{\mu\nu}iD_m(x-y).\label{(1.5.11)}\ee
where only the massless $D$-distributions must be substituted by massive ones.

As in the massless case
we would like to characterize the physical subspace with help of a
nilpotent gauge charge $Q$. The old definition given by Eq. (\ref{(1.4.6)}) does not work
because nilpotency is violated:
\be 
Q^2=\eh\{ Q,Q\}=\eh i\int d^3x\,(\sq u){\dl}_0u=-\eh im^2\int d^3x\,
u{\dl}_0u\ne 0.\ee
To restore it we modify the expression for $Q$ by
introducing a scalar field $\Phi(x)$ with the same mass $m$ as the gauge
field $A^\nu(x)$
\be 
Q=\int d^3x\,(\d_\nu A^\nu+m\Phi){\dl}_0u.\label{(1.5.31)}\ee
All fields satisfy the Klein-Gordon equation
\be (\sq+m^2)\Phi=0\ee
\be (\sq+m^2)u=0,\label{(1.5.33)}
\ee
but, while $u(x)$ is a Fermi field, $\Phi(x)$ is quantized with
commutation relations
\be [\Phi(x),\Phi(y)]=-iD_m(x-y),\label{(1.5.34)}\ee
and all other commutators are the same as before. Now we can check
the nilpotency:
\be Q^2=-\eh\int d^3x\,[\d_\nu A^\nu+m\Phi,Q]=0,\ee
because the first term in the commutator gives $-i\sq u$ and the second
one $-im^2u$, so that the sum is zero by Eq. (\ref{(1.5.33)}). The infinitesimal
gauge transformations or gauge variations in the massive case are now given by
\be d_Q A^\mu(x)=[Q,A^\mu(x)]=i\d^\mu u(x)\ee
\be d_Q\Phi(x)=[Q,\Phi(x)]=imu(x)\label{(1.5.37)}\ee
\be d_Qu(x)=\{Q,u(x)\}=0\ee
\be d_Q\tilde u(x)=\{Q,\tilde u(x)\}=-i(\d_\mu A^\mu+m\Phi(x)).
\label{(1.5.39)}\ee
The last equation follows from Eq. (\ref{(1.5.31)}); using $Q^2=0$ from Eq. (\ref{(1.5.39)})
implies Eq. (\ref{(1.5.37)}).

Let us stress the difference between our approach to massive gauge
fields and the conventional one. In the
usual approach one starts with massless gauge fields and the scalar
field $\Phi$ is the so-called Goldstone boson. The fields become massive
after "spontaneous breaking" of the gauge symmetry. We start directly
with massive vector fields. To define a gauge variation with a nilpotent
$Q$, we are forced to introduce the scalar field $\Phi$, spontaneous
symmetry breaking plays no immediate r\^{o}le. There is a \emph{common misconception}
in the literature, that the Higgs field "gives mass" to the particles.
One could also argue that if particles are massive, then a consistent theory requires
additional degrees of freedom, i.e. a Higgs sector.

\subsubsection{\it Spin-2 Fields \label{sec:gravitonfield}}
Finally, we comment on spin-2 quantum gauge theories which can be analyzed on the
same footing as spin-1 theories. We only consider the massless case which is relevant for
quantum gravity. We start from a symmetric tensor field
$h^{\alpha\beta}(x)$ with arbitrary trace which
is assumed to satisfy the wave equation
\be \sq h^{\alpha\beta}(x)=0\label{(1.7.2)}\ee
The gauge transformation similar to Eq. (\ref{(1.4.1)}) is of the form
\be \tilde h^{\alpha\beta}=h^{\alpha\beta}+\lambda(u^ 
{\alpha{\displaystyle,}\beta}+u^{\beta{\displaystyle,}\alpha}-
g^{\alpha\beta}u^\mu,_\mu),\label{(1.7.5)}\ee
where the comma denotes partial derivatives. This transformation leaves
the so-called Hilbert condition $h^{\alpha\beta},_\beta=0$ unchanged,
if $u^\alpha$ fulfills the wave equation
\be \sq u^\alpha =0.\label{(1.7.7)}\ee
The Hilbert gauge condition is analogous to the Lorentz condition in the 
spin-1 case. The
corresponding gauge charge can immediately be written down in
analogy to Eq. (\ref{(1.4.6)}):
\be Q=\int d^3x\,h^{\alpha\beta},_\beta{\dl}_0u_\alpha.\label{(1.7.8)}\ee
The vector field $u_\alpha$
must be quantized with anticommutators, in order to get $Q$ nilpotent.
The operator $Q$ given by Eq. (\ref{(1.7.8)}) is the right
infinitesimal generator for Eq. (\ref{(1.7.5)}) if it has the following commutator
\be 
[Q,h^{\alpha\beta}(x)]=-{i\over 2}\Bigl(u^{\alpha{\displaystyle,} 
\beta}+u^{\beta{\displaystyle,}\alpha}-g^{\alpha\beta}u^\mu,_\mu
\Bigl)(x)
\=d -ib^{\alpha\beta\mu\nu}u_\mu,_\nu=d_Q h^{\alpha\beta}.\label{(1.7.9)}\ee
The factor $-i/2$ is convention and the $b$-tensor
\be 
b^{\alpha\beta\mu\nu}=\eh(g^{\alpha\mu}g^{\beta\nu}+g^{\alpha\nu} 
g^{\beta\mu}-g^{\alpha\beta}g^{\mu\nu})\label{(1.7.10)}\ee
often appears in connection with tensor fields. It is also the four-dimensional
extension of DeWitt's supermetric \cite{DeWitt}.
The commutator Eq. (\ref{(1.7.9)}) implies the following commutation relation for
the tensor field
\be 
[h^{\alpha\beta}(x),h^{\mu\nu}(y)]=-ib^{\alpha\beta\mu\nu}D_0(x-y)\ee
\be =-{i\over 2}(g^{\alpha\mu}g^{\beta\nu}+g^{\alpha\nu} 
g^{\beta\mu}-g^{\alpha\beta}g^{\mu\nu})D_0(x-y).\label{(1.7.11)}\ee

The vector field $u^\mu$ must again be quantized with anticommutators
\be 
\{u^\mu(x),\tilde u^\nu(y)\}=ig^{\mu\nu}D_0(x-y)\label{(1.6.5)}\ee
and the anticommutators between two $u$'s or two $\tilde u$'s vanish.
$u^\mu, u^\nu$ are called vector-ghost fields, respectively.
An explicit representations of the ghost fields is given by
\be
\begin{split}
u^{\nu}(x)&=(2\pi)^{-3/2}\int\!\frac{d^{3}p}{\sqrt{2\omega}}\Big( +c_2^{\nu}(\vec{p})e^{-ipx}
-g^{\nu\nu}{c_1^{\nu}(\vec{p})}^{\dagger} e^{ipx} \Big)\, ,\\
\tilde{u}^{\nu}(x)&=(2\pi)^{-3/2}\int\!\frac{d^{3}p}{\sqrt{2\omega}}\Big( -c_1^{\nu}(\vec{p})e^{-ipx}
-g^{\nu\nu}{c_2^{\nu}(\vec{p})}^{\dagger} e^{ipx} \Big)\, ;
\end{split}
\ee
where the absorption and creation operators satisfy the commutation relations
\be
\big\{ c_j^{\mu}(\vec{p}),c_k^{\nu}(\vec{k})^{\dagger}\big\} =
\delta_{jk} \, \delta^{\mu\nu} \,  \delta^{(3)} (\vec{p} - \vec{k}).
\ee

The gauge variation of the vector ghost fields now follows from Eq. (\ref{(1.7.8)})
\be d_Qu^\alpha=\{Q,u^\alpha\}=0\label{(1.7.13)}\ee
\be 
d_Q\tilde u^\alpha=\{Q,\tilde u^\alpha\}=ih^{\alpha\mu},_\mu.
\label{(1.7.14)}\ee

\subsection{\it The Causal Structure of the Perturbative S-Matrix \label{sec:causal}}
\subsubsection{\it General Construction of the S-Matrix \label{sec:generalconst}}
Perturbation theory relies strongly on the axiom of causality, as shown by
H.Epstein and V.Glaser \cite{Epstein} after previous work of St\"uckelberg,
Bogoliubov and Shirkov \cite{Bogol}. The S-matrix is constructed inductively
order by order as a formal power series of operator valued distributions
\be 
S(g)=  \sum_{n=0}^\infty  S^{(n)} (g) =
{\bf 1}+\sum_{n=1}^\infty {1\over n!}\int d^4x_1\ldots d^4x_n  
T_n(x_1,\ldots x_n) g(x_1)\ldots g(x_n) \label{(2.2.1)}
\ee 
where $g(x)$ is a tempered test function that switches the interaction.
The operator-valued distributions $T_n$ act in the Fock space of some
collection of free fields. They are called time-ordered or chronological
products and should verify the so-called Bogoliubov axioms.

(1) It is clear from Eq. (\ref{(2.2.1)}) that $T_n$ can be assumed to be
completely symmetrical in all variables $x_1,\ldots x_n$.

(2) We must have Poincar\'e invariance:
\be 
U_{a,\Lambda}T_n(x_1,\ldots x_n)U_{a,\Lambda}^{-1}=T_n(\Lambda\cdot x_1+a,
\ldots\Lambda\cdot x_n+a)\label{(2.4.3)}\ee 
for all proper Lorentz transformations $\Lambda$. In particular, translation
invariance is essential in the causal approach.

(3) The central axiom is the requirement of causality which can be written
compactly as follows. If $X=\{x_1,\ldots x_m\}\in \mathds{R}^{4m}$ and
$y=\{y_1,\ldots y_n\}\in \mathds{R}^{4n}$ are such that $x_i-y_j\not\in\overline 
{V^-}$ for all $i$ and $j$, we say $X$ is later than $Y$, $X\ge Y$.
We use the compact notation $T_n(X)=T_n(x_1,\ldots x_n)$ and by $X\cup Y$
we mean the union of the elements of $X$ and $Y$. In particular, the
expression $T_{n+m}(X\cup Y)$ makes sense because of the symmetry
property (1). Now the causality axiom expresses causal factorization:
\be T_{n+m}(X\cup Y)=T_m(X)T_n(Y),\quad\forall X\ge Y.\label{(2.4.4)}
\ee
Physically this means that later action does not influence what has happened
before.

Like $S(g)$ given by Eq. (\ref{(2.2.1)}), the inverse $S(g)^{-1}$ can be expressed by a
perturbation series
\be 
S(g)^{-1}={\bf 1}+\sum_{n=1}^\infty {1\over n!}\int d^4x_1\ldots d^4x_n 
\tilde T_n(x_1\ldots x_n) g(x_1)\ldots g(x_n)\label{(2.2.4)}\ee
The corresponding $n$-point distributions $\tilde T_n$, called anti-chronological 
products follow from Eq. (\ref{(2.2.1)}) as formal inversion of a power series 
\be 
\tilde T_n(X)=\sum_{r=1}^n (-)^r\sum_{P_r} T_{n_1}(X_1)\ldots T_{n_r}
(X_r),\label{(2.2.5)}\ee
where the second sum runs over all partitions $P_r$ of $X$ into $r$
disjoint subsets
\be X=X_1\cup\ldots\cup X_r,\quad X_j\ne\emptyset ,\quad |X_j|=n_j.\ee
All products of distributions in Eq. (\ref{(2.2.5)}) are well-defined, because the
arguments are disjoint sets of points such that the products are direct
products of distributions.

(4) Unitarity of the S-matrix
\be 
S(g)^{-1}=S(g)^\dagger.\label{(2.2.14)}\ee 
can now be expressed by means of the time-ordered products in the form
\be \tilde T_n(X)=T_n(X)^\dagger.\label{(2.2.15)}\ee 
It is one aim of QFT to prove unitarity for the physically interesting theories.
In the inductive construction unitarity is not used.

Now we are ready to turn to the inductive construction of $T_n(x_1\ldots x_n)$
starting from $T_1(x)$ which is a given interaction Lagrangian
or coupling. Suppose all $T_m(x_1,\ldots ,x_m)$ for $1\le m\le n-1$
are known and have the above properties (1)-(3). Then, according to 
Eq. (\ref{(2.2.5)}), the $\tilde T_m(X)$ can
be calculated for all $1\le m=|X|\le n-1$. From this it is possible to
form the following distributions
\be A_n ' (x_1\ldots x_n)=\sum_{P_2}\tilde
T_{n_1}(X)T_{n-n_1}(Y,x_n)\label{(2.2.31)}\ee
\be R_n ' (x_1\ldots x_n)=\sum_{P_2}T_{n-n_1}(Y,x_n)\tilde T_{n_1}(X),
\label{(2.2.32)}\ee
where the sums run over all partitions
\be P_2:\>\{ x_1,\ldots ,x_{n-1}\}=X\cup Y,\quad X\ne\emptyset\label{(2.2.33)}
\ee
into disjoint subsets with $|X|=n_1\ge 1$, $|Y|\le n-2$. We also introduce
\be
D_n(x_1,\ldots x_n )=R_n ' (x_1,\ldots x_n ) -A_n ' (x_1,\ldots x_n ) .
\label{(2.2.34)}\ee
If the sums are extended over all partitions $P_2^0$, including the
empty set $X=\emptyset $, then we get the distributions
\be A_n(x_1,\ldots x_n)=\sum_{P_2^0}\tilde T_{n_1}(X)T_{n-n_1}(Y,x_n)\ee
\be =A_n ' +T_n(x_1\ldots x_n),\label{(2.2.35)}\ee
\be R_n(x_1,\ldots x_n )=\sum_{P_2^0}T_{n-n_1}(Y,x_n)\tilde T_{n_1}(X)\ee
\be =R_n ' +T_n(x_1\ldots x_n).\label{(2.2.36)}\ee 
These two distributions $A_n, R_n$ are not known by the induction assumption 
because they contain the unknown $T_n(x_1,\ldots x_n)$.
Only the difference
\be
D_n=R_n ' -A_n ' =R_n-A_n \label{(2.2.37)}
\ee
is known according to Eq. (\ref{(2.2.34)}).
What remains to be done is to determine $R_n$ (or
$A_n$) in Eq. (\ref{(2.2.37)}) separately. This is achieved by investigating the
support properties of the various distributions.

We recall the definition Eq. (\ref{(38)}) of the $n$-dimensional generalizations 
of the forward and backward light-cones with respect to the point $x$.
Then we have
\be \supp R_{n_1+1}(Y,\, x)\subseteq\Gamma_{n_1+1}^+(x)\label{(2.2.52)}\ee
and 
\be \supp A_{n_1+1}(Y,\, x)\subseteq\Gamma _{n_1+1}^-(x).\label{(2.2.53)}\ee
Because of these support properties, $R$ and $A$ are called retarded and
advanced distributions, respectively. The distribution $D$, which can be expressed by
Eq. (\ref{(2.2.37)}), then has a causal support:
\be \supp D_n(x_1,\ldots x_{n-1},x_n)\subseteq\Gamma_n^+(x_n)\cup\Gamma_n
^-(x_n).\label{(2.2.55)}\ee
We do not present the proof here (see \cite{Scharf}, sect. 3.1) but we can indicate the essential
reason for this important causal support property: According to Eq. (\ref{(2.2.31)}) and
Eq. (\ref{(2.2.32)}) $D_n$ is a sum of commutators
\be D_n ' (x_1\ldots x_n)=\sum_{P_2}[T_{n-n_1}(Y,x_n),\,\tilde T_{n_1}(X)].
\label{(2.2.56)}\ee
Since all $T$'s are products of free fields, the commutators contain Jordan-Pauli
distributions which have causal support according to Eq. (\ref{(32)}).
 
Now we see the inductive construction clearly before us: From the known 
$T_m(x_1,\ldots ,x_m),\, m\le n-1$ one computes $A_n$ given by Eq. (\ref{(2.2.31)})
and $R_n$ from Eq. (\ref{(2.2.32)}), and then $D_n =R_n ' -A_n'$.
One decomposes $D_n$ with respect to the supports Eq. (\ref{(2.2.55)})
\be
D_n(x_1,\ldots ,x_n)=R_n(x_1,\ldots ,x_n)-A_n(x_1,\ldots ,x_n),
\label{(2.2.65)}
\ee
\be
\supp R_n\subseteq\Gamma _{n-1}^+(x_n)\quad ,\quad\supp A_n\subseteq\Gamma
_{n-1}^-(x_n).
\ee 
Finally, $T_n$ is found from Eq. (\ref{(2.2.35)}) or Eq. (\ref{(2.2.36)})
\be
T_n(x_1,\ldots ,x_n)=R_n(x_1,\ldots ,x_n)-R_n'(x_1,\ldots ,x_n)
\label{(2.2.66a)}
\ee
\be
=A_n(x_1,\ldots ,x_n)-A_n ' (x_1,\ldots ,x_n).\label{(2.2.66b)}
\ee 

The only non-trivial step in this construction is the distribution
splitting Eq. (\ref{(2.2.65)}). In sect. 2.3 we have discussed the splitting
of causal numerical distributions. The transformation of the operator-
valued distribution $D_n$ to numerical distributions is achieved by means
of Wick expansion. $D_n$ can be written in a unique way in terms of
normally ordered products of free field operators $\Psi$
\be
D_n=\sum d_{k_1\ldots k_n} (x_1,\ldots x_n)\,:\Psi_{k_1}(x_1) \ldots \Psi_{k_n}(x_n): \, ,
\label{(1.9.33)}
\ee
where $k_1, \ldots ,k_n$ are indices specifying the individual field operator types
and the corresponding field components related to external symmetries
(e.g., Lorentz indices) and inner symmetries (e.g., color indices), depending
on the theory under study.

In the normal product between double dots all absorption operators stand
to the right of all emission operators. Consequently, the vacuum expectation
value of a normal product vanishes. This allows us to write the normal
ordering of, e.g., $n$ scalar fields $\fii$ in the compact form
\be
\fii(x_1)\ldots\fii(x_n)=\sum_{s_1,\ldots s_n\atop s_j=0,1}(\Omega,
\fii^{1-s_1}(x_1)\ldots\fii^{1-s_n}(x_n)\Omega):\fii^{s_1}(x_1)\ldots
\fii^{s_n}(x_n):,\label{(1.9.34)}
\ee
and the generalization to products of general free field operators is straightforward.
Here $\Omega$ is the vacuum in Fock space and the brackets $(\cdot,\cdot)$
mean the scalar product.
Note that all terms with an odd number of $s_j=0$ are zero because the
vacuum expectation value vanishes.

The splitting of $D_n$ in Eq. (\ref{(1.9.33)}) can now be carried out by splitting
all numerical distributions $d_k$. We have learned in sect. 2.3 that the
splitting solutions may be not unique. Then the free but finite local terms
must be chosen such that all properties required for the S-matrix are true.
This is a subtle problem for gauge theories.

\subsubsection{\it Example}
As a simple illustration of the causal method we consider the coupling
\be T_1(x)=-i\lambda :\fii^\dagger(x)\fii(x):\Phi(x) \label{5.1} \ee 
between a charged scalar field $\fii$ of mass $m$ and a neutral scalar
$\Phi$ of mass $M$. This theory has the same structure as quantum 
electrodynamics: $\fii$ is a spin-0 electron and $\Phi$ a scalar massive 
photon. To perform the normal ordering according to Eq. (\ref{(1.9.34)}) we
need the commutators, or contractions
\be
\contracted{}{\fii}{(x)}{\fii}{^\dagger(y)}=
[\fii(x)^{(-)},\fii^\dagger(y)^{(+)}]=(\Omega,\,\fii(x)\fii^\dagger(y))=-i
D^{(+)}_m(x-y)\label{3.14}\ee 
and similarly for $\Phi$.

In the inductive step from $T_1$ to $T_2$ we must first compute
\be R'_2(x_1,x_2)=-T_1(x_2)T_1(x_1)\ee 
by normal ordering:
$$ =\lambda^2\Bigl(:\fii^\dagger(x_1)\fii(x_1)\fii^\dagger(x_2)\fii(x_2)::\Phi(x_1)
\Phi(x_2):$$ 
$$ -iD_m^{(+)}(x_2-x_1):\fii^\dagger(x_1)\fii(x_2)::\Phi(x_1)\Phi(x_2):$$
$$ -iD_m^{(+)}(x_2-x_1):\fii(x_1)\fii^\dagger(x_2)::\Phi(x_1)\Phi(x_2):$$
$$ -iD_M^{(+)}(x_2-x_1):\fii^\dagger(x_1)\fii(x_1)\fii^\dagger(x_2)\fii(x_2):$$
$$ -D_m^{(+)}(x_2-x_1)D_m^{(+)}(x_2-x_1):\Phi(x_1)\Phi(x_2):$$
$$ -D_m^{(+)}(x_2-x_1)D_M^{(+)}(x_2-x_1):\fii^\dagger(x_1)\fii(x_2):$$
$$ -D_m^{(+)}(x_2-x_1)D_M^{(+)}(x_2-x_1):\fii(x_1)\fii^\dagger(x_2):$$
\be -iD_m^{(+)}(x_2-x_1)D_m^{(+)}(x_2-x_1)D_M^{(+)}(x_2-x_1)\Bigl).\label{5.8}\ee
We emphasize that here the product of two $D^{(+)}(x)D^{(+)}(x)$ is well
defined, in contrast to the product of two Feynman propagators $D_F(x)$ 
in sect. 2.1. The reason is that in the Fourier-transformed expression of, e.g.,
\be
{1\over (2\pi)^2}\int d^4q\,\hat D^{(+)}_m(p-q)\hat D^{(+)}_m(q)
\label{(2.6.13)}
\ee 
the intersection of the supports of the two $\hat D^{(+)}_m$ is a
compact set. This can be easily understood if one remembers that
the support in momentum space of the two individual Pauli-Jordan distributions
in Eq. (\ref{(2.6.13)}) is contained in a forward and a backward light-cone, respectively.

In the same way $A'_2$ can be computed and then $D_2=R'_2-A'_2$. From the
third line in Eq. (\ref{5.8}) we get the following contribution to $D_2$: 
\be D_2^1=-i\lambda^2 D_m(x_2-x_1):\fii(x_1)\fii^\dagger(x_2)::\Phi(x_1)\Phi(x_2):.
\label{5.10}\ee
Here $D_m$ can be trivially split. The retarded part with respect to $x_2$
contains $D^{av}_m(x_2-x_1)$. Adding $R_2\sim D^{(+)}_m$ and using
$D^{av}_m+D^{(+)}_m=D_F^m$ we finally obtain
\be T_2^1=i\lambda^2 D^m_F(x_2-x_1):\fii(x_1)\fii^\dagger(x_2)::\Phi(x_1)\Phi(x_2):.
\label{5.16}\ee
This gives "electron-photon" scattering in this model. The one contraction
between the vertices $x_1$ and $x_2$ is represented by the Feynman propagator.
So in tree graphs the usual Feynman rules hold.

Now let us consider a loop graph with two contractions, for example
"vacuum polarization" which comes from the fifth term in Eq. (\ref{5.8}).
The corresponding causal distribution is given by
\be D_2^2(x_1,x_2)=\lambda^2[D_m^{(+)}(y)D_m^{(+)}(y)-D_m^{(+)}(-y)D_m^{(+)}(-y)] :\Phi(x_1)\Phi(x_2):,\label{5.19}\ee 
where $y=x_1-x_2$. We calculate
\be d_+(-y)\=d D_m^{(+)}(y)D_m^{(+)}(y)\label{5.23}\ee 
in momentum space:
\be \hat d_+(k)=-(2\pi)^{-4}\int d^4p\,\Theta(k_0-p_0)\delta((k-p)^2-m^2)
\Theta(p_0)\delta(p^2-m^2).\label{5.24}\ee 
It is easy to evaluate this for time-like $k$ in the special Lorentz frame
such that $k=(k_0,\vec 0)$. The result for arbitrary $k$ is then
\be  \hat d_+(k)=-{(2\pi)^{-3}\over 4}\Theta(k_0)\Theta(k^2-4m^2)
\sqrt{1-{4m^2\over k^2}}.\label{5.31}\ee 
The total result for the Fourier transform of the square bracket in Eq. (\ref{5.19})
denoted by $\hat d(k)$ is simply obtained by substituting $\Theta(k_0)$ 
by $\sgn(k_0)$.

Since $\hat d(k)$ has a constant quasi-asymptotics it has singular order
$\omega=0$. Consequently, the distribution splitting is non-trivial.
Therefore, the two internal lines in the vacuum-energy graph cannot be
represented by Feynman propagators; the Feynman rules are not true for
loop graphs. Instead we must use the dispersion integral Eq. (\ref{(67)})
with $\omega=0$:
\be \hat r(k_0)={i\over 2\pi}k_0\int\limits_{-\infty}^{+\infty}
{dp_0\over (p_0-i0)(k_0-p_0+i0)}\ee 
\be\cdot\Bigl[-{(2\pi)^{-3}\over 4}\Theta(p^2-4m^2)\sqrt{1-
{4m^2\over p^2}}\sgn (p_0)\Bigl].\label{5.34}\ee 
For the time-ordered product we have to calculate $\hat t(k_0)=\hat r(k_0)
-\hat r'(k_0)$. $r'(k_0)\sim \hat d_+(-k_0)$ nicely combines with Eq. (\ref{5.34})
so that
\be \hat t(k_0)=-{i\over (2\pi)^4}{k_0^2\over 4}\int\limits_{4m^2}^{+\infty}
ds {1\over s(k_0-s+i0)}\sqrt{1-{4m^2\over s}}.\label{5.37}\ee 
This integral is elementary, the final result for time-like momentum
$k_0=\sqrt{-k^2}$ is equal to
\be \hat t(k_0)={i\over (2\pi)^4}\Biggl[1+{1+\Bigl({ik_0\over 2m}-\sqrt{1- 
{k_0^2\over 4m^2}}\Bigl)^2\over 1-\Bigl({ik_0\over 2m}-\sqrt{1-{k_0^2\over 4m^2}}
\Bigl)^2}\log\Bigl({ik_0\over 2m}-\sqrt{1-{k_0^2\over 4m^2}}\Bigl)\Biggl].\label{5.46}
\ee
The result for space-like $k$ is obtained by analytic continuation.
We leave it as an exercise to the reader to compare this result to the
alternative outcome given below at the end of sect. \ref{sec:scalaroneloop}.

\section{Regularization Methods}
\subsection{\it Basic remarks}
It would not be worthwhile to recapitulate the well-known details of the different regularization
methods which are on the market. While the Pauli-Villars regularization can be considered
as an ad hoc method to solve the apparent problem of infinities in perturbative
quantum field theory, dimensional regularization is treated by many introductory texts
like \cite{Narison,Jegerlehner}, including the original works on the topic \cite{Hooft1972,Bollini}.

Dimensional regularization regularizes Feynman diagrams by analytic continuation to
$4-\epsilon$ (complex) space-time dimensions and isolates infrared and ultraviolet divergences as poles
in $\epsilon$. From a technical point of view, the main question is how to evaluate Feynman
diagrams in $n$ dimensions, i.e. it is necessary to know the usual Feynman rules of
the theory, properties of Dirac matrices in $n$ dimensions, and techniques like the Feynman
parametrization for performing the momentum integrals in $n$ dimensions.
Difficulties may arise when one has to deal with topological quantities which exist only in integer
dimensions.

The same applies to the causal method, however, there is a big conceptual difference
between the causal and dimensional regularization approach to perturbative quantum field
theory. Whereas the causal approach is a mathematically fully understood perturbative method,
is is hard to give a precise meaning to the idea of physics in an arbitrary complex number
of space-time dimensions. Still, the method has many interesting technical advantages,
and although a proof of the physical equivalence of the causal and dimensional regularization
method is lacking due to the technical complexity of the problem,
one should not be too pessimistic about that issue.

In the forthcoming section, we will illustrate to conceptual differences by presenting
some specific examples in the light of the different approaches. There, the diagrams will
turn out to be finite in most cases, but this is not the central issue since
infinities can always be removed in one or the other way. In the present short
section, we present a direct comparison of the treatment of the scalar one-loop integral
in the causal and dimensional regularization approach. 

\subsection{\it Scalar One-Loop Diagram in $n$ Dimensions \label{sec:scalaroneloop}}
The positive frequency part of the Pauli-Jordan distribution in $n$ dimensions is given
in the causal framework by
\be
D^{(+)}_m(x)={\frac{i}{(2\pi)^{n-1}}}\int d^n p \, \delta(p^2-m^2)\Theta(p_0)e^{-ipx}.
\label{(PJn)}
\ee
Strictly speaking, the expression above is well-defined in integer dimensions.
In order to obtain an analytic expression for the scalar one-loop diagram
in arbitrary dimensions, we generalize Eq. (\ref{5.24}) to
\begin{displaymath}
\hat d_+^{(n)}(k)=\frac{i^2}{(2 \pi)^{n/2+2n-2}} \int d^n k e^{ikx}
\int d^n p_1 e^{-i p_1 x} \int d^n p_2 e^{-i p_2 x} \,
\Theta(p_1^0) \delta(p_1^2-m^2) \Theta(p_2^0) \delta(p_2^2-m^2)
\end{displaymath}
\be
=-\frac{1}{(2 \pi)^{3n/2-2}}
\int d^n p \, \Theta(k^0-p^0) \delta((k-p)^2-m^2) \Theta(p^0) \delta(p^2-m^2). \label{scalar_dim}
\ee
We exploit the last $\Theta$- and $\delta$-distributions in Eq. (\ref{scalar_dim})
and evaluate the integral over space-like momenta ($E=\sqrt{\vec{p}^{\, 2}+m^2}$, $k^0=k_0$)
\be
I^{(n)}=\int \frac{d^{n-1} p}{2 E} \delta(k_0^2-2k^0E) \Theta(k^0-E)
\ee
and obtain a radial integral, using $E=k^0/2$, $|\vec{p} \, |=\sqrt{k_0^2/4-m^2}$
\be
I^{(n)}= \frac{2 \pi^{(n-1)/2}}{\Gamma((n-1)/2)} \Theta(k_0^2-4 m^2) 
\int d |\vec{p} \, | \frac{|\vec{p} \, |^{ n-2}}{2 E} \delta(2 k^0(k^0/2-E)) \Theta(k^0-E),
\ee
since the $(n-2)$-dimensional surface of an $(n-1)$-dimensional unit ball
is given by $(2 \pi^{(n-1)/2})/\Gamma((n-1)/2)$.
Substituting $ |\vec{p} \, |  d |\vec{p} \, |=E dE$, $I^{(n)}$ can be written
\begin{displaymath}
I^{(n)}= \frac{\pi^{(n-1)/2}}{\Gamma((n-1)/2)} \Theta(k_0^2-4 m^2)
\int dE |\vec{p} \, |^{n-3} \frac{\delta(E-k^0/2)}{2 k^0} \Theta(k^0/2)
\end{displaymath}
\be
=\frac{\pi^{(n-1)/2}}{\Gamma((n-1)/2)} \Theta(k_0^2-4 m^2) \Theta(k^0)
\frac{\sqrt{k_0^2/4 -m^2}^{n-3}}{2 k^0}
\ee
or
\be
\hat d_+^{(n)}(k)=-\frac{\pi^{(n-1)/2}}{(2 \pi)^{3n/2-2}
\Gamma((n-1)/2)} \Theta(k^2-4 m^2) \Theta(k^0)
\frac{\sqrt{k^2/4 -m^2}^{n-3}}{2 \sqrt{k^2}}. \label{causal_dim_scal}
\ee
This result can be compared directly to Eq. (\ref{5.24}) for $n=4$.

Now, the interesting point is that the $(1-t+i0)$-term in the central
splitting formula Eq. (\ref{(68)}) generates the real part of the scalar loop amplitude,
denoted here by $\hat t^{(n)}(k)$, since
\be
\frac{1}{1-t+i0}=P\frac{1}{1-t}-i\pi \delta(1-t),
\ee
where the symbol $P$ denotes the Cauchy principal value in the sense of distributions.
We obtain for $k$ in the forward light-cone
\be
\mathfrak{Re}(\hat t^{(n)}(k)) \sim \Theta(k^2-4 m^2)
\frac{\sqrt{k^2/4 -m^2}^{n-3}}{2 \sqrt{k^2}}, \label{real_part}
\ee
where we have omitted numerical factors. In fact, this result is valid for arbitrary
momenta $k$.

Now the scalar loop integral in $n$ dimensions is given by the expression
\be
\mathcal{I}^{(n)}(p) = \int \frac{d^{n}k}{\left( k^2-m^{2} + i0 \right)
\left[ \left( k-p \right)^{2}-m^{2} + i0 \right]}
\ee
Feynman parametrization
\be
\frac{1}{AB} = \int\limits_{0}^{1} d\alpha \left[\alpha A + (1-\alpha)B\right]^{-2}
\ee
and a subsequent momentum translation
$k^{\mu} \mapsto k^{\mu} + \alpha p^{\mu}$ leads to the formal integral
\be
\mathcal{I}^{(n)}(p) = \int d^{n}k \int\limits_{0}^{1} d\alpha  \left[ k^{2} - m^{2} + \alpha (1- \alpha) p^{2} + i0 \right]^{-2} \: .
\ee
To evaluate this integral, we may use the relation
\be
\int \frac{d^{n}k}{\left(k^{2}-a^{2}+i0\right)^{m}} = i \: \pi^{\frac{n}{2}} \: \frac{\Gamma \left(2-\frac{n}{2}\right)}{\left(a^{2}-i0\right)^{2-\frac{n}{2}}} \: , \label{rel_int}
\ee
which is divergent for $n \ge 4$. In the present case, we have
\be
a^2=m^2-\alpha(1-\alpha) p^2.
\ee

The calculation of the finite integral in the case
$n=3$ is discussed in detail in sect. \ref{sec:sQED}. We focus here on the case $n=4$.
Note that the following manipulations are formal to some extent.
Since
\be
\Gamma \left(2-\frac{n}{2}\right)= \frac{2}{4-n}- \gamma -
\biggl(\frac{\pi^2}{24}+\frac{\gamma^2}{4} \biggr)(n-4)+ \ldots
\ee
and
\be
a^{n-4}=1+(n-4) \log a + \ldots ,
\ee
where $\gamma$ is the Euler-Mascheroni constant, we obtain in the limit $n \rightarrow 4$
\be
\mathcal{I}^{(4)}(p) \rightarrow \frac{2i \pi^2}{4-n} -i \pi^2 \int \limits_{0}^{1}
d \alpha \log (m^2-\alpha (1-\alpha)p^2-i0) -i \pi^2 \gamma -i \pi^2 \log(\pi).
\ee
A relevant finite part of the integral above is given by
\begin{displaymath}
\mathcal{I}^{(4)}_{reg}(p) =
-i \pi^2 \int \limits_{0}^{1} d \alpha \log \biggl( \frac{m^2-\alpha(1-\alpha) p^2-i0}{m^2} \biggr)
\end{displaymath}
\be
= -i \pi^2 \int \limits_{0}^{1} d \alpha 
\biggl( \frac{\alpha(2\alpha-1)}{m^2-\alpha(1-\alpha) p^2-i0} \biggr),
\ee
where an integration by parts was performed. For $p^2>4m^2$, the $-i0$-term
generates a real part of $\mathcal{I}^{(4)}_{reg}(p)$,
which is obtained in a trivial manner from the partial fraction decomposition
\be
\frac{1}{\alpha-\alpha_1-i0}+\frac{1}{\alpha-\alpha_2+i0}=\frac{2 \alpha-1}{\alpha^2-\alpha+m^2/p^2-i0}
\ee
with
\be
\alpha_{1,2}=\frac{1}{2} \pm \frac{1}{2} \sqrt{1-\frac{4 m^2}{p^2}}, \quad
0 < \alpha_{1,2} < 1,
\ee
leading to
\be
\mathfrak{Re} (\mathcal{I}^{(n)}_{reg}(p)) \sim (\alpha_1-\alpha_2) \Theta(p^2-4m^2)=
\Theta(p^2-4m^2) \sqrt{1-\frac{4 m^2}{p^2}},
\ee
in accordance with Eq. (\ref{real_part}).
For $p^2>4m^2$, one can also write
\be
\mathcal{I}^{(4)}_{reg}(p)=i \pi^2 \sqrt{1-\frac{4 m^2}{p^2}} \Biggl(
\log \biggl( \frac{1-\sqrt{1-4m^2/p^2}}{1+\sqrt{1-4m^2/p^2}} \biggr) +i \pi \Biggr)+2i \pi^2.
\ee
This solution is normalized according to
\be
\lim_{p \rightarrow 0} \mathcal{I}^{(4)}_{reg, an}(p)=0,
\ee
i.e. it corresponds to the central splitting solution in the causal approach
when continued analytically to arbitrary $p$.

We observe that the real part of the scalar loop diagram coincides both for the
causal and the dimensional approach, and it is straightforward to show that this
result holds in arbitrary dimensions. Furthermore, Eq. (\ref{causal_dim_scal}) provides
a kind of "dimensional" generalization of the causal method. From the real part,
the imaginary part of the amplitude is obtained from the dispersive splitting formula in the
causal approach or by direct computation according to the rules of dimensional
regularization. Up to finite renormalizations, the finite parts of the results also agree.

\section{Comparison of Regularization Methods to the Causal Approach:\\
Specific Examples}

\subsection{\it Axial Anomalies}

Axial or triangle anomalies are a subtle problem because their treatment by
regularization of divergent Feynman integrals is unsatisfactory. On gets the
impression that the anomalies are a consequence of the ultraviolet
regularization.  Then the question
remains whether by some other method of calculating the divergent integral
the anomaly might disappear. The causal method is free of such uncertainties
as we are going to show.

We consider QED with pseudovector and pseudoscalar couplings
$$T_1(x)=i\,c_V\,j_V^\mu (x)A_\mu (x)+i\,c_A\,j_A^\mu (x)B_\mu (x)$$
\be +ic_\pi j_\pi(x)\Pi(x).\label{5.3.1}\ee 
Here
\be j_V^\mu =\,:\,\psq\gamma^\mu\psi\,:\quad ,\quad j_A^\mu =\,:\,\psq
\gamma^\mu\gamma^5\psi\, :\label{5.3.2a}\ee 
are the vector and axial vector currents and
\be j_\pi =i\,:\,\psq\gamma^5\psi\, :\label{5.3.2b}\ee 
is a pseudoscalar, all being formed from a free massive Dirac field
$\psi (x)$ with mass $m$. The vector-, axial vector and pseudoscalar
vertices defined by Eq. (\ref{5.3.1}) will be abbreviated by $V$, $A$ and $\Pi$ in
the following. The fields $A_\mu$, $B_\mu$ and $\Pi(x)$ play
no essential r\^ole and are, therefore, assumed as classical external
fields.

\begin{figure}
\begin{minipage}[b]{0.5\linewidth} % A minipage that covers half the page
\centering
\includegraphics[width=8cm]{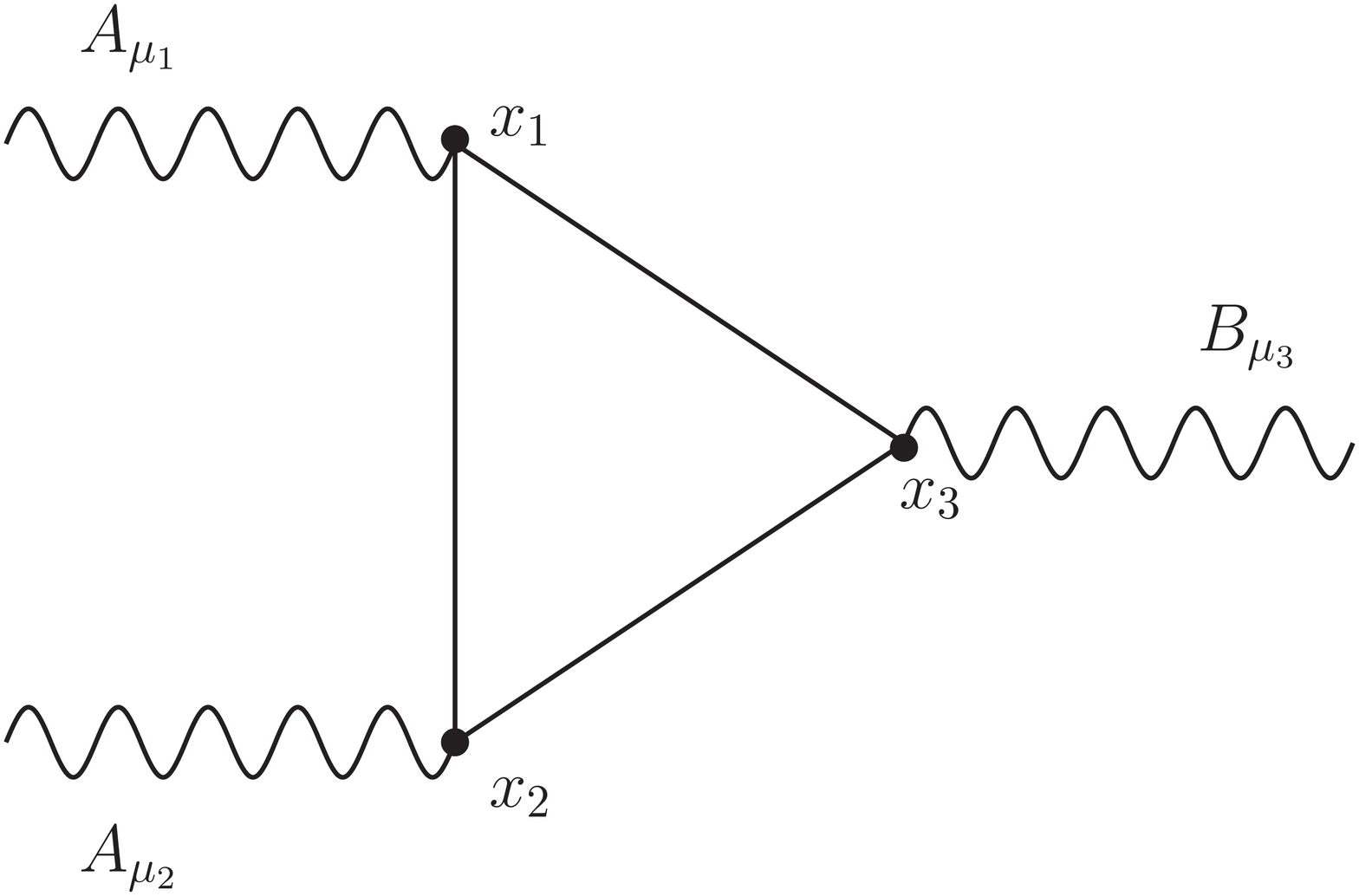}
\caption{Axial vector triangular graph}
\label{triangle1}
\end{minipage}
\hspace{0.5cm} % To get a little bit of space between the figures
\begin{minipage}[b]{0.5\linewidth}
\centering
\includegraphics[width=8cm]{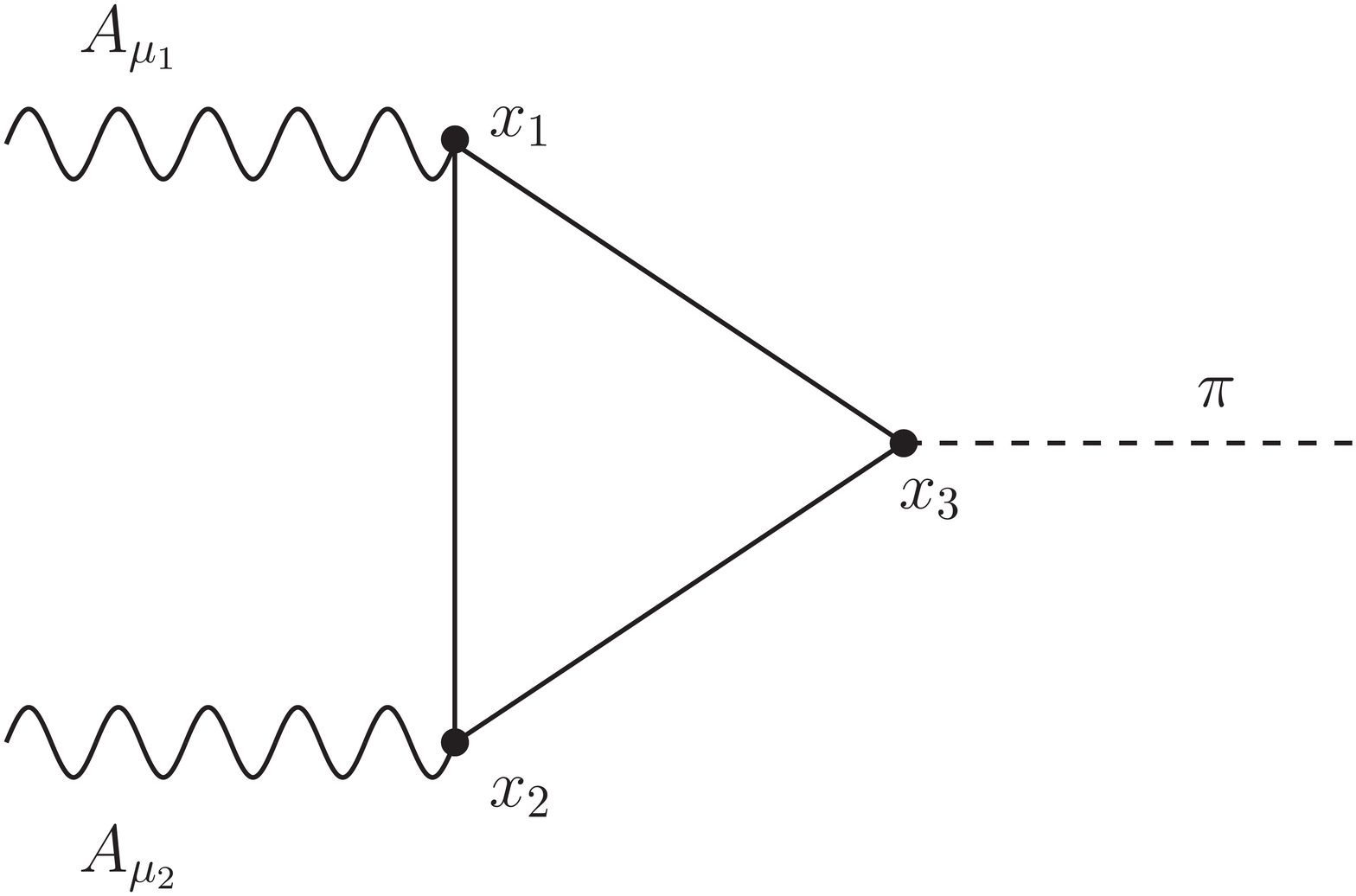}
\caption{Pseudoscalar triangular graph}
\label{triangle2}
\end{minipage}
\end{figure}

From Eq. (\ref{5.3.2a}) and Eq. (\ref{5.3.2b}) we have the following divergence 
relations for the free currents
\be \d_\mu j_V^\mu=0,\quad \d_\mu j_A^\mu=2mj_\pi,\label{5.3.3}\ee
as a consequence of the Dirac equation.
Our problem is whether similar divergence relations hold at higher
orders, in particular for the two triangular graphs with vertices 
$VVA-$ and $VV\Pi$ of Figs. (\ref{triangle1}) and (\ref{triangle2})
which contribute to the 3-point function $T_3$. 
To compute the latter, we must first calculate
$$D_3(x_1,x_2,x_3)=T_2(x_1,x_3)T_1^\dagger(x_2)+T_2(x_2,x_3)T_1^\dagger(x_1)
+T_1(x_3)T_2^\dagger(x_1,x_2)$$
\be -T_1^\dagger(x_1)T_2(x_2,x_3)-T_1^\dagger(x_2)T_2(x_1,x_3)-T_2^\dagger(x_1,x_2)
T_1(x_3),\label{5.3.4}\ee 
where we have used unitarity to express the $T$-distributions of
the inverse S-matrix.
Concerning the triangle graphs, the 2-point distributions
which contribute, come from Compton scattering. For these distributions
the usual divergence relations in Eqns. (\ref{5.3.3}) still hold, so that
\be {\d\over\d x_3^{\mu_3}}d_B^{\mu_1\mu_2\mu_3}(x_1,x_2,x_3)=2m
{c_A\over c_\pi}d_\pi^{\mu_1\mu_2}(x_1,x_2,x_3).\label{5.3.5}\ee
Here $d_B, d_\pi$ are the numerical 3-point distributions corresponding to 
the triangle graphs displayed in Figs. (\ref{triangle1}) and (\ref{triangle2})
without the external fields $A, B, \Pi$. The question 
is whether the same relation remains true after splitting for the 
retarded distributions. One therefore defines the anomaly by
\be a^{\mu_1\mu_2}={\d\over\d x_3^{\mu_3}}r_B^{\mu_1\mu_2\mu_3}(x_1,x_2,x_3)
-2m{c_A\over c_\pi}r_\pi^{\mu_1\mu_2}(x_1,x_2,x_3).\label{5.3.6}\ee 
The $t$-distributions have the same anomaly because the
$r'$-distributions are anomaly-free.

Since we work with massive Fermi fields, we can perform the splitting in
momentum space by means of the central solution
\be \hat r_B(p,q)={i\over 2\pi}\int\limits_{-\infty}^{+\infty} 
{\hat d_B(tp,tq)\over (1-t+i0)t^{\omega_B+1}}\, dt\label{5.3.7}\ee 
where $p$, $q$ are assumed to be in the forward cone $V^+$, and
similarly for $\hat r_\pi$. $\omega$ is the singular order of the
$d$-distributions. The Fourier transformation is carried out in the
difference variables $y_1=x_1-x_3$, $y_2=x_2-x_3$, taking translation
invariance into account. From Eq. (\ref{5.3.5}) we then get
\be i(p_{\mu_3}+q_{\mu_3})\hat d_B^{\mu_1\mu_2\mu_3}(p,q)=2m{c_A\over
c_\pi}\hat d_\pi^{\mu_1\mu_2}(p,q),\label{5.3.8}\ee
and the anomaly in Eq. (\ref{5.3.6}) becomes
\be \hat a^{\mu_1\mu_2}(p,q)=i(p_{\mu_3}+q_{\mu_3})\hat r_B^{\mu_1\mu_2\mu_3}
-2m{c_A\over c_\pi}\hat r_\pi^{\mu_1\mu_2}.\label{5.3.9}\ee
Substituting Eq. (\ref{5.3.7}) and the analogous equation for $\hat r_\pi$ herein,
and using Eq. (\ref{5.3.8}), we arrive at the following formula for the anomaly
\be \hat a^{\mu_1\mu_2}(p,q)={i\over \pi}{c_A\over c_\pi}m \int \limits_{-\infty}^{\infty} dt\, 
{\hat d_\pi^{\mu_1\mu_2}(tp,tq)\over 1-t+i0}\biggl({1\over
t^{\omega_B+2}}-{1\over t^{\omega_\pi+1}}\biggl).\label{5.3.10}\ee 
Hence, the anomaly is due to the fact that $\omega_\pi-\omega_B\ne 1$.

To evaluate Eq. (\ref{5.3.10}) we only need the pseudoscalar $d$-distribution.
From the first three terms in Eq. (\ref{5.3.4}) we find
$$r^{\prime\mu_1\mu_2}(y_1,y_2)=c_V^2c_\pi\tr\bigr[i\gamma_5S^{(+)}_m(x_3-x_2)
\gamma^{\mu_2}S_{AF}^m(x_2-x_1)\gamma^{\mu_1}S^{(-)}_m(x_1-x_3)$$
$$+i\gamma_5S^{(+)}_m(x_3-x_2)\gamma^{\mu_2}S^{(-)}_m(x_2-x_1)\gamma^{\mu_1}
S_F(x_1-x_3)^m$$ 
\be +i\gamma_5 S_F^m(x_3-x_2)\gamma^{\mu_2} S^{(+)}_m(x_2-x_1)\gamma^{\mu_1} 
S^{(-)}_m(x_1-x_3)\bigr]+\tr\bigr[x_1\leftrightarrow x_2,\mu_1 
\leftrightarrow\mu_2\bigr].\label{5.3.11}\ee 
Here $S_{AF}^m$ denotes the anti-Feynman propagator which is obtained from
the Feynman propagator in momentum space by changing $+i0$ into $-i0$:
\be \hat S_{AF}^m(p)=(2\pi)^{-2}{\ps+m\over p^2-m^2-i0}.\label{5.3.12}\ee 
It comes from the adjoint in Eq. (\ref{5.3.4}). If one replaces $S^{(+)}_m$ by
$S^{(-)}_m$ and vice versa without changing the arguments, one gets
$a^{\prime\mu_1\mu_2}$. The difference $r'-a'$ gives $d^{\mu_1\mu_2}$.

Expressing the spinor distributions by scalar ones, we see that the
terms with three $\ds$ contain $\gamma_5$ plus five other
$\gamma$-matrices. Then the trace vanishes. In the non-vanishing terms
one has at least one factor $m$ instead of $\ds$. This lowers $\omega$
by one, so that we get $\omega_\pi=0$, instead of the power-counting
estimate 1. But we will see below that the splitting 
with $\omega = 0$
or 1 gives the same result. If we replace $i\gamma_5$ by $\gamma^{\mu_3}
\gamma_5$ in Eq. (\ref{5.3.12}), we get the $r'$-distribution for the axial vector
graph: $r^{\prime\mu_1\mu_2\mu_3}$. Then the terms with three $\ds$
contain $\gamma_5$ plus six $\gamma^\mu$ matrices and the trace does not
vanish. In this case we have the power-counting result $\omega_B=1$.
 
After Fourier transformation
\be \hat r^{\prime\mu_1\mu_2}(p,q)=(2\pi)^{-4}\int r^{\prime\mu_1\mu_2}
(y_1,y_2)e^{i(py_1+qy_2)}dy_1\, dy_2\label{5.3.13}\ee
we shall obtain
$$\hat r^{\prime\mu_1\mu_2}(p,q)={c_V^2c_\pi\over (2\pi)^2}\int dk\biggl\{
\tr\bigr[-i\gamma_5\hat S^{(+)}_m(-P+k)
\gamma^{\mu_2}\hat S_{AF}^m(-p+k)\gamma^{\mu_1}\hat S^{(-)}_m(k)$$
$$ +i\gamma_5\hat S^{(+)}_m(-P+k)\gamma^{\mu_2}\hat S^{(-)}_m(-p+k)\gamma^{\mu_1}
\hat S_F(k)^m$$ 
\be +i\gamma_5\hat S_F^m(-P+k)\gamma^{\mu_2}\hat S^{(+)}_m(-p+k)\gamma^{\mu_1} 
\hat S^{(-)}_m(k)\bigr]+\tr\bigr[p\lra q,\mu_1\lra\mu_2\bigr]\biggl\}
.\label{5.3.14}\ee 
Here we have introduced $P=p+q$. Computing the trace we get
$$\hat r^{\prime\mu_1\mu_2}(p,q)=-{4mc_V^2c_\pi\over (2\pi)^6}
\eps^{\mu_1\mu_2\alpha\beta}p_\alpha q_\beta\times$$
\be \times\>\bigr\{[I_-(P,p)+I_+(q,-p)+I_+(p,P)]
+[p\lra q]\bigr\},\label{5.3.15}\ee 
where the Lorentz invariant integrals $I_\pm$ are given by
$$I_\pm (p,q)\=d\int d^4k\,\Theta(-k^0)\delta(k^2-m^2)\times$$
\be \times\>\Theta(k^0-p^0)\delta[(k-p)^2-m^2]{1\over (k-q)^2-m^2\pm 
i0}.\label{5.3.16}\ee 

Owing to the two $\delta$- and $\Theta$-functions, these integrals
vanish if $p$ is not in the region $p^2\ge 4m^2$, $p_0<0$. But if $p$ is
in this region, one can use $\lo$ invariance
\be I_\pm(\Lambda p,\Lambda q)=I_\pm(p,q),\quad\forall\Lambda\in\lo,
\label{5.3.17}\ee 
to choose $\Lambda p=(-\sqrt{p^2},\vec 0)$. Then the integration is done
as follows: first we integrate over $k^0$, using the first $\delta$. In
the spatial integration $d^3\vec k$, we use polar coordinates with $\vec
q$ as polar axis. Integration over $|\vec k|$ kills the second $\delta$,
while the integral over the azimuth $\fii$ gives trivially $2\pi$. The
remaining integration over $\cos\te=\vec k\cdot\vec q/(|\vec k||\vec
q|)$ is elementary. The result in an arbitrary Lorentz system is
equal to
$$I_\pm(p,q)={\pi\over 4}\Theta(-p^0)\Theta(p^2-4m^2){1\over
\sqrt{N}}\times$$
\be \times\>\log\Bigl({-pq+q^2+\sqrt{(1-4m^2/p^2)N}\pm i0\over 
-pq+q^2-\sqrt{(1-4m^2/p^2)N}\pm i0}\Bigl),\label{5.3.18}\ee
where
\be N=N(p,q)=(pq)^2-p^2q^2.\label{5.3.19}\ee

The final result for Eq. (\ref{5.3.15}) is now given by
\be \hat r^{\prime\mu_1\mu_2}(p,q)=\eps^{\mu_1\mu_2\alpha\beta}p_\alpha
q_\beta\hat r'(p,q),\label{5.3.24}\ee 
$$\hat r'(p,q)={mc_V^2c_\pi\over (2\pi)^5\sqrt{N}}\Bigl[-\Theta(-p^0)
\Theta(p^2-4m^2)\log_1$$
\be -\Theta(-q^0)\Theta(q^2-4m^2)\log_2-\Theta(-P^0)\Theta(P^2-4m^2)
\log_3\Bigl],\label{5.3.25}\ee 
where
\be \log_1=\log\Bigl({q^2+pq+\sqrt{(1-4m^2/p^2)N}+i0\over 
q^2+pq-\sqrt{(1-4m^2/p^2)N}+i0}\Bigl),\label{5.3.26}\ee
\be \log_2=\log\Bigl({p^2+pq+\sqrt{(1-4m^2/q^2)N}+i0\over 
p^2+pq-\sqrt{(1-4m^2/q^2)N}+i0}\Bigl),\label{5.3.27}\ee
\be \log_3=\log\Bigl({-pq+\sqrt{(1-4m^2/P^2)N}-i0\over 
-pq-\sqrt{(1-4m^2/P^2)N}-i0}\Bigl).\label{5.3.28}\ee 
The expressions for $a'$ and $d$ are similar:
\be \hat a^{\prime\mu_1\mu_2}=\eps^{\mu_1\mu_2\alpha\beta}p_\alpha
q_\beta\hat a'(p,q),\label{5.3.29}\ee 
$$\hat a'(p,q)={mc_V^2c_\pi\over (2\pi)^5\sqrt{N}}\Bigl[-\Theta(p^0)
\Theta(p^2-4m^2)\log_1$$
\be -\Theta(q^0)\Theta(q^2-4m^2)\log_2-\Theta(P^0)\Theta(P^2-4m^2)\log_3
\Bigl],\label{5.3.30}\ee 
\be \hat d^{\mu_1\mu_2}=\eps^{\mu_1\mu_2\alpha\beta}p_\alpha
q_\beta\hat d(p,q),\label{5.3.31}\ee 
$$\hat d(p,q)={mc_V^2c_\pi\over (2\pi)^5\sqrt{N}}\Bigl[\sgn(p^0)
\Theta(p^2-4m^2)\log_1$$
\be +\sgn(q^0)\Theta(q^2-4m^2)\log_2+\sgn(P^0)\Theta(P^2-4m^2)\log_3
\Bigl].\label{5.3.32}\ee 
Since the scaling limit in Eq. (\ref{5.3.32}) is equal to
$$\lim_{\lambda\to\infty}\hat d^{\mu_1\mu_2}(\lambda p,\lambda q)=
\hat d^{\mu_1\mu_2}_{m=0}(p,q),$$
we conclude that $\omega_\pi=0$. However, the central splitting solution
is independent of choosing $\omega=1$ or $\omega=0$, respectively. To
see this, we calculate the difference
$$\hat r^{\mu_1\mu_2}_{\omega=1}-\hat r^{\mu_1\mu_2}_{\omega=0}=
{i\over 2\pi} \int \limits_{-\infty}^{\infty} dt\,{\hat d^{\mu_1\mu_2}(tp,tq)\over 1-t+i0}
\Bigl({1\over t^2}-{1\over t}\Bigl)$$
$$={i\over 2\pi} \int \limits_{-\infty}^{\infty} dt\,{\hat d^{\mu_1\mu_2}(tp,tq)\over t^2}=0,
$$
because the denominator is an odd function of $t$.

To calculate the anomaly, we now insert Eq. (\ref{5.3.31}) into Eq. (\ref{5.3.10})
\be \hat a^{\mu_1\mu_2}(p,q)=\eps^{\mu_1\mu_2\alpha\beta}p_\alpha q_\beta
a(p,q),\label{5.3.35}\ee 
where
\be a(p,q)={im\over\pi}c_V^2c_A \int \limits_{-\infty}^{\infty} {dt\over t}\,\hat
d(tp,tq),\label{5.3.36}\ee 
for all $p,q\in V^+$. In Eq. (\ref{5.3.32}) we introduce
\be f_i(p^2,q^2,P^2)={m\over (2\pi)^5\sqrt{N}}\log_i,\quad i=1,2,3,
\label{5.3.37}\ee 
and combine the integrals from $-\infty$ to 0 and from 0 to $\infty$,
taking the sign- functions into account. Substituting $t^2=\tau$ we get
$$a(p,q)={i\over\pi}c_V^2c_A m\int\limits_0^\infty{d\tau\over 
\tau}\,\Bigl[\Theta(\tau p^2-4m^2)f_1(\tau p^2,\tau q^2,\tau P^2)$$
\be +\Theta(\tau q^2-4m^2)f_2(\tau p^2,\tau q^2,\tau P^2)+ 
\Theta(\tau P^2-4m^2)f_3(\tau p^2,\tau q^2,\tau P^2)\Bigl].\label{5.3.38}\ee 

Since the anomaly is a polynomial of degree $\omega_B+1=2$, $a(p,q)$
must be a pure number independent of $p, q$, we can take the limit
$p^2\to 0$ and $q^2\to 0$ in Eq. (\ref{5.3.38}), while keeping $P^2>0$. Then only
the last term contributes. Substituting $\tau P^2=s$, we obtain
\be a(p,q)={i\over\pi}c_V^2c_A m\int\limits_{4m^2}^\infty{ds\over 
s}\,f_3(0,0,s).\label{5.3.39}\ee 
We have for $P^2\ge 4m^2$:
$$f_3(p^2=0,q^2=0,P^2)={2m\over (2\pi)^5P^2}\log{1-\sqrt{1-4m^2/P^2}
\over 1+\sqrt{1-4m^2/P^2}},$$
which implies
$$a(p,q)=i{4m^2\over (2\pi)^6}c_V^2c_A\int\limits_{4m^2}^ 
\infty{ds\over s^2}\,\log{1-\sqrt{1-4m^2/s}
\over 1+\sqrt{1-4m^2/s}}.$$
Substituting $x=4m^2/s$, we get
$$a(p,q)={2i\over (2\pi)^6}c_V^2c_A\int\limits_0^1 
dx\,\log{1-\sqrt{1-x}\over 1+\sqrt{1-x}},$$
which shows the mass independence of the anomaly. The further
substitution $\sqrt{1-x}=z$ makes the integral elementary and we get
\be a(p,q)=-{2i\over (2\pi)^6}c_V^2c_A.\label{5.3.43}\ee 
Summing up, according to Eq. (\ref{5.3.35}) the axial anomaly  
for the triangle graphs
is equal to
\be a^{\mu_1\mu_2}(p,q)=-{2i\over (2\pi)^6}c_V^2c_A 
\eps^{\mu_1\mu_2\alpha\beta}p_\alpha q_\beta.\label{5.3.44}\ee 

We have still to investigate whether there exist other splitting
solutions which do not have an anomaly while preserving all desired
properties of the theory. $\hat t^{\mu_1\mu_2}(p,q)$ is a pseudotensor
of rank two. The lowest order normalization polynomial with this
property is $\sim\eps^{\mu_1\mu_2\alpha\beta} p_\alpha q_\beta$. But
this has already $\omega=2$, in contrast to $\omega_\pi=0$. Hence,
renormalization of the $VV\Pi$ triangle does not help. There seems to be
a better chance with $\hat t^{\mu_1\mu_2\mu_3}(p,q)$, which is a pseudotensor
of rank three with $\omega_B=1$. The most general normalization
polynomial which preserves unitarity and the symmetry in the two $V$ vertices
is now given by
$$P^{\mu_1\mu_2\mu_3}(p,q)=C\eps^{\mu_1\mu_2\mu_3\alpha}(p_\alpha-
q_\alpha),$$
where $C$ is a real constant. But this would destroy vector gauge
invariance
$$p_{\mu_1}P^{\mu_1\mu_2\mu_3}(p,q)=-C\eps^{\mu_1\mu_2\mu_3\alpha}p_{\mu_1}
q_\alpha\not\equiv 0,$$
which we do not allow for. That means that the axial anomaly cannot be
removed by renormalization. We have to live with it. In the electroweak
theory the anomalies cancel by compensation between leptons and quarks
(\cite{Scharf2}, sect. 4.9).

The analysis of the axial anomaly presented above may appear technical
at first sight for a reader which is not yet familiar with the causal method.
However, working with divergent Feynman integrals has some \emph{ad hoc} character,
and since the causal method works without divergent, i.e. ill-defined expressions,
the computation of the axial anomaly presented above is more rigorous than in other
approaches and serves as an unambiguous consistency check.

\subsection{\it Schwinger Model}
The Schwinger model \cite{Schwinger1962,Walther1998} is a popular laboratory
for quantum field theoretical methods. As a soluble quantum field
theoretical model, its nonperturbative
properties and relations to confinement \cite{Casher1973,Casher1974}
have always been of greatest interest. It is also possible to
discuss the model perturbatively in a straightforward way.
The interesting features of the model, originally designed to describe
QED with massless fermions in 1+1-dimensional spacetime, are related to
the fact that the massless fermions and the photon field actually disappear
from the physical spectrum, whereas a "physical" massive scalar field
appears with the so-called Schwinger mass $m_s^2=e^2/\pi$, which
depends on the coupling constant $e$. For a full discussion of the model
we refer to the literature (see \cite{Strocchi} and references therein).
In the following, we focus on the calculation of the vacuum polarization (VP)
diagram at second order, where the appearance of a mass term in the Schwinger model
can be traced in the photon propagator by resummation.
It turns out that the correct treatment of the VP is a delicate task, where the
careful discussion of the scaling behavior of distributions given above
becomes very useful.

\subsubsection{\it The Causal Approach}
As previously discussed, the $S$-matrix is constructed inductively order by order
as an operator valued functional in the 1+1-dimensional case
\be
S(g)=1+\sum_{n=1}^\infty{1\over n!}\int d^2x_1\ldots d^2x_n\,
T_n(x_1,\ldots x_n)g(x_1)\ldots g(x_n),
\ee
where $g(x)$ is a tempered test function that switches the interaction.
The first order interaction term for QED given in terms of
asymptotic free fields is
\be
T_1 (x) = ie:\bar{\Psi}(x) \gamma^\mu \Psi(x): A_\mu(x).
\ee
We note here that the so-called adiabatic limit $g(x) \rightarrow 1$ has been
shown to exist in purely massive theories at each order of the perturbative
expansion of the $S$-matrix \cite{Epstein}. We therefore keep a mass term
for the fermion fields in the following, and consider the limit $m \rightarrow 0$
when appropriate. We further note that, of course,
the properties of 1+1-dimensional fermion fields do not have much in common with the
corresponding counterparts in 3+1-dimensional spacetime, both from a
physical and mathematical point of view. 

The interesting second order distribution $T_2$ is constructed by first considering
the causal distribution $D_2(x,y)$
\be
D_2 (x,y) = [T_1(x),T_1(y)] \quad ,
\ee
\be
{\mbox {supp}} \, D_2 =\{(x-y)\>|\>(x-y)^2\ge 0\} \quad,
\ee
which has causal support.
Then $D_2$ is split into a retarded and an advanced part $D_2=R_2-A_2$,
with
\be
{\mbox {supp}} \, R_2 =\{(x-y)\>|\>(x-y)^2\ge 0,\> (x^0-y^0)\ge 0\} \quad,
\ee
\be
{\mbox {supp}} \, A_2 =\{(x-y)\>|\>(x-y)^2\ge 0,\> -(x^0-y^0)\ge 0\} \quad.
\ee
Finally $T_2$ is given by
\be
T_2(x,y) = R_2(x,y) + T_1(y)T_1(x) = A_2(x,y) - T_1(x)T_1(y) \quad .
\ee

For the massive Schwinger model with fermion mass m,
the part in the Wick ordered distribution
$D_2$ corresponding to VP
\be
D_2(x,y) = e^2 [ d_2^{\mu \nu} (x-y)-d_2^{\nu \mu}(y-x) ]
:A_\mu(x) A_\nu (y): + ...
\ee
then becomes after a short calculation
\begin{displaymath}
{\hat d}_2^{\mu \nu} (k) := \frac{1}{2 \pi} \int d^2z \,
[d_2^{\mu \nu} (z)-d_2^{\mu \nu} (-z)] e^{ikz},
\end{displaymath}
\be
{\hat d}_2^{\mu \nu} (k) = \Bigl(g_{\mu \nu} - \frac{k_\mu k_\nu}{k^2}\Bigr)
\frac{4m^2}{2 \pi} \frac{1}{k^2 \sqrt{1-4m^2/k^2}} {\mbox {sgn}}(k^0) \Theta(k^2-4m^2) \quad . \label{ip}
\ee

Obviously, ${\hat d}_2^{\mu \nu}$ has a naive power counting degree
$\omega_p = -2$ \cite{Weinberg1960}. But the singular order of the distribution is
$\omega=0$ \cite{Vladimirov}.
Indeed, applying the definitions from sect. \ref{sec:regdistr}
to ${\hat d}_2^{\mu \nu} (k)$, we obtain the quasi-asymptotics
\be
\lim_{\delta \rightarrow 0} {\hat d}_2^{\mu \nu}(k/\delta) =
\frac{1}{2 \pi} \Bigl(g^{\mu \nu} k^2 - k^\mu k^\nu \Bigr)
\delta(k^2) {\mbox {sgn}}(k^0) \quad , \label{res}
\ee
and we have $\rho(\delta)=1$, hence $\omega=0$.
The quasi-asymptotics \emph{differs} from the naively expected formal result
\be
{\hat d}_2^{\mu \nu} (k) = \Bigl(g_{\mu \nu} - \frac{k_\mu k_\nu}{k^2}\Bigr)
\frac{4m^2}{2 \pi} \frac{\Theta(k^2) \mbox{sgn}(k^0) }{k^2} ,
\ee
which would be ill-defined as a distribution in 2 dimensions.
Note that the
$g^{\mu \nu}$-term in Eq. (\ref{res}) does not contribute to the quasi-asymptotics.
The reason for the result Eq. (\ref{res}) can be explained by the existence of a
{\em {sum rule}} \cite{Adam1992}
\be
\int \limits_{4m^2\delta^2}^{\infty} d(q^2) \frac{\delta^2 m^2}
{q^4 \sqrt{1-\frac{4m^2\delta^2}{q^2}}} = \frac{1}{2} \quad , \label{asy}
\ee
so that the l.h.s. of Eq. (\ref{res}) is weakly
convergent to the r.h.s.
In spite of ${\mbox{sgn}} (k^0)$,
the r.h.s. of Eq. (\ref{res}) is a well-defined tempered distribution
due to the factor $(g^{\mu \nu} k^2 - k^\mu k^\nu)$.

This has the following consequence: The (Fourier transformed)
retarded part $r_2^{\mu \nu}$ of $d_2^{\mu \nu}$ would be given in
the case $\omega < 0$ by the unsubtracted splitting formula
\begin{displaymath}
\hat{r}_2^{\mu \nu}(k) = \frac{i}{2 \pi} \int \limits_{-\infty}^{\infty}
\frac{dt}{1-t+i0}d_2^{\mu \nu}(tk)
\end{displaymath}
\be
=\frac{im^2}{\pi^2} \Bigl( g^{\mu \nu} - \frac{k^\mu k^\nu}{k^2} \Bigr)
\frac{1}{k^2 \sqrt{1-4m^2/k^2}} \log \frac{\sqrt{1-4m^2/k^2}+1}
{\sqrt{1-4m^2/k^2}-1} \quad ,
\quad k^2>4m^2,k^0>0. \label{unsubtracted}
\ee
This distribution will vanish in the limit $m \rightarrow 0$, and
the photon would remain massless. But since we have $\omega=0$,
we must use the subtracted splitting formula
\begin{displaymath}
\hat{r}_2^{\mu \nu}(k) = \frac{i}{2 \pi} \int \limits_{-\infty}^{\infty}
\frac{dt}{(t-i0)^{\omega+1}(1-t+i0)} \hat{d}_2^{\mu \nu}(tk)
\end{displaymath}
\be
=\frac{im^2}{\pi^2} \Bigl( g^{\mu \nu} - \frac{k^\mu k^\nu}{k^2} \Bigr)
\Bigl( \frac{1}{k^2 \sqrt{1-4m^2/k^2}} \log \frac{\sqrt{1-4m^2/k^2}+1}
{\sqrt{1-4m^2/k^2}-1} + \frac{1}{2m^2} \Bigr) \quad ,
\quad k^2>4m^2,k^0>0 \, . \label{subtracted}
\ee
The new local term survives in the limit $m \rightarrow 0$. After a consistent resummation
of the second order VP diagrams, which we will not discuss here, the well-known Schwinger
mass term $m_s^2=e^2/\pi$ appears.
Consequently, the difference between simple power-counting and the
correct determination of the singular order is by no means a mathematical
detail, it is crucial for the proper description of the dynamics of the model.
The singular order $\omega=0$ of the distribution ${\hat d}_2^{\mu \nu}$ implies
that a local renormalization is admissible, and even necessary to preserve
the gauge structure of the theory.

An further important property of Eq. (\ref{subtracted}) is its behavior for
$k^2 \rightarrow 0$. From $\hat{r}_2^{\mu \nu}$ one obtains the corresponding VP amplitude
$\hat{t}_2^{\mu \nu}$ by the replacement $m^2 \rightarrow m^2-i0$ for arbitrary $k$.
It is then straightforward to show that the logarithmic term in Eq. (\ref{subtracted})
behaves for $0>k^2 \rightarrow 0$ like
\be
\log \frac{\sqrt{1-4m^2/k^2}+1}{\sqrt{1-4m^2/k^2}-1} \sim
\log \frac{1+\sqrt{-k^2/4m^2}}{1-\sqrt{-k^2/4m^2}} \sim \sqrt{-k^2/m^2},
\ee
where $k^2<0$ and $\log(1+z)=z$ for $|z| \ll 1$ was used and the case $0<k^2 \rightarrow 0$
behaves accordingly.
Therefore, we obtain the VP amplitude
\be
\hat{t}_2^{\mu \nu}(k)=\Bigl( g^{\mu \nu} - \frac{k^\mu k^\nu}{k^2} \Bigr) \hat{t}_2(k),
\ee
where $\hat{t}_2=\hat{t}_{2, \mu}^{\quad \mu}$ with $\hat{t}_2 (k) \rightarrow 0$
for $k^2 \rightarrow 0$.
Of course, this observation is a direct consequence of the central splitting solution
used in Eq. (\ref{subtracted}).

The causal method provides the most unambiguous guide to the construction of
every order of the perturbative $S$-matrix.

\subsubsection{\it Dimensional Regularization}
We start with the traditional expression for the VP in 1+1 dimensions
given by
\be
\tilde{t}_{\mu \nu}(k) = \int  \frac{d^2 p}{(2 \pi)^2} \mbox{tr} \,
\gamma_\mu \frac{1}{\slash{\! \! \! p}-m} \gamma_\nu
\frac{1}{\slash{\! \! \! p-\slash \! \! \! k}-m}, \label{vp_dim}
\ee
where $m^2$ is used synonymously for $m^2-i0$.
According to the recipes of the dimensional regularization procedure,
we consider the trace of Eq. (\ref{vp_dim}) in $n$ dimensions
\be
\tilde{t}_\mu^{\, \mu}(k)=-2^{n/2}  \int \frac{d^n p}{(2 \pi)^n}
\frac{(2-n)(p^2-pk)+n m^2}{(p^2-m^2)[(p-k)^2-m]}. \label{dim_trace}
\ee
We will now proceed in two different ways. First, we perform a naive
dimensional regularization by taking the limit $(2-n) \rightarrow 0$
in Eq. (\ref{dim_trace}) before performing the integral. Then only the term
\be
\tilde{t}_\mu^{\, \mu}(k)=2^{n/2} i \int \frac{d^n p}{(2 \pi)^n}
\frac{n m^2}{(p^2-m^2)[(p-k)^2-m]} \label{dim_trace_1}
\ee
remains. Inserting the Feynman parameter integral
\be
\frac{1}{ab}=\int \limits_{0}^{1} \frac{dx}{[ax+b(1-x)]^2}
\ee
and using the 't Hooft-Veltman formula \cite{Hooft1972}
\be
I_0:=\int \frac{d^n p}{(p^2-2 pk -m^2)^\alpha}=i^{1-2 \alpha} \pi^{n/2}
\frac{\Gamma(\alpha-n/2)}{\Gamma(\alpha)} \frac{1}{(k^2+m^2)^{\alpha-n/2}}
\ee
for $\alpha=2$ and $n=2$, we obtain
\be
I_0:=- \frac{ i \pi}{x(1-x) k^2-m^2}
\ee
and hence
\be
\tilde{t}_\mu^{\, \mu}(k)=\frac{i}{\pi} \int \limits_{0}^{1} \frac{m^2 dx}{x(1-x) k^2 -m^2}.
\ee
The important observation is that obviously
\be
\tilde{t}_\mu^{\, \mu}(k) \rightarrow -i/\pi \quad \mbox{for} \, \,  k \rightarrow 0,
\label{unsub_dim}
\ee
i.e. we are left with the same problem as in Eq. (\ref{unsubtracted}) that
Eq. (\ref{unsub_dim}) reproduces only "half" the VP amplitude.
Note that the result differs by a factor $2 \pi$ from the result derived in our
causal framework, since there a symmetric definition of the (inverse) Fourier
transform has been used.

We now renormalize Eq. (\ref{vp_dim}) properly according to the prescriptions
of 't Hooft and Veltmann and include all terms in the integral.
The full tensor structure is given by
\be
\tilde{t}_{\mu \nu}(k)=-2^{n/2} \int  \frac{d^n p}{(2 \pi)^n} \int \limits_{0}^{1}
\frac{[(p^\alpha p^\beta - p^\alpha k^\beta)(g_{\mu \alpha} g_{\nu \beta}
-g_{\mu \nu} g_{\alpha \beta} + g_{\mu \beta} g_{\nu \alpha}) +m^2 g_{\mu \nu}]}
{[p^2-2 pk (1-x) + k^2(1-x) -m^2]^2}.
\label{vp_dim_full}
\ee
To perform the integrals, we use the 't Hooft-Veltman integrals
\be
\int \frac{d^n p \, p^\mu}{(p^2-2pk-m^2)^\alpha}=I_0 k^\mu,
\ee
\be
\int \frac{d^n p \, p^\mu p^\nu}{(p^2-2pk-m^2)^\alpha}=
\Bigl( k^\mu k^\nu +\frac{k^2+m^2}{n+2-2 \alpha} g^{\mu \nu} \Bigr) I_0,
\ee
for $\alpha=2$ and $n \rightarrow 2$ and obtain
\begin{displaymath}
\tilde{t}_{\mu \nu}(k)= \frac{2^{n/2} i \pi}{(2 \pi)^n} \int \limits_{0}^{1}
\frac{dx}{x(1-x) k^2 -m^2}
\{m^2 g_{\mu \nu} -(1-x) k^\alpha k^\beta ( g_{\mu \alpha} g_{\nu \beta} -
g_{\mu \nu} g_{\alpha \beta}+g_{\mu \beta} g_{\nu \alpha})
\end{displaymath}
\be
+[(1-x)^2 k^\alpha k^\beta +\frac{1}{2-n}(x(1-x)k^2 - m^2) g^{\alpha \beta}]
(g_{\mu \alpha} g_{\nu \beta}-g_{\mu \nu} g_{\alpha \beta} +
g_{\mu \beta} g_{\nu \alpha}) \}.
\ee
Interestingly, the integral above is finite due to a cancellation of
the dimensional pole $(2-n)^{-1}$ by a factor generated by
\be
g_{\mu \alpha} g_{\nu \beta}-g_{\mu \nu} g_{\alpha \beta} +
g_{\mu \beta} g_{\nu \alpha}=(2-n) g_{\mu \nu},
\ee
since $g_\mu^{\, \mu}=n$.
This way we arrive at the \emph{gauge invariant} amplitude
\be
\tilde{t}_{\mu \nu}(k)=(k^2 g_{\mu \nu} - k_\mu k_\nu) \tilde{t}(k),
\ee
\be
\tilde{t}(k)=\frac{i}{\pi} \int \limits_{0}^{1}
\frac{x(1-x) dx}{x(1-x) k^2-m^2}
\ee
with vanishing trace in the limit $k^2 \rightarrow 0$
\be
\tilde{t}_{\mu}^{\, \mu}(k)=k^2 \tilde{t}(k)=-\frac{i}{\pi} \frac{k^2}{6 m^2} +
o \Bigl( \frac{k^4}{m^4} \Bigr) \rightarrow 0.
\ee
Both results in the dimensional and causal regularization scheme are consistent,
however, we observe that the regularization of distributions must be performed
with due care.

\subsection{\it Scalar QED in 2+1 Dimensions} {\label{sec:sQED}}
In this section we illustrate how gauge invariance
is automatically preserved by dimensional regularization
by using scalar quantum electrodynamics (sQED) in one time and two space dimensions as
a an example. As expected, both the causal method and
dimensional regularization lead to compatible results,
although the underlying premises on which the two methods
are based and the resulting perturbative description of
the model theory display a rather different behavior.

The traditional starting point of any quantum field theory is a
Lagrangean containing coupled classical fields describing the interaction.
After quantization, $S$-matrix elements or Greens functions are
constructed with the help of, e.g., Feynman rules. One should accept that
this point of view is obsolete to some extent within the causal approach, which is
based on a description of the $S$-matrix by the help of free fields and well-defined
interaction terms which can be expressed as a sum of normal-ordered products of
free fields. However, the Lagrangean provides a formal tool to express the classical
structure of the theory, but one should keep in view that physical theories
are always subject to quantization.

The scalar $\mbox{sQED}_{2+1}$ Lagrangean is given by
\be
\mathcal{L} = -\frac{1}{4} F_{\mu \nu} F^{\mu \nu} +(\partial_\mu -ie A_\mu) \varphi^\dag
(\partial^\mu + ie A_\mu) \varphi - m^2 \varphi^\dag \varphi,
\ee
where $A_\mu$ describes the electromagnetic field with the field strength tensor
$F_{\mu \nu} = \partial_\mu A_\nu -\partial_\nu A_\mu$ and $\varphi$ a charged scalar meson field
with mass $m$ and electric charge $e$. The coupling constant $e$ has the dimension of an
energy in three-dimensional spacetime, and consequently $\mbox{sQED}_{2+1}$ is superrenormalizable
by naive power counting.

The Lagrangean can be decomposed according to
\be
\mathcal{L} = \mathcal{L}^0_{em}+\mathcal{L}^0_{matter}+\mathcal{L}_{int},
\ee
where the interaction part $\mathcal{L}_{int}$ is given by the minimal coupling of the
electromagnetic current to the electromagnetic potential
\be
\mathcal{L}_{int} = - j_\mu A^\mu, \quad j_\mu= ie \varphi^\dag
\stackrel{\leftrightarrow}{\partial}_\mu \varphi -e^2 \varphi^\dag \varphi A_\mu.
\ee
It is now straightforward to construct the Hamiltonian interaction density
\be
\mathcal{H}_{int} = ie \varphi^\dag \stackrel{\leftrightarrow}{\partial}_\mu \varphi A^\mu
-e^2 \varphi^\dag \varphi A_\mu A^\mu +e^2 \varphi^\dag \varphi (A^0)^2.
\ee
Obviously, this expression is not manifestly covariant. It has been shown in
\cite{Rohrlich} that the non-covariant term $-e^2 \varphi^\dag \varphi (A^0)^2$
is canceled in the full perturbative quantum field theory by a local normalization term
appearing in the so-called seagull graph at second order in the coupling constant which is
sesquilinear in the meson and bilinear in the photon field. This observation applies to
scalar sQED in $n+1$ dimensions in general, and a detailed discussion of $\mbox{sQED}_{3+1}$
with respect to gauge invariance can be found in \cite{Duetsch_sQED,Scharf}.

It must also be mentioned that scalar sQED has a pathological infrared behavior both in
$3+1$ and in $2+1$ dimensions. E.g., the contraction of the two photon lines in the seagull
graph with the two photon lines in another seagull graph gives rise to a
$\varphi^\dag \varphi^\dag \varphi \varphi$-interaction which is highly singular at short
distances, and it is generally accepted that this leads to a transmutation of the original
underlying perturbative theory. However, in this work we will focus on strictly perturbative
aspects of the theory, which are well-defined at every order of the coupling constant $e$
in our case.

\subsubsection{\it Causal Approach}
The crucial difference between the two approaches discussed in this work
is the following. In the causal approach, the interaction Hamiltonian density
is given by the \emph{normally ordered} product of free quantized fields,
whereas in the dimensional regularization ansatz, all kinds of UV divergences
including terms which stem from contractions of fields at the same space-time
point are taken into account.

In the Feynman gauge, the free photon field $A_\mu(x)$ fulfills the wave equation
\be
\Box A_\mu(x)=0
\ee
and the translation invariant distributional commutation relations
\be
[A_\mu(x),A_\nu(y)]=[A_\mu(x-y),A_\nu(0)]= i g_{\mu \nu} D^{(+)}_0 (x-y),
\ee
where
\be
D^{(+)}_0 (x)=\frac{i}{(2 \pi)^2} \int d^3 p \delta (p^2) \Theta(p^0) e^{-ipx}
\ee
applies for photonic contractions without time-ordering.
For the scalar field, we have the contractions
\be
\contracted{}{\varphi}{(x)}{\varphi}{^\dagger(y)} = -i D_m^{(+)}(x-y),
\ee
\be
\contracted{}{\varphi}{(x)}{^\dagger \varphi}{(y)} = +i D_m^{(+)*}(x-y) = +i D_m^{(-)}(y-x)=-iD_m^{(+)}(x-y) ,
\ee
where
\be
D^{(+)}_m (x)=\frac{i}{(2 \pi)^2} \int d^3 p \delta (p^2-m^2) \Theta(p^0) e^{-ipx}.
\ee

One may choose as a starting point a
Hamiltonian density which is given by the normally ordered products of free fields
\be
\mathcal{H}_{int}(x) = -(ie \varphi^\dag (x) \stackrel{\leftrightarrow}{\partial}_\mu \varphi(x)
+e^2 \varphi^\dag (x) \varphi (x) A_\mu (x)) A^\mu (x) \label{Hdensity}
\ee
and $x$ is an element of $2+1$-dimensional Minkowski space. The perturbative
$S$-Matrix is then constructed according to the expansion
\be
S={\bf 1}+\sum \limits_{n=1}^{\infty} \frac{(-i)^n}{n !} \int dx^3_1 ... dx^3_n
T\{{\cal{H}}_{int}(x_1) {\cal{H}}_{int}(x_2) \cdot ... \cdot {\cal{H}}_{int}(x_n)\}, \label{stoer1}
\ee
where $T$ is the time-ordering operator.

In the causal approach, we use only the first order term (in the coupling constant) of
Eq. (\ref{Hdensity})
(which is motivated by first order interaction term appearing in the corresponding Lagrangian)
given by
\be
T_{1}(x) =
e :   \varphi^{\dag}(x) \stackrel{\leftrightarrow}{\partial}_{\mu} \varphi(x)  :  A^{\mu}(x)  .
\ee Thus, the primed distributions can be written as (taking into account
that $\widetilde{T}_{1} = - T_{1}$)
\begin{eqnarray}
A^{\prime}_{2}(x,y) = -e^{2}: \varphi^{\dag}(x) \stackrel{\leftrightarrow}{\partial}_{\mu} \varphi(x) : : \varphi^{\dag}(y) \stackrel{\leftrightarrow}{\partial}_{\nu} \varphi(y)  :  : A^{\mu}(x) A^{\nu}(y) :  , \\
R^{\prime}_{2}(x,y) = -e^{2}:  \varphi^{\dag}(y) \stackrel{\leftrightarrow}{\partial}_{\mu} \varphi(y)  :  :  \varphi^{\dag}(x) \stackrel{\leftrightarrow}{\partial}_{\nu} \varphi(x)  :  : A^{\mu}(x) A^{\nu}(y) :  ,
\end{eqnarray}
the causal distribution is again
\be
D_{2}(x,y) = R^{\prime}_{2}(x,y) - A^{\prime}_{2}(x,y) =[T_1(x),T_1(y)] .
\ee
As an example, from Wick's theorem we have
\begin{displaymath}
: \varphi^{\dag}(x)  \partial^{\mu}\varphi(x) :  : \varphi^{\dag}(y)  \partial^{\nu}\varphi(y) :
\end{displaymath}
\begin{displaymath}
 =  : \varphi^{\dag}(x)  \partial^{\mu}\varphi(x)  \varphi^{\dag}(y)  \partial^{\nu}\varphi(y) : + \varphi^{\dag}(x) \partial^{\nu}\varphi(y) :  i\partial^{\mu}_{x} D^{(+)}_m(x-y)
\end{displaymath}
\be
- : \partial^{\mu}\varphi(x)  \varphi^{\dag}(y) :  i\partial^{\nu}_{y} D^{(-)}_m(y-x)
+ \partial^{\mu}_{x} D^{(+)}_m(x-y)  \partial^{\nu}_{y} D^{(-)}_m(y-x)  ,
\ee
with $D^{(\pm)}_m$ the positive/negative frequency part of the Pauli-Jordan distribution.

We focus now on the vacuum polarization diagram, corresponding at second order to
\begin{eqnarray}
D_{2}(x,y) = -e^{2} : A_{\mu}(x) \: A_{\nu}(y) :\\
\cdot \left[ \partial^{\nu}_{y}D^{(+)}_m(y-x) \: \partial^{\mu}_{x}D^{(-)}(x-y) -
\partial^{\nu}_{y}D^{(-)}_m(y-x)  \partial^{\mu}_{x}D^{(+)}_m(x-y) \right. \\
+ \partial^{\mu}_{y}D^{(+)}_m(y-x) \: \partial^{\nu}_{y}D^{(-)}_m(x-y) -
\partial^{\mu}_{x}D^{(-)}_m(y-x) \partial^{\nu}_{y}D^{(+)}_m(x-y) \\
- \partial^{\mu}_{x}\partial^{\nu}_{y}D^{(+)}_m(y-x) \: D^{(-)}_m(x-y) + \partial^{\mu}_{x}\partial^{\nu}_{y}D^{(-)}_m(y-x) \: D^{(+)}_m(x-y) \\
\left. - D^{(+)}_m(y-x)  \partial^{\mu}_{x}\partial^{\nu}_{y}D^{(-)}_m(x-y) + D^{(-)}_m(y-x)
 \partial^{\mu}_{x}\partial^{\nu}_{y}D^{(+)}_m(x-y) \right] \: .
\end{eqnarray}
For further calculations, we get rid of external fields and change to momentum space
\be
D_{2}(x,y) =: d^{\mu\nu}(x-y) \: : A_{\mu}(x) A_{\nu}(y) : \mbox{ with } \hat{d}^{\mu\nu}(p) =: \frac{e^{2}}{4} \left( g^{\mu\nu} - \frac{p^{\mu}p^{\nu}}{p^{2}} \right)  \hat{d}(p)  .
\ee
In a straightforward manner, one derives ($p=\sqrt{p^2}$)
\be
\hat{d}(p) = \frac{p}{2} \Theta(p^{2}-4m^{2})  \mbox{sgn}(p_{0})  \left( 1 - \frac{4m^{2}}{p^{2}} \right)  ,
\ee
Obviously, this distribution has singular (and power counting) degree $\omega = 1$.

The retarded distribution $\hat{r}(p)$ follows directly from the splitting formula
(first for $p=(p^0,\vec{0})$, $p^0=p_0>0$)
\be
\hat{r}(p_{0}) =  i (2\pi)^{2} p_{0}^{\omega+1} \int\limits_{-\infty}^{\infty} dk_{0} \: \frac{\left|k_{0}\right| \Theta(k_{0}^{2}-4m^{2}) \: \mbox{sgn}(k_{0})}{(k_{0}-i0)^{\omega+1} \: (p_{0}-k_{0}+i0)} \cdot \frac{1}{2} \left( 1 - \frac{4m^{2}}{k_{0}^{2}} \right) 
\ee
This dispersion integral can be written
\begin{displaymath}
\hat{r}(p_{0}) = 2 i \pi^{2} p_{0}^{2} \int \limits_{4 m^2}^{\infty} \Biggl[
\frac{k^0}{k_0^2 (p^0-k^0+i0)}-\frac{k^0}{k_0^2 (p^0+k^0+i0)} \Biggl]
\left( 1 - \frac{4m^{2}}{k_{0}^{2}} \right) dk^0
\end{displaymath}
\begin{displaymath}
=2 i \pi^{2} p_{0}^{2} \int \limits_{4 m^2}^{\infty}
\frac{2 k^0 dk^0}{k_0^2 ((p^0+i0)^2-k_0^2)} \left( 1 - \frac{4m^{2}}{k_{0}^{2}} \right)
\end{displaymath}
\be
=  2 i \pi^{2} p_{0}^{2} \int\limits_{4m^{2}}^{\infty} \frac{ds}{s^{3/2}}
\frac{s-4m^{2}}{p_{0}^{2}-s+i0 p^{0}} \: ,
\ee
where a substitution $s=k_0^2$ was used.
For the C-number part $r^{\prime}_{2}$ of $R^{\prime}_{2}$ follows
\be
\hat{r}^{\prime}(p_{0}) = - \frac{p_{0}}{2} \Theta(p_{0}^{2}-4m^{2}) \Theta(-p_{0}) \left( 1 - \frac{4m^{2}}{p_{0}^{2}} \right)
\ee
and therefore
\be
\hat{t}(p_{0}) = \hat{r}(p_{0}) - \hat{r}^{\prime}(p_{0}) = 2 i \pi^{2} p_{0}^{2} \int\limits_{4m^{2}}^{\infty} \frac{ds}{s^{3/2}} \frac{s-4m^{2}}{p_{0}^{2}-s+i0} \: .
\ee
The integral can be evaluated by standard methods, and 
proper analytic continuation of $\hat{r}(p_{0})$ leads to the result
\be
\hat{t}^{\mu\nu}(p) = - i e^{2} \pi^{2} p \left( g^{\mu\nu} - \frac{p^{\mu}p^{\nu}}{p^{2}} \right) \left[ \frac{2m}{p} + \frac{p^{2}-4m^{2}}{2p^{2}} \log  \frac{2m+p}{2m-p}  \right] \: .
\ee

\subsubsection{\it Dimensional Regularization}
Applying the well-known Feynman rules to the vacuum polarization diagram leads to
\be
\tilde{t}^{\mu \nu} = 
\int d^{3}k \frac{2 k^{\mu}p^{\nu} + 2 k^{\nu} p^{\mu} - p^{\mu} p^{\nu}-4 k^{\mu}k^{\nu}}
{\left( k^2-m^{2} + i0 \right) \left[ \left( k-p \right)^{2}-m^{2} + i0 \right]} \: ,
\ee
using the definition of the Feynman propagator
\be
\left\langle 0 \right| \mathrm{T} \left[ \varphi(x) \varphi^{\dag}(y) \right] \left| 0 \right\rangle = -i D_{F}^m (x-y) \: .
\ee
The integral can be split into a scalar, vector and tensor part:
\be
\tilde{t}^{\mu \nu} = 2 p^{\nu} \mathcal{I}_{2}^{\mu}(p) + 2 p^{\mu} \mathcal{I}_{2}^{\nu}(p)
-p^{\mu}p^{\nu} \mathcal{I}_{1}(p) - 4 \mathcal{I}_{3}^{\mu\nu}(p) \: , \label{split}
\ee
with the definitions
\begin{eqnarray}
\mathcal{I}_{1}(p) & := &  \int d^{3}k \frac{1}{\left( k^2-m^{2} + i0 \right) \left[ \left( k-p \right)^{2}-m^{2} + i0 \right]} \\
\mathcal{I}_{2}^{\mu}(p) & := & \int d^{3}k \frac{k^{\mu}}{\left( k^2-m^{2} + i0 \right) \left[ \left( k-p \right)^{2}-m^{2} + i0 \right]} \\
\mathcal{I}_{3}^{\mu\nu}(p) & := & \int d^{3}k \frac{k^{\mu}k^{\nu}}{\left( k^2-m^{2} + i0 \right) \left[ \left( k-p \right)^{2}-m^{2} + i0 \right]} \: .
\end{eqnarray}
We first calculate $\mathcal{I}_{1}$. Using the Feynman parametrization
\be
\frac{1}{AB} = \int\limits_{0}^{1} d\alpha \left[\alpha A + (1-\alpha)B\right]^{-2}
\ee
and performing the momentum translation $k^{\mu} \mapsto k^{\mu} + \alpha p^{\mu}$, the integral can be written as
\be
\mathcal{I}_{1} (p) = \int d^{3}k \int\limits_{0}^{1} d\alpha  \left[ k^{2} - m^{2} + \alpha (1- \alpha) p^{2} + i0 \right]^{-2} \: .
\ee
Changing the integration dimension to $D = 3-2\epsilon$ and using the general relation
\be
\int d^{D}k \frac{\left(k^{2}\right)^{r}}{\left(k^{2}-a^{2}+i0\right)^{m}} = i (-1)^{r-m} \: \pi^{\frac{D}{2}} \: \frac{\Gamma \left(r+\frac{D}{2}\right) \Gamma \left(m-r-\frac{D}{2}\right)}{\Gamma\left(\frac{D}{2}\right) \Gamma(m) \left(a^{2}-i0\right)^{m-r-\frac{D}{2}}} \: , \label{dim_int}
\ee
the momentum integral can be carried out (with the trivial limit $\epsilon \rightarrow 0$), after which the $\alpha$-integral becomes simply
\be
\mathcal{I}_{1} (p) = i \pi^{2} \int\limits_{0}^{1} \frac{d\alpha}{\sqrt{\alpha(\alpha-1)p^{2} + m^{2}}} = \frac{i \pi^{2}}{p} \log \frac{2m+p}{2m-p} , \label{scalar_I1}
\ee
where $p = \sqrt{p^{2}}$.

Note that Eq. (\ref{scalar_I1}) is indeed valid for arbitrary $p$, when $m^2$ is substituted
by $m^2-i0$. For $p^2>4 m^2$, the logarithmic term contains both a real and an imaginary part,
for $0<p^2<4m^2$, the logarithm is real. For space-like momenta $p^2<0$, the logarithm becomes
purely imaginary, but also the prefactor $1/p = 1/\sqrt{p^2}=-i/\sqrt{-p^2}$, since the integrand in
Eq. (\ref{scalar_I1}) is real in this case.
For $p^2<0$, one can also write
\be
\mathcal{I}_{1} (p)  = \frac{i \pi^{2}}{\sqrt{p^2}} \log \frac{2m+\sqrt{p^2}}{2m-\sqrt{p^2}}
 = \frac{2 i \pi^{2}}{\sqrt{|p^2|}} \mbox{arcsin}
\Biggl( \frac{1}{1-4m^2/p^2} \Biggr)  ,
\ee
but we will maintain the shorthand used in Eq. (\ref{scalar_I1}) in the following.

The same procedure as above can be applied to $\mathcal{I}_{2}^{\mu}$, leading to
\be
\mathcal{I}_{2}^{\mu}(p) = \int d^{3}k \int\limits_{0}^{1} d\alpha \left( k^{\mu} + \alpha p^{\mu} \right) \left[ k^{2} - m^{2} + \alpha (1- \alpha) p^{2} + i0 \right]^{-2} \: .
\ee
Integrating symmetrically makes the integral proportional to $k^{\mu}$ disappear, leaving
\be
\mathcal{I}_{2}^{\mu}(p) = i \pi^{2} \int\limits_{0}^{1} d\alpha \frac{\alpha p^{\mu}}{\sqrt{\alpha (\alpha-1) p^{2} + m^{2}}} = \frac{i \pi^{2} p^{\mu}}{2p} \log  \frac{2m+p}{2m-p}  \: .
\ee

$\mathcal{I}_{3}^{\mu\nu}$
Finally, we use Feynman parametrization and momentum translation invariance
in order to obtain for $\mathcal{I}_{3}^{\mu\nu}$
\begin{displaymath}
\mathcal{I}_{3}^{\mu\nu}(p) = \int d^{3}k \int\limits_{0}^{1} d\alpha \left[ k^{\mu}k^{\nu} + \alpha \left( k^{\mu}p^{\nu} + p^{\mu}k^{\nu} \right) + \alpha^{2} p^{\nu}p^{\nu} \right]
\end{displaymath}
\be
\times \left[ k^{2} - m^{2} + \alpha (1- \alpha) p^{2} + i0 \right]^{-2} \: ,
\ee
where the integrals proportional to an odd power of $k$ vanish.\\
Rewriting $k^{\mu}k^{\nu} = g^{\mu\nu} k^{2} / D$ and performing $D = 3-2\epsilon$ dimensional integration, we arrive at a finite integral for $\epsilon \rightarrow 0$
\be
\mathcal{I}_{3}^{\mu\nu}(p) = i \pi^{2}  \int\limits_{0}^{1} d\alpha \left[ g^{\mu\nu} \sqrt{\alpha (\alpha-1) p^{2} + m^{2}} + p^{\mu}p^{\nu} \frac{\alpha^{2}}{\sqrt{\alpha (\alpha-1) p^{2} + m^{2}}} \right]  ,
\ee
which can be evaluated in a straightforward manner to give
\begin{displaymath}
\mathcal{I}_{3}^{\mu\nu}(p) = \frac{i\pi^{2}}{4} p \left\{ - \left( g^{\mu\nu} - \frac{p^{\mu}p^{\nu}}{p^{2}} \right)
\cdot \left[ \frac{2m}{p} + \frac{p^{2}-4m^{2}}{2p^{2}} \log  \frac{2m+p}{2m-p} \right]
\right.
\end{displaymath}
\begin{displaymath}
\left.
+ g^{\mu\nu} \frac{4m}{p} + \frac{p^{\mu}p^{\nu}}{p^{2}}  \log \frac{2m+p}{2m-p}  \right\}  .
\end{displaymath}

Now, all integrals can be combined in the manner expressed by Eq. (\ref{split}) to give the full
VP
\begin{displaymath}
\tilde{t}^{\mu\nu}
=  - i\pi^{2} p \cdot \left( g^{\mu\nu} - \frac{p^{\mu}p^{\nu}}{p^{2}} \right)
\cdot \left[ \frac{2m}{p} + \frac{p^{2}-4m^{2}}{2p^{2}} \log  \frac{2m+p}{2m-p} \right]
\end{displaymath}
\be
+ 4i \pi^{2} g^{\mu\nu} m \: . \label{loc_scatt}
\ee

\begin{figure}
\begin{minipage}[b]{0.5\linewidth} % A minipage that covers half the page
\centering
\includegraphics[width=7cm]{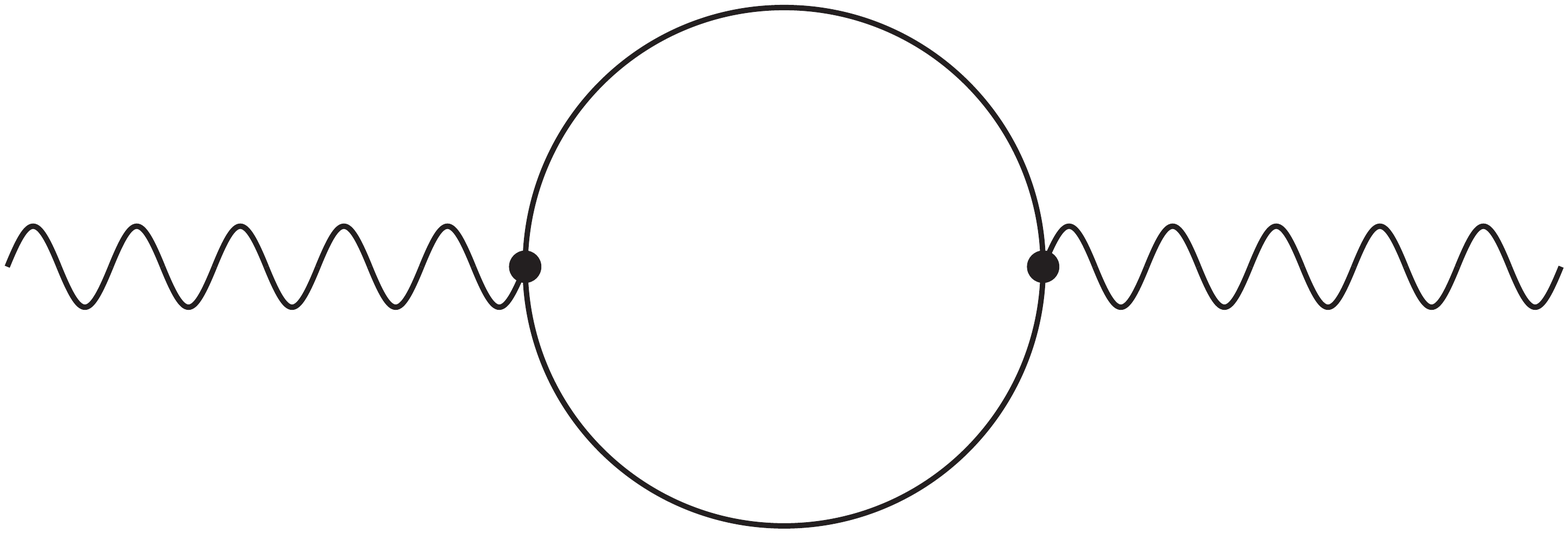}
\caption{Vacuum polarization}
\label{vp}
\end{minipage}
\hspace{0.5cm} % To get a little bit of space between the figures
\begin{minipage}[b]{0.5\linewidth}
\centering
\includegraphics[width=6cm]{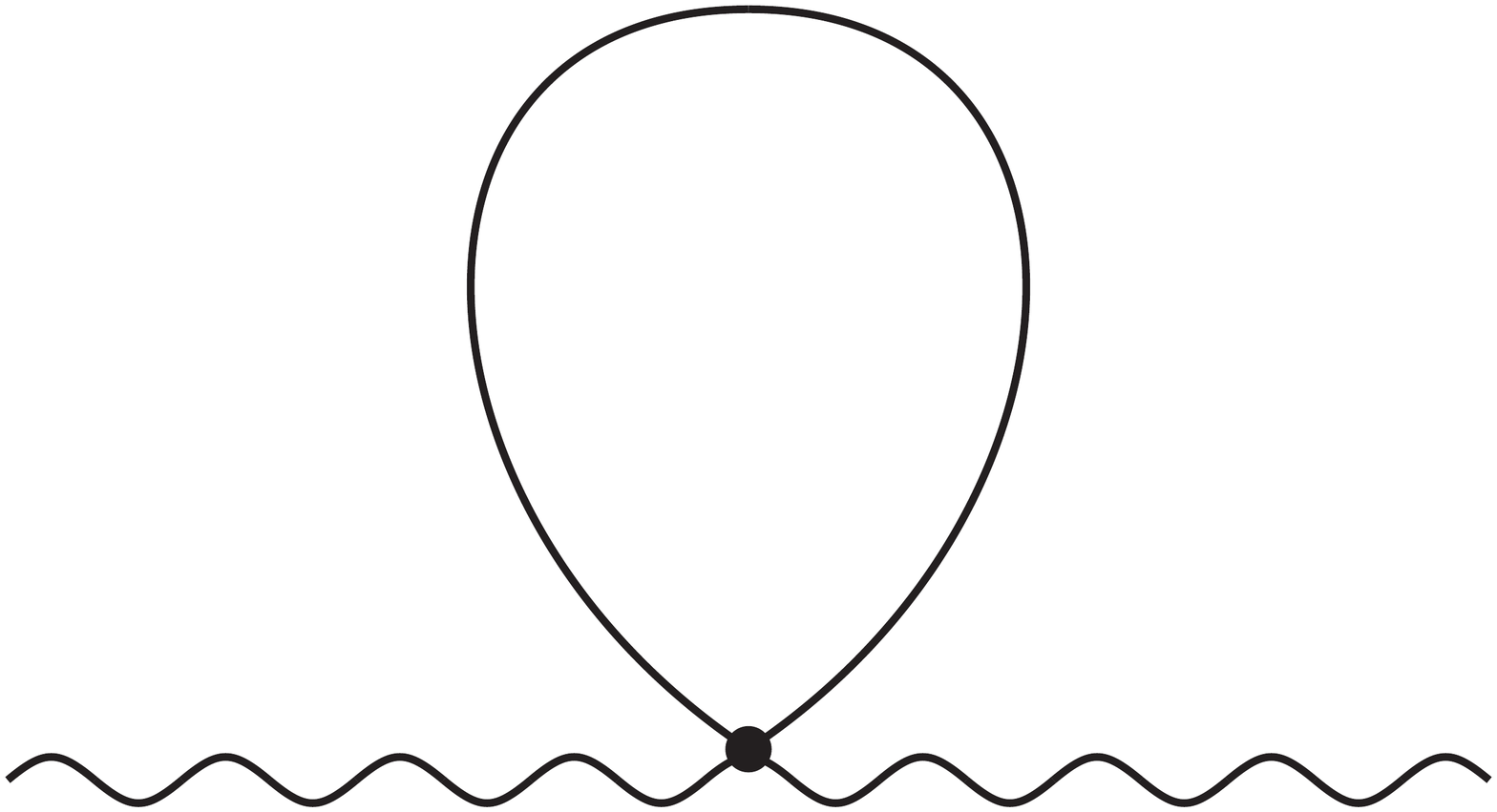}
\caption{One-loop contracted seagull graph}
\label{seagull}
\end{minipage}
\end{figure}

The interesting observation is given by the fact that the result obtained so far does not
exactly match the gauge invariant result derived in the framework of causal perturbation
theory. In fact, we have to include the one-loop contracted seagull graph displayed in Fig. (\ref{seagull}),
which contributes to the photon-photon transition amplitude like the VP diagram
displayed in Fig. (\ref{vp}) as well.

Calculating the formal contribution of the fermion-line self-contraction of the seagull graph gives
(with the correct normalization factor) by the help of Eq. (\ref{dim_int})
\be
\tilde{t}_{seagull}^{\mu \nu} = -2 g^{\mu \nu}
\int \frac{d^{3}k}{k^{2}-m^{2} + i0} = -4 g^{\mu \nu} i \pi^{2}  m  \: ,
\ee
which exactly cancels the local term in the scattering matrix element Eq. (\ref{loc_scatt})
of the vacuum polarization, which only appears if one uses dimensional regularization.

\subsubsection{\it Pauli-Villars Regularization \label{pauli21}}
$\mathcal{I}_{1}$, according to naive power counting, is convergent and can be calculated
using Feynman parameters
\be
\mathcal{I}_{1} (p) = \frac{i \pi^{2}}{p} \log \frac{2m+p}{2m-p} \: .
\ee
For $\mathcal{I}_{2}^{\mu}$, we change the propagators according to the procedure
\be
\frac{1}{q^{2}-m^{2}+i0} \mapsto \frac{1}{q^{2}-m^{2}+i0} - \frac{1}{q^{2}-\Lambda^{2}+i0},
\ee
leading to
\be
\mathcal{I}_{2}^{\mu} = \int d^{3}k \frac{k^{\mu} \left(m^{2}-\Lambda^{2} \right)^{2}}{\left(k^2-m^{2}\right) \left(k^2-\Lambda^{2}\right) \left[ (k-p)^{2}-m^{2} \right]\left[ (k-p)^{2}-\Lambda^{2} \right]} \: .
\ee
Subtraction at $p=0$ gives
\begin{displaymath}
\mathcal{I}_{2}^{\mu} (p)  =  \mathcal{I}_{2}^{\mu} (0) + \mathcal{\widetilde{I}}_{2}^{\mu} (p) \, ,
\end{displaymath}
\begin{displaymath}
\mathcal{I}_{2}^{\mu} (0)
=  \int d^{3}k \frac{k^{\mu} \left(m^{2}-\Lambda^{2} \right)^{2}}{\left(k^2-m^{2}\right)^{2} \left(k^2-\Lambda^{2}\right)^{2}}
\end{displaymath}
\begin{displaymath}
\mathcal{\widetilde{I}}_{2}^{\mu} (p) =  \int d^{3}k \frac{k^{\mu} \left(m^{2}-\Lambda^{2} \right)^{2}}{\left(k^2-m^{2}\right) \left(k^2-\Lambda^{2}\right) }
\end{displaymath}
\be
\left\{ \frac{-1}{\left(k^2-m^{2}\right) \left(k^2-\Lambda^{2}\right)} + \frac{1}{\left[ (k-p)^{2}-m^{2} \right] \left[ (k-p)^{2}-\Lambda^{2} \right]} \right\} \: .
\ee
Taking the limit $\Lambda \rightarrow \infty$ yields
\begin{eqnarray*}
\mathcal{I}_{2}^{\mu} (0) & = & 0 \, \\
\mathcal{\widetilde{I}}_{2}^{\mu} (p) & = & \int d^{3}k \frac{-k^{\mu} \left( p^{2}-2k \cdot p \right)}{\left(k^2-m^{2}\right)^{2} \left[ (k-p)^{2}-m^{2} \right]} \: .
\end{eqnarray*}
Using Feynman parametrization
\be
\frac{1}{ABC} = 2 \int\limits_{0}^{1} d\alpha \int\limits_{0}^{\alpha} d\beta \left[ (\alpha-\beta) A + \beta B + (1-\alpha) C \right]^{-3}
\ee
and momentum translation leads to
\begin{displaymath}
\mathcal{\widetilde{I}}_{2}^{\mu} = \int d^{3}k \; 2 \int\limits_{0}^{1} d\alpha \int\limits_{0}^{\alpha} d\beta
\end{displaymath}
\begin{displaymath}
\times \left[ 2 p \cdot k \: k^{\mu} - (1-2\beta)p^{2} k^{\mu} + 2\beta p^{\mu} p \cdot k - (1-2\beta)\beta p^{2} p^{\mu} \right]
\end{displaymath}
\be
\times \left[ k^{2} -\beta(\beta-1)p^{2} - m^{2} + i0 \right]^{-3} \: .
\ee
Again, the integrals proportional to an odd power of $k$ disappear and we find
\be
\mathcal{\widetilde{I}}_{2}^{\mu} = 2 p^{\mu} \int\limits_{0}^{1} d\alpha \int\limits_{0}^{\alpha} d\beta  \int d^{3}k \left[ \frac{2}{3} k^{2} - (1-2\beta)\beta p^{2} \right] \cdot \left[ k^{2} -\beta(\beta-1)p^{2} - m^{2} + i0 \right]^{-3} \: ,
\ee
which can now be Wick rotated and integrated over k to
\begin{displaymath}
\mathcal{\widetilde{I}}_{2}^{\mu} =  \frac{i\pi^{2}}{2}p^{\mu} \int\limits_{0}^{1} d\alpha \int\limits_{0}^{\alpha} d\beta \left\{ \frac{2}{\sqrt{\beta(\beta-1)p^{2} + m^{2}}} + \frac{(1-2\beta)\beta p^{2}}{\left[ \beta(\beta-1)p^{2} + m^{2}\right]^{3/2}} \right\}
\end{displaymath}
\begin{displaymath}
=  i \pi^{2}p^{\mu} \int\limits_{0}^{1} d\alpha \frac{\alpha}{\sqrt{\alpha(\alpha-1)p^{2} +m^{2} }}
\end{displaymath}
\be
=  \frac{i \pi^{2} p^{\mu}}{2p} \log  \frac{2m+p}{2m-p}  \;\; = \;\; \mathcal{I}_{2}^{\mu} (p) .
\ee

The same procedures may be applied to $\mathcal{I}_{3}^{\mu\nu}$, too. One has
\be
\mathcal{I}_{3}^{\mu\nu} = \frac{i \pi^{2}}{4}p \left\{ - \left( g^{\mu\nu} - \frac{p^{\mu}p^{\nu}}{p^{2}} \right) \left[ \frac{2m}{p} + \frac{p^{2}-4m^{2}}{2p^{2}} \log \frac{2m+p}{2m-p} \right] + \frac{p^{\mu}p^{\nu}}{p^{2}} \log \frac{2m+p}{2m-p} \right\} \: .
\ee

Combining all three integrals obtained by Pauli-Villars regularization, one obtains
\be
\mathcal{I}^{\mu\nu} = - i\pi^{2} p \cdot \left( g^{\mu\nu} - \frac{p^{\mu}p^{\nu}}{p^{2}} \right) \cdot \left[ \frac{2m}{p} + \frac{p^{2}-4m^{2}}{2p^{2}} \log  \frac{2m+p}{2m-p} \right] \: ,
\ee
which differs from the solution with dimensional regularization by a term $4i\pi^{2}g^{\mu\nu}m$,
i.e. gauge invariance is preserved "by hand" by proper normalization of the photon-photon
transition amplitude in the present case. However, this is only true if one starts
with a normally ordered first order coupling and neglects the one-loop contracted seagull
graph from the beginning.

Again, we observe that the calculations differ strongly in different approaches.
In the causal method, the non-trivial part of one-loop calculations is a one-dimensional,
\emph{finite} dispersion integral. 
As a general remark, we point out that the causal method, as well as dimensional
regularization, has some particular advantages for gauge theories
because it does not use a cutoff which breaks gauge invariance, an observation which is
illustrated on a rather basic level in this sect. \ref{pauli21}.
In a more general setting, it works in a fixed number of physical dimensions so that
problems originating from axial couplings, which are also related to the
'$\gamma_5$-problem' in dimensional regularization.

\section{Infrared Divergences}
The infrared structure in the causal approach differs strongly from other approaches.
Whereas infrared divergences show up as poles in dimensional regularization
or as divergences in the artificial mass parameter introduced for the originally
massless fields contained in the theory under consideration (a strategy commonly used e.g.
in the Pauli-Villars approach), they are automatically regularized by the test function
$g$ in Eq. (\ref{(2.2.1)}).
From a mathematical point of view, this is the most natural formulation of the
infrared problem, since the $T_n$'s are operator-valued distributions, and
therefore must be smeared out by test functions in $\mathcal{S}(\mathds{R}^{4n})$,
the Schwartz space of functions of rapid decrease. 
The test function $g \in \mathcal{S}(\mathds{R}^{4})$ plays the role of an
"adiabatic switching" and provides a cutoff in the long-range part of the interaction, which
can be considered as a natural infrared regulator.
An appropriate adiabatic limit $g \rightarrow 1$ must be performed at the end
of actual calculations in the right quantities (like cross sections)
where this limit exists. Strictly speaking, the perturbative S-matrix according to
Eq. (\ref{Smatrix_textbook}) does \emph{not} exist for $g \equiv 1$ for many
theories involving massless fields. This observation is also closely related to the notion of
\emph{infraparticles} \cite{Schroer1,Schroer2}. The existence of the adiabatic limit for
physical observables is a non-trivial issue of a theory and is not automatically
guaranteed at every order of the theory from the mere lowest order definition of the
interaction. Below, we will prove the existence of the adiabatic limit for a scattering
cross section at fourth order in $g$ for a model theory.

Note that introducing a mass as infrared regularizator for massless fields is a questionable
procedure, since it is unclear whether the original massless theory
is restored by taking the massless limit of the massive theory, which may suffer from
potential problems like, e.g., broken gauge invariance.

In order to demonstrate the causal approach to the infrared problem
we consider a theory in 3+1 spacetime dimensions, called totally scalar QED in the following,
where a massive scalar charged field is coupled to a massless
scalar field, in close analogy to Eq. (\ref{5.1}).
The corresponding scalar particles will be called meson and photon in the following.
The theory is defined by the first order coupling term
\be
T_1(x)= -ie :  \varphi^\dagger(x) \varphi(x) : [A_0(x)+A^{ext}(x)]
=-ie :  \varphi^\dagger(x) \varphi(x) : A(x),
\ee
where $A^{ext}(x)$ denotes an external C-number field and
$A_0(x)$ the quantized massless neutral scalar photon field. For dimensional reasons,
the coupling constant $e$ has the dimension of an energy or an inverse length.

\begin{figure}
\begin{minipage}[b]{0.5\linewidth} % A minipage that covers half the page
\centering
\includegraphics[width=5.5cm]{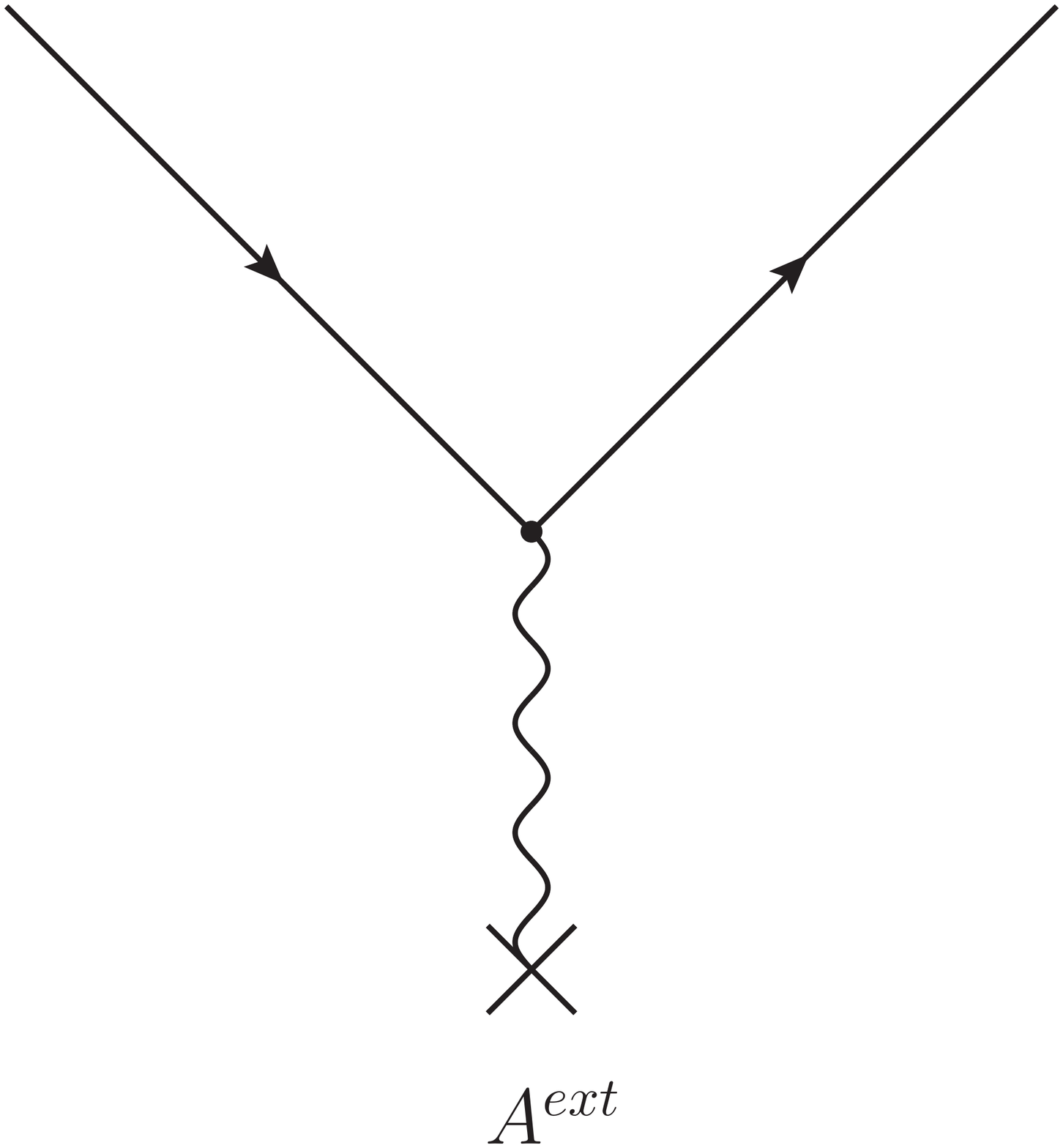}
\caption{Lowest order scattering}
\label{tree}
\end{minipage}
\begin{minipage}[b]{0.5\linewidth} % A minipage that covers half the page
\centering
\includegraphics[width=5.5cm]{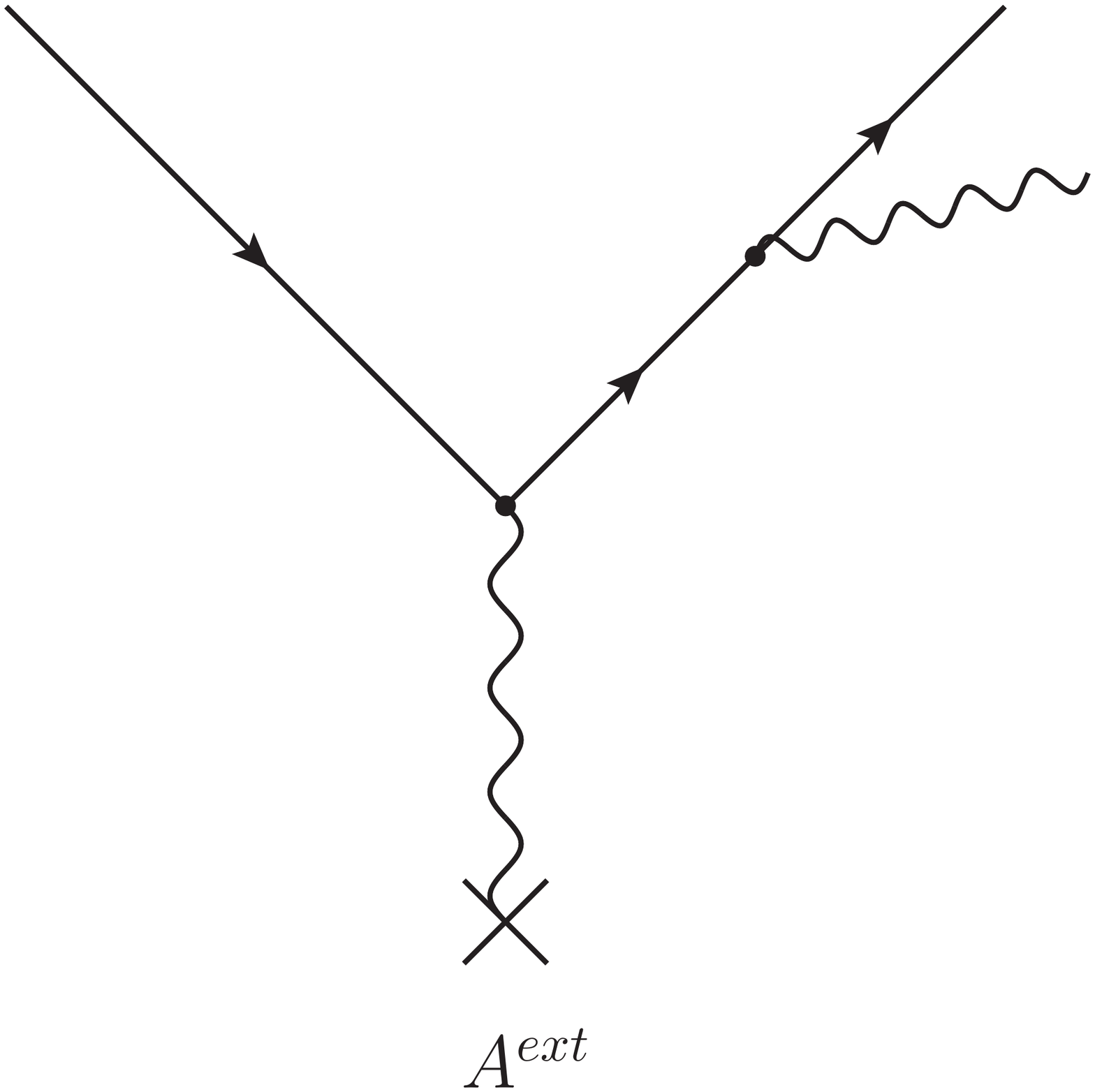}
\caption{Bremsstrahlung diagram}
\label{bremsstrahlung}
\end{minipage}
\end{figure}

In the following, we consider the scattering process of the meson off the
external field, according to Fig. (\ref{tree}).
At first order in the coupling constant, the matrix element for
the scattering of a meson with mass $m$ and initial state momentum $\vec{q}$ 
and a different final state momentum $\vec{p}$ is given by
\be
S^{(1)}_{fi}=\langle f | S^{(1)} | i \rangle = 
\langle 0 | a(\vec{p}) S^{(1)} a^\dagger (\vec{q})| 0 \rangle,
\ee
where $S^{(1)} = -i e \int d^4 x_1 :\varphi^\dagger (x_1) \varphi (x_1): A^{ext}(x_1)$
and $| i \rangle =a^\dagger (\vec{q})| 0 \rangle$ and
$| f \rangle =a^\dagger (\vec{p})| 0 \rangle$, i.e. $a$ and $a^\dagger$ denote the
annihilation and creation operators for one charge type of particles of the meson field,
and the vacuum shall be denoted by $| 0 \rangle$ in this section. One obtains ($k^0=(\vec{k}^{\, 2}+m^2)^{1/2}$ etc)
\be
S^{(1)}_{fi} = -ie \int  \frac{d^4x_1}{(2\pi)^3}\int\frac{d^3kd^3k'}
{2\sqrt{k_0k'_0}}e^{-ikx_1}e^{ik'x_1} \langle0|a(\vec{p}):a(\vec{k})a^\dagger(\vec{k'}):a^\dagger(\vec{q})|0\rangle
A^{ext}(x_1).
\ee
Exploiting the commutation relations and the
distributional identity $\int d^n x e^{ikx}=(2 \pi)^n \delta(k)$ leads to
\begin{eqnarray}
S_{fi}& = &-ie\int d^4x_1 \frac{1}{(2\pi)^3}\int\frac{d^3kd^3k'}{2\sqrt{k_0k'_0}}e^{-ikx_1}e^{ik'x_1} \delta^{(3)}(\vec{q}-\vec{k})\delta^{(3)}(\vec{k'}-\vec{p})  A^{ext}(x_1)
\nonumber\\ & &
=\frac{-ie}{(2\pi)^3}\int d^4x_1 \frac{1}{2\sqrt{p_0q_0}}e^{-iqx_1}e^{ipx_1}  A^{ext}(x_1)
=\frac{-ie}{2(2\pi)^3 \sqrt{p_0q_0}}\int d^4x_1 e^{-iqx_1}e^{ipx_1}  A^{ext}(x_1)
\end{eqnarray}
Introducing the Fourier transform of $A^{ext}(x_1)$, we have
\begin{eqnarray}
S^{(1)}_{fi}&=&\frac{-ie}{2(2\pi)^3\sqrt{p_0q_0}}\int d^4x_1 e^{-iqx_1}e^{ipx_1} \frac{1}{(2\pi)^2}\int d^4k''e^{-ik''x_1} \hat{ A}^{ext} (k'')\nonumber\\
&=&
\frac{-ie}{2(2\pi)^3\sqrt{p_0q_0}}\frac{1}{(2\pi)^2}\int d^4x_1 \int d^4k''
e^{-iqx_1}e^{ipx_1}e^{-ik''x_1} \hat{ A}^{ext} (k'')\nonumber\\
&=&
\frac{-ie}{2(2\pi)^3\sqrt{p_0q_0}}\frac{1}{(2\pi)^2}\int d^4k''(2\pi)^4 
\delta(-q+p-k'') \hat{ A}^{ext} (k''),
\end{eqnarray}
and the first order matrix element becomes
\begin{eqnarray}
S^{(1)}_{fi}&=&\frac{-ie}{2 (2\pi) \sqrt{p_0q_0}} \hat{ A}^{ext} (p-q)
\end{eqnarray}

We assume for the moment that $ A^{ext}(x)$ is a Coulomb potential
\be
A^{ext}(x)=-\frac{1}{|\vec{x}|}.
\ee
For the sake of completeness, we calculate the corresponding cross section in detail.
In the present case, we have
\be
S^{(1)}_{fi}=\frac{ie}{(2\pi)2\sqrt{p_0q_0}}\bigg[\frac{1}{(2\pi)^2}\int d^4x
\frac{e^{i(p-q)x}}{|\vec{x}|} \bigg]
=\frac{ie}{(2\pi)^32\sqrt{p_0q_0}}\int dx_0 e^{i(p_0-q_0)x_0}\int d^3x
\frac{e^{-i(\vec{p}-\vec{q})\vec{x}}}{|\vec{x}|} .
\ee
The space integral can be evaluated as follows
\begin{displaymath}
\int d^3x e^{-i(\vec{p}-\vec{q})\vec{x}}\frac{1}{|\vec{x}|}=-\frac{1}{(\vec{p}-\vec{q})^2}\int d^3x \frac{1}{|\vec{x}|}\triangle e^{-i(\vec{p}-\vec{q})\vec{x}}\nonumber\\=-\frac{1}{(\vec{p}-\vec{q})^2}
\int d^3x\Big(\triangle \frac{1}{|\vec{x}|} \Big)e^{-i(\vec{p}-\vec{q})\vec{x}}
\end{displaymath}
\be
=-\frac{1}{(\vec{p}-\vec{q})^2}\int d^3x\Big(-4\pi\delta^{(3)}
(\vec{x})\Big)e^{-i(\vec{p}-\vec{q})\vec{x}} = \frac{4\pi} {(\vec{p}-\vec{q})^2} ,
\ee
and thus
\be
S^{(1)}_{fi} = \frac{ie}{(2\pi)^32\sqrt{p_0q_0}}(2\pi\delta(p_0-q_0))
\frac{4\pi}{(\vec{p}-\vec{q})^2}.
\ee
The transition rate $dR$ from the initial state to a final state within an infinitesimal
phase space volume $d^3p$ is given by ($T$ denotes a large time interval)
\be
dR=\frac{dW}{T}=\frac{|S_{fi}|^2d^3p}{T}=\frac{e^2}{(2\pi)^4p_0q_0T}
(2\pi\delta(p_0-q_0))^2\frac{d^3p}{(\vec{p}-\vec{q})^4} .
\ee
Of course, the square of the $\delta$-distribution above is ill-defined, since
one should work with wave packets in order to get well-defined expressions.
However, for the moment we content ourselves with Fermi's trick and perform some
\emph{formal} manipulations, starting from
\be
2\pi\delta(p_0-q_0)=\lim_{T\to\infty}\int\limits^{\frac{T}{2}}_{-\frac{T}{2}}
dt e^{i(p_0-q_0)t}.
\ee
The intuitive argument is that this expression is non-zero for $p_0=q_0$,
so that one may replace one $\delta$-distribution by
\be
2\pi\delta(p_0-q_0)=\lim_{T\to\infty}\int\limits^{\frac{T}{2}}_{-\frac{T}{2}} dt
\ee
and for large $T$ one has $2\pi\delta(p_0-q_0)=T$.
Then $dR$ becomes
\be
dR=\frac{e^2}{(2\pi)^4p_0q_0}(2\pi\delta(p_0-q_0))\frac{d^3p}{(\vec{p}-\vec{q})^4} .
\ee
The cross section is given by the ratio of the transition rate $dR$ and the
flux of incoming particles, given by the initial state expectation value of the operator
\be
j^{\mu}=\frac{i}{2} :\varphi^\dagger(x)\partial^{\mu}\varphi(x)-\varphi(x)
\partial^{\mu}\varphi^\dagger(x):,
\ee
leading to
\be
|\vec{j}|=\frac{|\vec{q}|}{(2\pi)^3q_0} 
\ee
for the chosen initial state.
The differential cross section is therefore
\begin{displaymath}
d\sigma =\frac{dR}{|\vec{j}|}=\frac{e^2}{(2\pi)^4\Big(\frac{|\vec{q}|}{(2\pi)^3q_0}\Big)
p_0q_0(\vec{p}-\vec{q})^4}(2\pi\delta(p_0-q_0))|\vec{p}|^2d|\vec{p}|d\Omega
\end{displaymath}
\be
=\frac{e^2}{|\vec{q}|p_0(\vec{p}-\vec{q})^4}\delta(p_0-q_0)|\vec{p}|^2d|\vec{p}|
d\Omega .
\ee
Since $p_0dp_0 = |\vec{p}|d|\vec{p}|$ and $|\vec{p}|=|\vec{q}|$, we obtain
\begin{eqnarray}
d\sigma &=& \frac{e^2}{(\vec{p}-\vec{q})^4}\delta(p_0-q_0)dp_0 d\Omega
\end{eqnarray}
and finally, performing the integral over $p_0$,
\be
\frac{d\sigma}{d\Omega}=\int \frac{e^2}{(\vec{p}-\vec{q})^4}\delta(p_0-q_0)dp_0
=\frac{e^2}{(\vec{p}-\vec{q})^4}\Big|_{p_0=q_0},
\ee
i.e. we basically recover the Rutherford cross section.

\subsection{\it Bremsstrahlung}
We now consider the case where the scattered meson emits a soft photon
with four-momentum $k$ according to Fig. (\ref{bremsstrahlung}).
The scattering matrix at second order is given by
\begin{eqnarray}
\contraction{S^{(2)}=\frac{(-ie)^2}{2}\int d^4x_1 d^4x_2:\varphi(x_1)}
{\varphi^\dagger}{(x_1)A(x_1)::}{\varphi}
S^{(2)}=\frac{(-ie)^2}{2}\int d^4x_1 d^4x_2:\varphi(x_1)\varphi^\dagger(x_1)A(x_1)::
\varphi(x_2)\varphi^\dagger(x_2)A(x_2):g(x_1)g(x_2) \label{2nd_order}
\end{eqnarray}
where the contraction symbol denotes one possible Wick contraction of massive fields.
Since one has two possibilities to contract the massive fields, 
the relevant bremsstrahlung term becomes
\begin{eqnarray}
S^{(2)}=(-ie)^2\int d^4x_1 d^4x_2 (-i) D_F^m (x_1-x_2):\varphi(x_1)A(x_1)::
\varphi^\dagger(x_2)A(x_2):g(x_1)g(x_2) + \ldots
\end{eqnarray}
and the dots denote other terms from the Wick ordering of $S^{(2)}$ which are irrelevant
for the present case. The external field operators have the form
\be
\varphi(x_1)=\frac{1}{(2\pi)^{3/2}}\int\frac{d^3k^{(4)}}{\sqrt{2k^{(4)}_0}}
[a(\vec{k}^{(4)})e^{-ik^{(4)}x_1}+b^\dagger(\vec{k}^{(4)})e^{ik^{(4)}x_1}],
\ee
\be
\varphi^\dagger(x_2)=\frac{1}{(2\pi)^{3/2}}\int\frac{d^3k'}{\sqrt{2k'_0}}
[a^\dagger(\vec{k}')e^{ik'x_2}+b(\vec{k}')e^{-ik'x_2}],
\ee
\be
A(x_1)=\frac{1}{(2\pi)^{3/2}}\int\frac{d^3k''}{\sqrt{2k''_0}}[c(\vec{k''})
e^{-ik''x_1}+c^\dagger(\vec{k''})e^{ik''x_1}]+A^{ext}(x_1),
\ee
\be
A(x_2)=\frac{1}{(2\pi)^{3/2}}\int\frac{d^3k'''}{\sqrt{2k'''_0}}[c(\vec{k'''})
e^{-ik'''x_2}+c^\dagger(\vec{k'''})e^{ik'''x_2}]+A^{ext}(x_2),
\ee
and we obtain the bremsstrahlung matrix element
$S_{fi}^{brems} = \langle0|a(p)c(k)\hat{S}a^\dagger(q)|0\rangle$ after some calculation
\bd
S_{fi}^{brems} = (-ie)^2 \int
d^4x_1d^4x_2 (-i) D_F^m (x_1-x_2)\frac{1}{(2\pi)^{(9/2)}}g(x_1)g(x_2)
\ed
\bd
\bigg[\int\frac{d^3k^{(4)}}{\sqrt{2k_0^{(4)}}}\frac{d^3k'}{\sqrt{2k'_0}}\frac{d^3k''}
{\sqrt{2k''_0}}e^{-ik^{(4)}x_1}e^{ik'x_2}e^{ik''x_1} A^{ext}(x_2)\delta(\vec{p}-\vec{k'})\delta(\vec{q}-\vec{k^{(4)}})\delta(\vec{k}-\vec{k''})
\ed
\be
+\int\frac{d^3k^{(4)}}{\sqrt{2k_0^{(4)}}}\frac{d^3k'}{\sqrt{2k'_0}}\frac{d^3k'''}
{\sqrt{2k''_0}}e^{-ik^{(4)}x_1}e^{ik'x_2}e^{ik'''x_2} A^{ext}(x_1)\delta(\vec{p}-\vec{k'})\delta(\vec{q}-\vec{k^{(4)}})\delta(\vec{k}-\vec{k'''})\bigg]
\ee
\be
= \frac{ie^2}{(2\pi)^{9/2}}\int d^4x_1d^4x_2D_F^m (x_1-x_2)g(x_1)g(x_2)
\left[ \frac{e^{-iqx_1}e^{ipx_2}e^{ikx_1}}{\sqrt{8q_0p_0k_0}} A^{ext}(x_2)+\frac{e^{-iqx_1}
e^{ipx_2}e^{ikx_2}}{\sqrt{8q_0p_0k_0}} A^{ext}(x_1) \right].
\ee
Inserting the Fourier transforms of $g(x_1), g(x_2),  A^{ext}(x_1),  A^{ext}(x_2)$
\bd
g(x_1)=\frac{1}{(2\pi)^{2}}\int d^4k_1e^{-ik_1x_1}\hat{g}(k_1), \quad
g(x_2)=\frac{1}{(2\pi)^{2}}\int d^4k_2e^{-ik_2x_2}\hat{g}(k_2),
\ed
\be
A^{ext}(x_1)=\frac{1}{(2\pi)^{2}}\int d^4k_1e^{-ik_1x_1} \hat{ A}^{ext} (k_1), \quad
A^{ext}(x_2)=\frac{1}{(2\pi)^{2}}\int d^4k_2e^{-ik_2x_2} \hat{ A}^{ext} (k_2)
\ee
and the Fourier transform of the Feynman propagator leads to
\bd
S_{fi}^{brems} = \frac{-ie^2}{(2\pi)^{13/2}\sqrt{8q_0p_0k_0}(2\pi)^4(2\pi)^4}\int d^4k_1d^4k_2
\int d^4k''d^4k\int d^4x_1d^4x_2
\ed
\bd
\bigg[\frac{e^{-ik''x_1}e^{-iqx_1}e^{ikx_1}e^{-ik_1x_1}e^{ik''x_2}e^{ipx_2}e^{-ik'x_2}
e^{-ik_2x_2}}{k''^2-m^2} A^{ext}(k')
\ed
\be
\frac{e^{-ik''x_1}e^{-iqx_1}e^{-ik'x_1}e^{-ik_1x_1}e^{ik''x_2}e^{ipx_2}e^{ikx_2}
e^{-ik_2x_2}}{k''^2-m^2} A^{ext}(k')\bigg]
\hat{g}(k_1)\hat{g}(k_2).
\ee
Note that we sometimes omit the $i0$-term of the Feynman propagators for the sake
of brevity.
We evaluate all trivial integrals and arrive at
\bd
S_{fi}^{brems} = \frac{-ie^2}{2(2\pi)^{13/2}\sqrt{2q_0p_0k_0}}\int d^4k_1d^4k_2 
\bigg[\frac{1}{(q-k+k_1)^2-m^2}+\frac{1}{(p+k-k_2)^2-m^2}\bigg] 
\ed
\be
A^{ext}(p-q+k-k_1-k_2)\hat{g}(k_1)\hat{g}(k_2).
\ee
$k_1$ can be replaced by $k_2$
\bd
S_{fi}^{brems} = \frac{-ie^2}{2(2\pi)^{13/2}\sqrt{2q_0p_0k_0}}\int d^4k_1d^4k_2 
\bigg[\frac{1}{(q-k+k_2)^2-m^2}+\frac{1}{(p+k-k_2)^2-m^2}\bigg]
\ed
\be
A^{ext}(p-q+k-k_1-k_2)\hat{g}(k_1)\hat{g}(k_2).
\ee

Now we investigate the adiabatic limit by first replacing $\hat{g}(k_1)\hat{g}(k_2)$ by
\be
\frac{1}{\epsilon^4}\hat{g}_0\Big(\frac{k_1}{\epsilon}\Big)\frac{1}
{\epsilon^4}\hat{g}_0\Big(\frac{k_2}{\epsilon}\Big)
\ee
corresponding to the replacement $g(x) \rightarrow g_0(\epsilon x)$ in real space,
and taking the limit $\epsilon \rightarrow 0$. $g_0(x)$ is a fixed test function in
$\mathcal{S}(\mathds{R}^4)$ with $g_0(0)=1$, so that $g_0(\epsilon x) \rightarrow 1$
for $\epsilon \rightarrow 0$. Note, however, that $1 \notin \mathcal{S}(\mathds{R}^4)$.
Thus we have
\bd
S_{fi}^{brems} = \frac{-ie^2}{2(2\pi)^{13/2}\sqrt{q_0p_0}\sqrt{2k_0}}\int \frac{d^4k_1}{\epsilon^4}\frac{d^4k_2}{\epsilon^4} 
\bigg[\frac{1}{(q-k+k_2)^2-m^2}
\ed
\be
+\frac{1}{(p+k-k_2)^2-m^2}\bigg]
 A^{ext}(p-q+k-k_1-k_2) \hat{g}_0\Big(\frac{k_1}{\epsilon}\Big)\hat{g}_0\Big(\frac{k_2}
{\epsilon}\Big)
\ee
or
\bd
S_{fi}^{brems} = \frac{-ie^2}{2(2\pi)^{13/2}\sqrt{q_0p_0}\sqrt{2k_0}}\int d^4k_1d^4k_2 
\bigg[\frac{1}{(q-k+\epsilon k_2)^2-m^2}
\ed
\be
+\frac{1}
{(p+k-\epsilon k_2)^2-m^2}\bigg]  A^{ext}(p-q+k-\epsilon k_1-\epsilon k_2)
\hat{g}_0(k_1)\hat{g}_0(k_2).
\ee
Envisaging the limit $\epsilon\rightarrow 0$, we can neglect the $\epsilon$-dependent
term in the argument of $\hat{A}^{ext}$ and perform one trivial integral.
We obtain the result
\be
S_{fi}^{brems} = S^{(1)}_{fi}\frac{e}{(2\pi)^{7/2}\sqrt{2k_0}}\int d^4k_2
\bigg[\frac{1}{2p(k-\epsilon k_2)}- \frac{1}{2q(k-\epsilon k_2)}\bigg]\hat{g}_0(k_2)
\ee
where we used the first order matrix element $S^{(1)}_{fi}$.
The cross section follows, using the lowest order cross section
$\frac{d\sigma^{(1)}}{d\Omega}$
\bd
\frac{d\sigma (k)}{d\Omega}  =  \frac{d\sigma^{(1)}}{d\Omega}\frac{e^2}
{(2\pi)^{7}2|\vec{k}|}\bigg|\int d^4k_2
\bigg[\frac{1}{(-2q(k-\epsilon k_2))}
+\frac{1}{2p(k-\epsilon k_2)}\bigg]\hat{g}_0(k_2)\bigg|^2
\ed
\bd
=  \frac{d\sigma^{(1)}}{d\Omega}
\frac{e^2}{4(2\pi)^{7}2|\vec{k}|}\int d^4k_1d^4k_2 
\bigg[\frac{1}{(q(k-\epsilon k_1))(q(k-\epsilon k_2))}
\ed
\be
-\frac{1}{p(k-\epsilon k_1))q(k-\epsilon k_2)}-\frac{1}{q(k-\epsilon k_1))p(k-\epsilon k_2)}
+\frac{1}{p(k-\epsilon k_1)p(k-\epsilon k_2)}\bigg]\hat{g}_0(k_1)\hat{g}_0(k_2).
\ee
For well-known physical reasons, one has to integrate this cross section over the photon
momenta up to a cutoff $\omega_0$, assuming that photons with momenta $< \omega_0$
are not measured. The four integrands can be rewritten by the help of the Feynman trick,
leading to
\bd
\frac{d\sigma}{d\Omega} =  \int\limits_{|\vec{k}|<\omega_0}d^3k \frac{d\sigma^{(1)}}{d\Omega}\frac{e^2}{4(2\pi)^{7}2|\vec{k}|}\int d^4k_1d^4k_2
\ed
\bd
\bigg[\int\limits^1_0dx\frac{1}{[q(k-\epsilon k_1)x+q(k-\epsilon k_2)(1-x)]^2}
-\int\limits^1_0dx\frac{1}{[p(k-\epsilon k_1)x+q(k-\epsilon k_2)(1-x)]^2}
\ed
\be
-\int\limits^1_0dx\frac{1}{[q(k-\epsilon k_1)x+p(k-\epsilon k_2)(1-x)]^2} +\int\limits^1_0dx\frac{1}{[p(k-\epsilon k_1)x+p(k-\epsilon k_2)(1-x)]^2}
\bigg]\hat{g}_0(k_1)\hat{g}_0(k_2). \label{four_integrals}
\ee
We first consider the fist part of the integral above, the last term can be treated in
an analogous manner.
\bd
I_1=\int\limits^1_0 dx\int d^4k_1d^4k_2
\int\limits_{|\vec{k}|<\omega_0}\frac{d^3k}{2|\vec{k}|}\frac{1}{[q(k-\epsilon k_1)x+q(k-\epsilon k_2)(1-x)]^2}\hat{g}_0(k_1)\hat{g}_0(k_2)
\ed
\bd
=\int\limits^1_0 dx\int d^4k_1d^4k_2
\int\limits_{|\vec{k}|<\omega_0}\frac{|\vec{k}|d|\vec{k}|(-d\cos\theta)d\phi}{2[q_0|\vec{k}|
-|\vec{q}||\vec{k}|\cos\theta-(q\epsilon k_1+(1-x)q\epsilon k_2)]^2}
\hat{g}_0(k_1)\hat{g}_0(k_2)
\ed
\bd
=\int\limits^1_0 dx\int d^4k_1d^4k_2
\int\limits^{\omega_0}_0\frac{|\vec{k}|d|\vec{k}|}{2}\int\limits^1_{-1}
\frac{2\pi d\cos\theta }{[-|\vec{q}||\vec{k}|\cos\theta+q_0|\vec{k}|-(q\epsilon k_1+(1-x)q
\epsilon k_2)]^2}\hat{g}_0(k_1)\hat{g}_0(k_2)
\ed
\bd
=2\pi\int\limits^1_0 dx\int d^4k_1d^4k_2
\int\limits^{\omega_0}_0\frac{d|\vec{k}|}{2|\vec{q}|}
\bigg[\frac{1}{(q_0-|\vec{q}|)|\vec{k}|-(q\epsilon k_1+(1-x)q\epsilon k_2)}
\ed
\bd
 -\frac{1}{(q_0+|\vec{q}|)|\vec{k}|-(q\epsilon k_1+(1-x)q\epsilon k_2)}\bigg]\hat{g}_0(k_1)
\hat{g}_0(k_2)
\ed
\bd
=2\pi\int\limits^1_0 dx\int d^4k_1d^4k_2
\frac{1}{2|\vec{q}|}
\bigg[\frac{1}{(q_0-|\vec{q}|)}\log\Big|(q_0-|\vec{q}|)|\vec{k}|-(q\epsilon k_1+(1-x)q
\epsilon k_2)\Big|^{\omega_0}_0
\ed
\be
-\frac{1}{(q_0+|\vec{q}|)}\log\Big|(q_0-|\vec{q}|)|\vec{k}|-(q\epsilon k_1+(1-x)q
\epsilon k_2)\Big|^{\omega_0}_0\bigg]\hat{g}_0(k_1)\hat{g}_0(k_2). \label{infra_tree}
\ee
Focusing on the divergent part in the expression above, i.e. examining the logarithms
in Eq. (\ref{infra_tree}) for $|\vec{k}|=0$ only results in
\bd
I'_1 = 2\pi\int\limits^1_0 dx\int d^4k_1d^4k_2
\frac{1}{2|\vec{q}|}
\bigg[\frac{1}{(q_0+|\vec{q}|)}-\frac{1}{(q_0-|\vec{q}|)}\bigg]
\log\Big|(q\epsilon k_1+(1-x)q\epsilon k_2)\Big|\hat{g}_0(k_1)\hat{g}_0(k_2)
\ed
\bd
= 2\pi\int\limits^1_0 dx\int d^4k_1d^4k_2\bigg(-\frac{1}{q^2}\bigg)
\Big[\log|\epsilon|+O(1)\Big]\hat{g}_0(k_1)\hat{g}_0(k_2)
\ed
\be
= -2\pi\int\limits^1_0 dx\int d^4k_1d^4k_2\frac{1}{q^2}
\log|\epsilon|\hat{g}_0(k_1)\hat{g}_0(k_2)
= -\frac{(2\pi)^5}{m^2} \log|\epsilon|,
\ee
such that we obtain the divergent contributions to the cross section
from the first and the fourth part in the integral Eq. (\ref{four_integrals})
\be
I'_1+I'_4=-\frac{2(2\pi)^5}{m^2} \log|\epsilon|.
\ee
Finally we calculate the second integral appearing in Eq. (\ref{four_integrals}),
the third integral can be calculated analogously. We have
\bd
I_2=-\int\limits^1_0 dx\int d^4k_1d^4k_2
\int\limits_{|\vec{k}|<\omega_0}\frac{d^3k}{2|\vec{k}|}\frac{1}{[p(k-\epsilon k_1)x+q(k-\epsilon k_2)(1-x)]^2} \hat{g}_0(k_1)\hat{g}_0(k_2)
\ed
\bd
= -\int\limits^1_0 dx\int d^4k_1d^4k_2
\int\limits_{|\vec{k}|<\omega_0}\frac{|\vec{k}|d|\vec{k}|(-d\cos\theta)d\phi}
{2[p(k-\epsilon k_1)x+q(k-\epsilon k_2)(1-x)]^2} \hat{g}_0(k_1)\hat{g}_0(k_2)
\ed
\be
=-2\pi\int\limits^1_0 dx\int d^4k_1d^4k_2
\int\limits^{\omega_0}_0\frac{|\vec{k}|d|\vec{k}|}{2} \int\limits^1_{-1}\frac{d\cos\theta}{[(x(p-q)+q)k-(xp\epsilon k_1+(1-x)q\epsilon k_2)]^2}\hat{g}_0(k_1)\hat{g}_0(k_2).
\ee
We abbreviate $x(p-q)+q$ by $Q$ below.
\bd
I_2=-2\pi\int\limits^1_0 dx\int d^4k_1d^4k_2
\int\limits^{\omega_0}_0\frac{|\vec{k}|d|\vec{k}|}{2} 
\int\limits^1_{-1}\frac{d\cos\theta}{[Q_0|\vec{k}|-|\vec{Q}||\vec{k}|\cos\theta -(xp\epsilon k_1+(1-x)q\epsilon k_2)]^2}\hat{g}_0(k_1)\hat{g}_0(k_2)
\ed
\bd
=-2\pi\int\limits^1_0 dx\int d^4k_1d^4k_2\int\limits^{\omega_0}_0 \frac{d|\vec{k}|}{2|\vec{Q}|}
\bigg[\frac{1}{[(Q_0-|\vec{Q}|)|\vec{k}|
-(xp\epsilon k_1+(1-x)q\epsilon k_2)]}
\ed
\bd
-\frac{1}{[(Q_0+|\vec{Q}|)|\vec{k}|
-(xp\epsilon k_1+(1-x)q\epsilon k_2)]}\bigg]\hat{g}_0(k_1)\hat{g}_0(k_2)
\ed
\bd
=-2\pi\int\limits^1_0 dx\int d^4k_1d^4k_2
\frac{1}{2|\vec{Q}|}
\bigg[\frac{1}{(Q_0-|\vec{Q}|)}\log\Big|(Q_0-|\vec{Q}|)|
\vec{k}|-(xp\epsilon k_1+(1-x)q\epsilon k_2)\Big|^{\omega_0}_0
\ed
\be
-\frac{1}{(Q_0+|\vec{Q}|)}\log\Big|(Q_0-|\vec{Q}|)|\vec{k}|-(xp\epsilon k_1+(1-x)q
\epsilon k_2)\Big|^{\omega_0}_0\bigg] \hat{g}_0(k_1)\hat{g}_0(k_2).
\ee
Again, we evaluate only the infrared divergent part of the expression above.
\be
I'_2=-2\pi\int\limits^1_0 dx\int d^4k_1d^4k_2
\frac{1}{2|\vec{Q}|}
\bigg[\frac{1}{(Q_0+|\vec{Q}|)}-\frac{1}{(Q_0-|\vec{Q}|)}\bigg]
\log\Big|(xp\epsilon k_1+(1-x)q\epsilon k_2)\Big| \hat{g}_0(k_1)\hat{g}_0(k_2).
\ee
Performing the integral over $k_1$ und $k_2$ results in
\bd
I'_2=2\pi\int\limits^1_0 dx\int d^4k_1d^4k_2\bigg(\frac{1}{Q^2}\bigg)
\Big[\log|\epsilon|+O(1)\Big]\hat{g}_0(k_1)\hat{g}_0(k_2)
\ed
\be
=(2\pi)^{5}\log|\epsilon|\int\limits^1_0 dx\frac{1}{Q^2}
=(2\pi)^{5}\log|\epsilon|\int\limits^1_0 dx\frac{1}{[x^2(p^2-2pq+q^2)+
2x(p-q)q+q^2]^2}.
\ee
Now we use the fact that $pq=m^2+\frac{1}{2}\vec{P}^2$ for $P=p-q$
\be
pq=p_0q_0-\vec{p}\vec{q}=E^2-\vec{p}\vec{q}=m^2 +\frac{\vec{p}^2}{2}+\frac{\vec{q}^2}{2}-\vec{p}\vec{q}=m^2+\frac{\vec{P}^2}{2},
\ee
\be
I'_2=(2\pi)^{5}\log|\epsilon|\int\limits^1_0 dx\frac{1}{[-\vec{P}^2x^2+
\vec{P}^2x+m^2]^2}
=-(2\pi)^{5}\log|\epsilon|\frac{1}{\vec{P}^2}\int\limits^1_0 dx\frac{1}{[(x-\frac{1}{2})^2-(\frac{m^2}{\vec{P}^2}+\frac{1}{4})]^2}.
\ee
We substitute $y=x-\frac{1}{2} $ und $ a=\frac{m^2}{\vec{P}^2}+\frac{1}{4}$ and obtain
in a straightforward manner
\bd
I'_2=-(2\pi)^{5}\log|\epsilon|\frac{2}{\vec{P}^2}\int\limits^\frac{1}{2}_0 dy
\frac{1}{[y^2-a]}
\ed
\be
=(2\pi)^{5}\log|\epsilon|\frac{1}{\frac{|\vec{P}|}{2m}m^2
\sqrt{1+\frac{\vec{P}^2}{4m^2}}}\log\Bigg|\frac{|\vec{P}|}{2m}+\sqrt{1+\frac{\vec{P}^2}{4m^2}}\Bigg|.
\ee
For the sake of convenience, we set $b:=\frac{|\vec{P}|}{2m}$, so that
\be
I'_2=(2\pi)^{5}\log|\epsilon|\frac{1}{m^2b\sqrt{1+b^2}}\log\Big|b+\sqrt{1+b^2}\Big|.
\ee
Combining $I'_2$ with the third integral in Eq. (\ref{four_integrals}), we have
\be
I'_2+I'_3=(2\pi)^{5}\log|\epsilon|\frac{2}{m^2b\sqrt{1+b^2}}\log\Big|b+\sqrt{1+b^2}\Big|,
\ee
and the full infrared divergent part of the bremsstrahlung cross section
from $I'_1$, $I'_2$, $I'_3$ und $I'_4$ is
\be
\displaystyle \frac{d\sigma^{brems}}{d\Omega_{div}}
= \frac{d\sigma^{(1)}}{d\Omega}\frac{e^2}{2(2\pi)^{2}m^2}
\Big[-1+\frac{1}{b\sqrt{1+b^2}}\log\Big|b+\sqrt{1+b^2}\Big|\Big]\log|\epsilon|.
\ee

\subsection{\it Self Energy}
\begin{figure}
\begin{minipage}[b]{0.5\linewidth} % A minipage that covers half the page
\centering
\includegraphics[width=6cm]{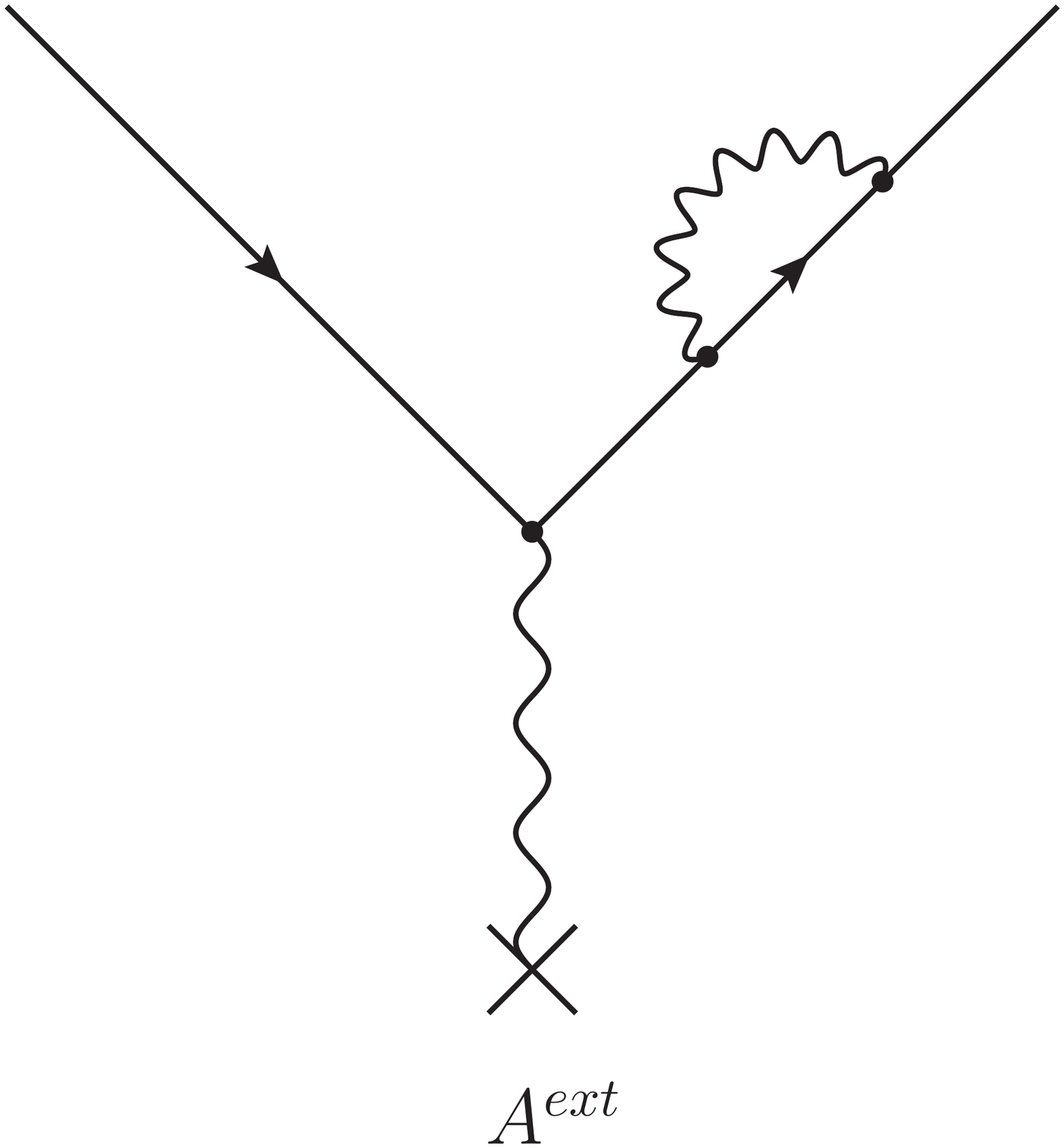}
\caption{Self energy diagram}
\label{selfenergy}
\end{minipage}
\begin{minipage}[b]{0.5\linewidth} % A minipage that covers half the page
\centering
\includegraphics[width=6cm]{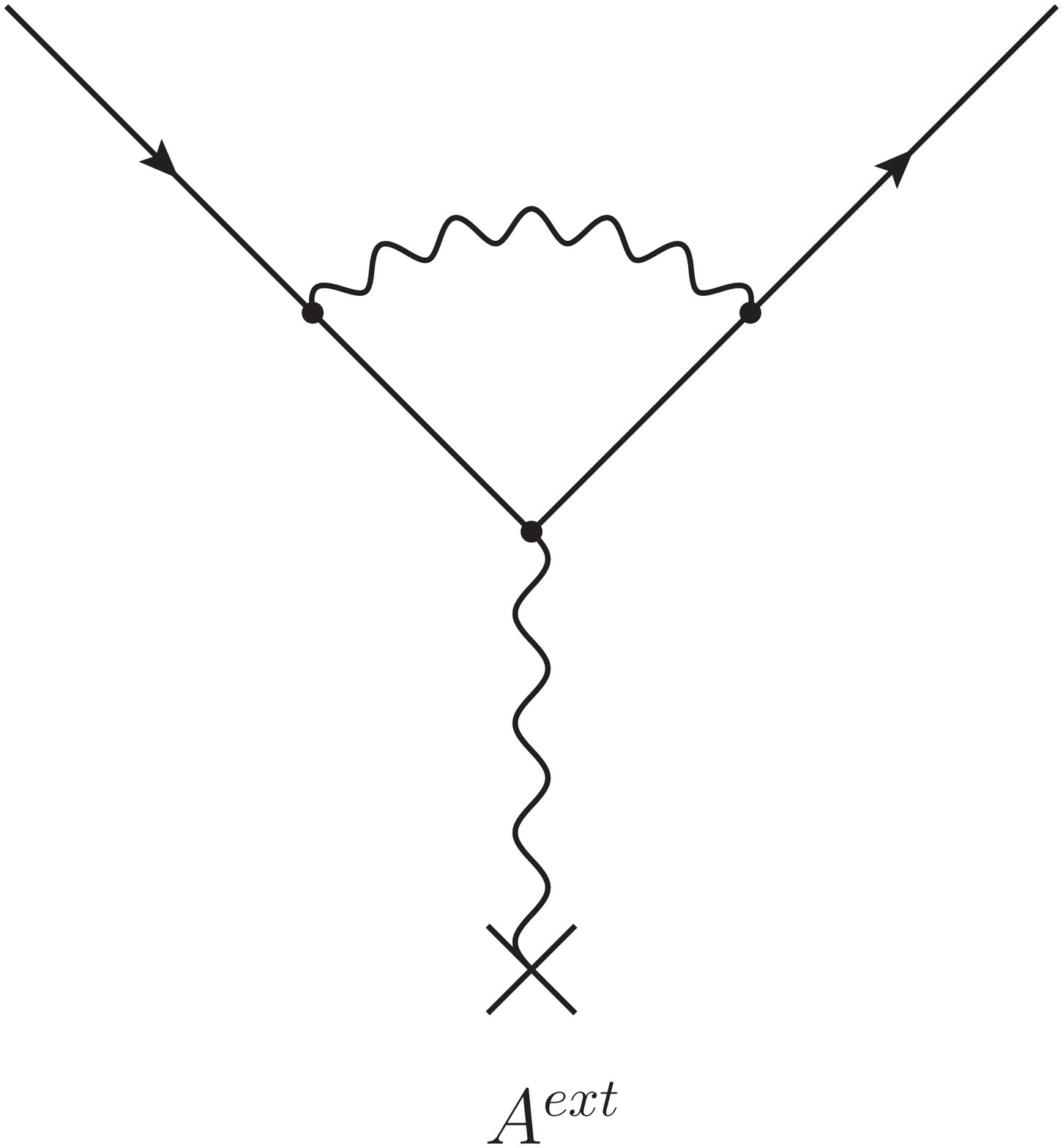}
\caption{Vertex diagram}
\label{vertex}
\end{minipage}
\end{figure}

So far we considered the infrared divergences in the bremsstrahlung cross section,
which represents a fourth order contribution in the coupling constant $e$ to the
inclusive cross section. To see how the logarithmic infrared divergences
$\sim \log(\epsilon)$ compensate, we have to investigate the third order Feynman diagrams
according to Figs. (\ref{selfenergy}) and (\ref{vertex}),
since they combine with the first order diagram in Fig. (\ref{tree}) according to
\begin{eqnarray}
\frac{d\sigma}{d\Omega}_{4th \, order} \sim |S_{fi}|^2_{4th \, order}=
|S_{fi}^{(2)}|^2+2Re|S_{fi}^{(1)}S_{fi}^{(3)}|.
\end{eqnarray}
We consider first the self energy diagram.
The relevant contributions to the $S$-matrix are given by six equivalent variants of the contraction
\bd
S^{(3)}  = \frac{(-ie)^3}{6}\int d^4x_1 d^4x_2d^4x_3g(x_1)g(x_2)g(x_3)
\ed
\be
\contraction{:\varphi(x_1)}{\varphi^\dagger(x_1)}{A(x_1)::}{\varphi(x_2)}
\contraction{:\varphi(x_1)\varphi^\dagger(x_1)A(x_1)::\varphi(x_2)}{\varphi^\dagger(x_2)}{A(x_2)::}
{\varphi(x_3)} \contraction[2ex]{:\varphi(x_1)\varphi^\dagger(x_1)A(x_1):
:\varphi(x_2)\varphi^\dagger(x_2)}{A(x_2)}{::\varphi(x_3)\varphi^\dagger(x_3)}{A(x_3)}
:\varphi(x_1)\varphi^\dagger(x_1)A(x_1)::\varphi(x_2)\varphi^\dagger(x_2)A(x_2):
:\varphi(x_3)\varphi^\dagger(x_3)A(x_3): .
\ee
The corresponding third order $S$-matrix contribution is therefore
\bd
S^{(3)}  = (-ie)^3\int d^4x_1 d^4x_2d^4x_3(-i)D_F^m(x_1-x_2)(-i)D_F^m(x_2-x_3)(-i)D_F^0(x_2-x_3)
\ed
\be
:\varphi(x_1) \varphi^\dagger(x_3): A(x_1) g(x_1)g(x_2)g(x_3)+ \ldots \, ,
\ee
the massless Feynman propagator $D_F^0$ coming from the photon propagator.
Calculating formally the matrix element with the corresponding initial and final meson states
scattering off the external photon field leads to
\bd
S_{fi}^{self}  =  \frac{e^3}{2(2\pi)^3\sqrt{q_0p_0}}
\int d^4k_1d^4k_2d^4k_3\bigg[\frac{\hat{A}^{ext} (p-q-k_1-k_2-k_3)}{(2\pi)^2 ((p-k_2-k_3)^2-m^2)}
\ed
\be
\int\frac{d^4k}{(2\pi)^6}\frac{1}{k^2((k-(k_3-p))^2-m^2)}
\bigg]\hat{g}(k_1)\hat{g}(k_2)\hat{g}(k_3).
\ee
However, the integral above contains a UV divergent part, which must be handled
properly.
The formal integral
\be
\Sigma^{div} (p^2,k_3,\epsilon)=\int d^4k\frac{1}{k^2((k-(\epsilon k_3-p))^2-m^2)}
\ee
must be regularized or treated within the causal framework.
The finite result for the self energy diagram is \cite{Aste2003}
\be
\displaystyle \Sigma(p^2)=i\pi^2\Big[1+\frac{m^2-p^2-i0}{p^2}
\log\Big(\frac{m^2-p^2-i0}{m^2}\Big)\Big] + C' \, ,
\ee
where $C'$ is a free normalization constant, and we must replace the
formal integral above by
\be
\Sigma((\epsilon k_3-p)^2)=i\pi^2\Big[1+\frac{m^2-(\epsilon k_3-p)^2}{(\epsilon k_3-p)^2}\log\Big(\frac{m^2-(\epsilon k_3-p)^2}{m^2}\Big)\Big] + C'.
\ee
In the following, we replace again $\hat{g}$ by $\hat{g}_0$ and perform all trivial
integrals
\bd
S_{fi}^{brems} =  S^{(1)}_{fi}\frac{-ie^2}{(2\pi)^2}\int d^4k_2d^4k_3\frac{1}{(2\pi)^2}
\frac{1}{-2p\epsilon (k_2+k_3)}
\ed
\be
\bigg[\frac{i\pi^2}{(2\pi)^4}\Big[1+\frac{m^2-(\epsilon k_3-p)^2}
{(\epsilon k_3-p)^2}\Big[\log\Big(\frac{m^2-(\epsilon k_3-p)^2}{m^2}\Big)
-i\pi\Theta((\epsilon k_3-p)^2-m^2)\Big]\Big]+C'
\bigg]\hat{g}_0(k_2)\hat{g}_0(k_3).
\ee
The integral over $k_2$ gives
\be
S_{fi}^{brems}  =  S^{(1)}_{fi}\frac{-ie^2}{(2\pi)^2}\int d^4k_3\frac{1}{-2p\epsilon k_3} \bigg[\frac{i\pi^2}{(2\pi)^4}\Big[1+\frac{2p\epsilon k_3}{m^2}
\Big[\log\Big(\frac{2p\epsilon k_3}{m^2}\Big)-i\pi
\Theta(-2p\epsilon k_3)\Big]\Big]+C' \bigg]\hat{g}_0(k_3).
\ee
Now we replace $C'$ by $\frac{i\pi^2}{(2\pi)^4}(C-1)$
\bd
S_{fi}^{brems}  =  S^{(1)}1_{fi}\frac{-ie^2}{(2\pi)^2}\int d^4k_3\frac{1}{-2p\epsilon k_3} \frac{i\pi^2}{(2\pi)^4}\Big[\frac{2p\epsilon k_3}{m^2}
\Big[\log\Big(\frac{2p\epsilon k_3}{m^2}\Big)-i\pi\Theta(-2p\epsilon k_3)\Big]+C\Big]
\hat{g}_0(k_3)
\ed
\bd
=  S^{(1)}_{fi}\frac{-e^2}{(2\pi)^2}\int d^4k_3\frac{1}{2p\epsilon k_3} \frac{\pi^2}{(2\pi)^4}\Big[\frac{2p\epsilon k_3}{m^2}\Big[\log|\epsilon|
+O(1)\Big]+C\Big] \hat{g}_0(k_3)
\ed
\be
=  S^{(1)}_{fi}\frac{-e^2}{(2\pi)^2} \frac{\pi^2}{(2\pi)^4m^2}
\log|\epsilon|\int d^4k_3\hat{g}_0(k_3)+S^1_{fi}
\frac{e^2}{(2\pi)^2}\frac{\pi^2}{(2\pi)^4\epsilon}
\int d^4k_3\frac{C}{2pk_3}\hat{g}_0(k_3).
\ee
The bremsstrahlung diagram contained only logarithmic infrared divergences.
For this reason, we choose $C=0$ in order to avoid a $1/\epsilon$-divergence
in the self energy diagram and obtain
\be
S_{fi}^{brems}  =  S^{(1)}_{fi}\frac{e^2}{(2\pi)^2} \frac{1}{4m^2}\log|\epsilon|,
\ee
and the corresponding contribution to the cross section follows from
\begin{eqnarray}
\frac{d\sigma}{d\Omega}_{4th \, order} \sim 2Re|S_{fi}^{(1)}S_{fi}^{(3)}|
\end{eqnarray}
and is given by
\begin{equation}
\displaystyle \frac{d\sigma^{brems}}{d\Omega_{div}} =
\frac{d\sigma^{(1)}}{d\Omega}\frac{e^2}{2(2\pi)^2m^2} \log|\epsilon|.
\end{equation}

\subsection{\it Vertex Function}
Finally, we consider the vertex diagram according to Fig. (\ref{vertex}).
Formally, one obtains the expression containing the UV divergent scalar vertex integral
\bd
S_{fi}^{vertex}=\frac{e^3}{(2\pi)^32\sqrt{p_0q_0}} \hat{ A}^{ext} (p-q)\frac{1}{(2\pi)^8}
\int d^4k_1d^4k_2d^4k_3
\ed
\be
\int d^4k\frac{1}{((k+q+k_1)^2-m^2)((k+p-k_3)^2-m^2)k^2}
\hat{g}(k_1)\hat{g}(k_2)\hat{g}(k_3).
\ee
Of course, we choose the causal approach to the problem.
The third order vertex contribution to the $S$-matrix can be written
\be
S^{(3)}=-e^3 \int d^4x_1d^4x_2d^4x_3:\varphi^{\dagger}(x_1) t_3^{vertex} (x_1,x_2,x_3)
\varphi(x_2): A(x_3)g(x_1)g(x_2)g(x_3)+ \ldots
\ee
The $S$-matrix element containing the external field is, correspondingly
\bd
S_{fi}^{vertex}=-e^3\frac{1}{(2\pi)^32\sqrt{p_0q_0}}\frac{1}{(2\pi)^2}
\frac{1}{(2\pi)^6}\int d^4x_1d^4x_2d^4x_3\int d^4k_1d^4k_2d^4k_3
\ed
\be
\int d^4k''t_3^{vertex}(x_1,x_2,x_3) e^{ipx_1}e^{-iqx_2}e^{-ik''x_3}e^{-ik_1x_1}e^{-ik_2x_2}e^{-ik_3x_3}
\hat{ A}^{ext} (k'')\hat{g}(k_1)\hat{g}(k_2)\hat{g}(k_3)
\ee
yielding
\bd
S_{fi}^{vertex}=-e^3\frac{1}{(2\pi)^3 2 \sqrt{p_0 q_0}}
\int d^4k_1d^4k_2d^4k_3 \hat{ A}^{ext} (p-q-\epsilon (k_1+k_2+k_3))
\ed
\be
\bigg[\frac{1}{(2\pi)^4}\int d^4y_1d^4y_2
\hat{t}_3^{vertex}(p-\epsilon k_1,q+\epsilon k_2)\bigg]
\hat{g}(\epsilon k_1)\hat{g}(\epsilon k_2)\hat{g}(\epsilon k_3),
\ee
where (note that ${t}_3^{vertex}(x_1,x_2,x_3)$ is translation invariant)
\be 
\hat{t}_3^{vertex}(p,q)=\frac{1}{(2 \pi)^4} \int d^4 y_1 d^4 y_2 {t}_3^{vertex}(x_1,x_2,x_3)
e^{ip y_1+ iq y_2} .
\ee
Again, one can factor out the first order scattering matrix element, and using the abbreviations
$p_1=p-\epsilon k_1$ and $q_1=q+\epsilon k_2$ leads to
\be
S_{fi}^{vertex}=S^{(1)}_{fi} \frac{-ie^2}{(2\pi)^2}\int d^4k_1d^4k_2d^4k_3
\hat{D}(p_1,q_1) \hat{g}(\epsilon k_1)\hat{g}(\epsilon k_2)\hat{g}(\epsilon k_3) .
\ee
In order to calculate $\hat{t}_3^{vertex}$, we must first construct the causal distribution
$\hat{d}_3^{vertex}(p_1,q_1)$, which is given by
\bd
\hat{d}_3^{vertex} (p_1,q_1)=\frac{1}{(2\pi)^2}\int d^4k
\ed
\bd
[\hat{D}^{(-)}_m(k-p_1)
\hat{D}^{av}_m(k-q_1)\hat{D}^{(+)}_0(k)-\hat{D}^{(-)}_m(k-q_1) \hat{D}^{av}_m(k-p_1)\hat{D}^{(+)}_0(k)
\ed
\bd
+\hat{D}^{(+)}_m(k-q_1)\hat{D}^{ret}_m(k-p_1)\hat{D}^{(-)}_0(k)-\hat{D}^{(+)}_m(k-p_1)
\hat{D}^{ret}_m(k-q_1)\hat{D}^{(-)}_0(k)
\ed
\be
+\hat{D}^{(+)}_m(k-q_1)\hat{D}^{(-)}_m(k-p_1)\hat{D}_F^0(k)-\hat{D}^{(+)}_m(k-p_1)
\hat{D}^{(-)}_m(k-q_1)\hat{D}^F_0(k)].
\ee
The full calculation of $\hat{d}_3^{vertex}$ can be found in \cite{Scharf,Twoloop}.
The result is
\bd
\hat{d}_3^{vertex}(p_1,q_1)= \frac{\pi}{4(2\pi)^6\sqrt{N}}
\ed
\bd
\bigg[\mbox{sgn}(p_{1_0})\Theta(p_1^2-m^2)\log\Bigg|\frac{q_1^2-m^2-p_1q_1(1-\frac{m^2}{p_1^2})+
\sqrt{N}(1-\frac{m^2}{p_1^2})}{q_1^2-m^2-p_1q_1(1-\frac{m^2}{p_1^2})-
\sqrt{N}(1-\frac{m^2}{p_1^2})}\Bigg|
\ed
\bd
-\mbox{sgn}(q_{1_0})\Theta(q_1^2-m^2)\log\bigg|\frac{p_1^2-m^2-p_1q_1(1-\frac{m^2}{q_1^2})+\sqrt{N}
(1-\frac{m^2}{q_1^2})}{p_1^2-m^2-p_1q_1(1-\frac{m^2}{q_1^2})-\sqrt{N}(1-\frac{m^2}{q_1^2})}\bigg|
\ed
\be
\mbox{sgn}(P_0)\Theta(P^2-4m^2)\log\Bigg|\frac{p_1q_1+m^2+\sqrt{N}\sqrt{1-\frac{4m^2}{P^2}}}
{p_1q_1+m^2-\sqrt{N}\sqrt{1-\frac{4m^2}{P^2}}}\Bigg|\bigg]. \label{d3vertex}
\ee
The time-ordered distribution $\hat{t}_3^{vertex}$ is obtained from
$\hat{d}_3^{vertex}$ by distribution
splitting, i.e. from a subtracted dispersion integral according to the vertex scaling degree $\omega=0$.
The infrared divergence is contained in the two first logarithmic terms of Eq. (\ref{d3vertex}),
and one may write
\be
J_{div}=\frac{i}{2\pi}\int\limits^{\infty}_{-\infty}dt\frac{I_{div}(tp_1,tq_1)}{t^2(1-t+i0)}
\ee
with a first term
\be
J_{div_1}=\frac{i}{8\sqrt{N}(2\pi)^6}\int\limits^{\infty}_{-\infty} dt\frac{sgn(t)
\Theta(t^2p_1^2-m^2)}{t^2(1-t+i0)}\log\bigg|\frac{\frac{m^2}{p_1^2}p_1q_1-m^2
+t^2(q_1^2-p_1q_1)-\frac{m^2}{p_1^2}\sqrt{N}+
\sqrt{N}t^2}{\frac{m^2}{p_1^2}p_1q_1-m^2+t^2(q_1^2-p_1q_1)+\frac{m^2}{p_1^2}\sqrt{N}-
\sqrt{N}t^2}\bigg|.
\ee
This integral can be evaluated in a straightforward manner and leads to expressions containing
Spence functions and logarithms. We restrict ourselves to the term which contains the infrared
divergence
\bd
J_{div_1} = \frac{i}{8\sqrt{N}(2\pi)^6}\log\bigg|\frac{m^2-p_1^2}
{p_1^2}\bigg| \log\bigg|\frac{\frac{m^2}{p_1^2}p_1q_1-m^2+q_1^2-p_1q_1+\sqrt{N}
(1-\frac{m^2}{p_1^2})}{\frac{m^2}{p_1^2}p_1q_1-m^2+q_1^2-p_1q_1-\sqrt{N}(1-\frac{m^2}{p_1^2})}\bigg|.
\ed
\bd
= \frac{i}{8\sqrt{N}(2\pi)^6} \log\bigg|\frac{m^2-p_1^2}
{p_1^2}\bigg| \log\bigg|\frac{p_1^2(q_1^2-m^2)-p_1q_1(p_1^2-m^2)+\sqrt{N}(p_1^2-m^2)}
{p_1^2(q_1^2-m^2)-p_1q_1(p_1^2-m^2)-\sqrt{N}(p_1^2-m^2)}\bigg|
\ed
\be
= -\frac{i}{8\sqrt{N}(2\pi)^6} \log\bigg|\frac{m^2-p_1^2}
{p_1^2}\bigg| \log\bigg|\frac{p_1^2\frac{(q_1^2-m^2)}{(p_1^2-m^2)}-p_1q_1-\sqrt{N}}
{p_1^2\frac{(q_1^2-m^2)}{(p_1^2-m^2)}-p_1q_1+\sqrt{N}}\bigg|.
\ee
Now we use the explicit form of $p_1$ and $q_1$ 
\be
p_1=p-\epsilon k_1 \rightarrow (p_1^2-m^2)=-2\epsilon pk_1 + O(1),
\ee
\be
q_1=q+\epsilon k_2 \rightarrow (q_1^2-m^2)=2\epsilon qk_2 + O(1),
\ee
and $\sqrt{N}=m|\vec{P}|\sqrt{1+\frac{\vec{P}^2}{4m^2}}$, leading to
\bd
J_{div_1} = -\frac{i}{(2\pi)^68m|\vec{P}|\sqrt{1+\frac{\vec{P}^2}
{4m^2}}} [\log|\epsilon|+O(1)]
\log\Bigg|\frac{-\frac{qk_2}{pk_1}m^2-m^2-\frac{\vec{P}^2}{2}-\frac{\vec{P}^2}{2}
\sqrt{1+\frac{4m^2}{\vec{P}^2}}}{-\frac{qk_2}{pk_1}m^2-m^2-\frac{\vec{P}^2}{2}+\frac{\vec{P}^2}{2}
\sqrt{1+\frac{4m^2}{\vec{P}^2}}}\Bigg|
\ed
\bd
= -\frac{i}{(2\pi)^68m|\vec{P}|\sqrt{1+\frac{\vec{P}^2}{4m^2}}}
[\log|\epsilon|+O(1)]
\ed
\bd
\log\Bigg|\frac{(1-\frac{qk_2}{pk_1})+(1+\frac{qk_2}{pk_1})
\sqrt{1+\frac{4m^2}{\vec{P}^2}}+(1-\frac{qk_2}{pk_1})\sqrt{1+\frac{4m^2}{\vec{P}^2}}+
(1+\frac{qk_2}{pk_1})(1+\frac{4m^2}{\vec{P}^2})}{(1-\frac{qk_2}{pk_1})-(1+\frac{qk_2}{pk_1})
\sqrt{1+\frac{4m^2}{\vec{P}^2}}-(1-\frac{qk_2}{pk_1})\sqrt{1+\frac{4m^2}{\vec{P}^2}}+(1+\frac{qk_2}{pk_1})
(1+\frac{4m^2}{\vec{P}^2})}\Bigg|
\ed
\bd
=-\frac{i}{(2\pi)^68m|\vec{P}|\sqrt{1+\frac{\vec{P}^2}{4m^2}}} 
[\log|\epsilon|+O(1)]
\ed
\be
\Bigg[\log\Bigg|\frac{1+\sqrt{1+\frac{4m^2}{\vec{P}^2}}}
{1-\sqrt{1+\frac{4m^2}{\vec{P}^2}}}\Bigg|+\log\Bigg|\frac{1-\frac{4m^2}{\vec{P}^2}+(1+\frac{qk_2}{pk_1})\sqrt{1+\frac{4m^2}{\vec{P}^2}}}{1-\frac{4m^2}{\vec{P}^2}-
(1+\frac{qk_2}{pk_1})\sqrt{1+\frac{4m^2}{\vec{P}^2}}}\Bigg].
\ee
$J_{div2}$ is calculated along the same lines
\bd
J_{div_2} =
-\frac{i}{(2\pi)^68m|\vec{P}|\sqrt{1+\frac{\vec{P}^2}{4m^2}}} 
[\log|\epsilon|+O(1)]
\ed
\be
\Bigg[\log \Bigg| \frac{1+\sqrt{1+\frac{4m^2}{\vec{P}^2}}}
{1-\sqrt{1+\frac{4m^2}{\vec{P}^2}}}\Bigg|-\log\Bigg|\frac{1-\frac{4m^2}
{\vec{P}^2}+(1+\frac{qk_1}{pk_3})\sqrt{1+\frac{4m^2}{\vec{P}^2}}}{1-\frac{4m^2}{\vec{P}^2}-
(1+\frac{qk_1}{pk_3})\sqrt{1+\frac{4m^2}{\vec{P}^2}}} \Bigg| \Bigg] .
\ee
The two results finally combine to
\be
J_{div} =-\frac{i}{(2\pi)^64m^2\frac{|\vec{P}|}{2m}
\sqrt{1+\frac{\vec{P}^2}{4m^2}}}\log\Bigg|\frac{|\vec{P}|}{2m}+
\sqrt{1+\frac{\vec{P}^2}{4m^2}}\Bigg|\log|\epsilon|
\ee
or
\be
J_{div}=-\frac{i}{(2\pi)^64m^2b\sqrt{1+b^2}}\log\Big|b+\sqrt{1+b^2}\Big|\log|\epsilon|.
\ee
The divergent vertex contribution to the cross section follows
\be
\frac{d\sigma^{vertex}}{d\Omega_{div}}=-\frac{d\sigma^{(1)}}
{d\Omega}\frac{e^2}{2(2\pi)^2m^2b\sqrt{1+b^2}}\log\Big|b+\sqrt{1+b^2}\Big| \log|\epsilon|.
\ee
Obviously, the self energy and vertex infrared divergences now cancel the divergence generated
by the bremsstrahlung process:
\be
\frac{d\sigma}{d\Omega}^{brems}_{div}  =   \frac{d\sigma^{(1)}}{d\Omega}\frac{e^2}{2(2\pi)^{2}m^2}\Big[-1+\frac{1}{b\sqrt{1+b^2}}
\log\Big|b+\sqrt{1+b^2}\Big|\Big]\log|\epsilon|,
\ee
\be
\frac{d\sigma}{d\Omega}^{self}_{div} = \frac{d\sigma^{(1)}}{d\Omega}
\frac{e^2}{2(2\pi)^2m^2} \log|\epsilon|,
\ee
\be
\frac{d\sigma}{d\Omega}^{vertex}_{div} = \frac{d\sigma^{(1)}}{d\Omega}\frac{e^2}{2(2\pi)^2m^2}\Big[-\frac{1}{b\sqrt{1+b^2}}
\log\Big|b+\sqrt{1+b^2}\Big|\Big] \log|\epsilon|.
\ee
This shows that the adiabatic limit $g \to 1$ exists in the causal formalism
for the inclusive cross-section. For a discussion of the
uniqueness of the adiabatic limit we refer to \cite{Scharf,Epstein_adiabatic}.

We conclude this section by highlighting the qualitative picture of the calculations
given above. The switching of the interaction with a test function $g$, which vanishes
for large space and time distances, corresponds to a gedanken experiment where
the charged mesons are liberated from their (scalar) "electromagnetic" field.
The non-perturbative description of interacting fields is highly non-trivial and
a hitherto unsolved problem. The good news is the fact that physical observables
can be constructed in our model theory in an unambiguous way in the limit $g \rightarrow 1$,
where the interaction becomes permanent.

\section{Gauge Theories}
\subsection{\it Spin 1}
As we have emphasized before all knows interactions in nature can be described by
quantum gauge theories. Gravity can be described within a very similar causal setting
as "ordinary" spin-1 gauge theory, as will be shown below. However, higher
order perturbative quantum gravity holds the highly non-trivial problem
of non-renormalizability, which may potentially show up in the causal framework
as a violation of perturbative quantum gauge invariance. At least, the theory is
still consistent at second order in the gravitational constant and may provide
an effective description of the interaction.

By quantum gauge theory we mean a theory which has a gauge invariant S-matrix.
This is different from classical gauge invariance where the classical Lagrangian is
gauge invariant. Instead we define gauge invariance for the time-ordered products
$T_n$ constructed by causal perturbation theory as described in sect. 3 using the
gauge variations of the free fields.

One is tempted to define perturbative gauge invariance simply by $d_QT_n=0$, 
but this is not correct. To find the
right definition let us consider QED where we certainly know what gauge
invariance means. Ordinary spinor quantum electrodynamics is constructed
from
\be T_1(x)=ie:\psq(x)\gamma^\mu\psi(x):A_\mu(x).\label{3.1.12}\ee
The $free$ Dirac fields $\psi, \psq$ have zero gauge variation
$d_Q\psi=0=d_Q\psq$, but $d_QA_\mu=i\d_\mu u$. Then we obtain
$$d_QT_1=-e:\psq\gamma^\mu\psi:\d_\mu u=ie\d_\mu(i:\psq\gamma^\mu
\psi:u).$$
Here we have used current conservation
$$\d_\mu:\psq\gamma^\mu\psi:=0$$
which follows from the $free$ Dirac equations. We see that $d_QT_1$ is
not zero, but a divergence
\be d_QT_1=i\d_\mu T^\mu_{1/1},\label{3.1.15}\ee
where
\be T^\mu_{1/1}=ie:\psq\gamma^\mu\psi:u\label{3.1.16}\ee
is called $Q$-vertex in the following and Eq. (\ref{3.1.15}) establishes the
first order gauge invariance.

It is not hard to generalize this to higher orders. If we freely
interchange $d_Q$ and the time ordering we can write
$$d_QT_n=d_QT\{T_1(x_1)\cdot\ldots\cdot T_1(x_n)\}$$
$$=\sum_{l=1}^nT\{T_1(x_1)\ldots d_QT_1(x_l)\cdot\ldots T_1(x_n)\}$$
\be =\sum_{l=1}^nT\{T_1(x_1)\ldots i\d_\mu T_{1/1}^\mu(x_l)\cdot\ldots 
T_1(x_n)\}.\label{3.1.17}\ee
The time ordered products herein have to be constructed correctly by the
causal method, using the $Q$-vertex from Eq. (\ref{3.1.16}) at $x_l$ instead of the
ordinary QED vertex Eq. (\ref{3.1.12}). Again formally taking the derivative out 
of the $T$-product we get
$$d_QT_n=i\sum_{l=1}^n{\d\over\d x_l^\mu}T\{T_1(x_1)\ldots T_{1/1}^\mu 
(x_l)\cdot\ldots T_1(x_n)\}$$
\be \=d i\sum_{l=1}^n{\d\over\d x_l^\mu}T_{n/l}^\mu(x_1\ldots x_n). 
\label{3.1.18}\ee
This equation certainly holds for $x_j\ne x_k$, for all $j\ne k$,
because there we can calculate with the $T$-product in the same way as 
with an ordinary product. But the extension to the diagonal 
$x_1=\ldots =x_n$ produces 
local terms in general, both in Eq. (\ref{3.1.17}) and in Eq. (\ref{3.1.18}). If it is
possible to absorb such local terms by suitable normalization of the
distributions $T_n$ and $T_{n/l}^\mu$,
then we call the theory gauge invariant to $n$-th order. We want to
emphasize that perturbative gauge invariance not only means that
$d_QT_n$ is a divergence, the divergence must also be of the specific form
Eq. (\ref{3.1.18}) involving the $Q$-vertex.

Now we check what perturbative gauge invariance defined by Eq. (\ref{3.1.18}) means
for the total S-matrix. Applying the gauge variation $d_Q$ to
the formal power series we obtain
$$d_QS(g)=\sum_{n=1}^\infty {i\over n!}\int d^4x_1\ldots d^4x_n\sum_{l=1}^n 
(\d_\mu^{x_l}T^\mu_{n/l}) g(x_1)\ldots g(x_n).$$
Since the test function $g(x)$ is in Schwartz space, we can
integrate by parts
$$=-\sum_{n=1}^\infty {i\over n!}\int d^4x_1\ldots d^4x_n\sum_{l=1}^n 
T^\mu_{n/l} g(x_1)\ldots (\d_\mu g)(x_l)\ldots g(x_n).$$ 
If it is possible to perform the so-called adiabatic limit $g\to 1$ here, then the
right-hand side goes to zero and we get the naive definition of gauge
invariance of the S-matrix
\be \lim_{g\to 1}d_Q S(g)=0.\label{3.1.25}\ee
The adiabatic limit exists if all gauge fields are massive. 
It does not
exist for the time-ordered products if some gauge field is massless. In
this case Eq. (\ref{3.1.25}) is meaningless and we must use the perturbative
definition in Eq. (\ref{3.1.18}). The latter is really at the heart of gauge theory
because it determines the possible couplings $T_1$. This we are now
going to show for massless spin-1 gauge fields.

We consider a collection of vector fields $A^\mu_a(x), a=1,\ldots N$ and
ghost fields $u_a(x)$ with anti-ghost fields $\tilde u_a(x)$ quantized
in the usual manner according to Eq. (\ref{(1.2.6)}). It is our goal to find all
possible gauge invariant self-couplings $T_1(x)$ of these fields. Since the gauge
variation $d_Q$ generates ghost fields from vector fields, it is pretty
clear that $T_1$ must involve ghost and anti-ghost fields as well. But
we assume ghost number =0, so that $u$ and $\tilde u$ must appear in
pairs. It is sufficient to consider trilinear couplings proportional to
a coupling constant $g$, quadrilinear ones proportional to $g^2$
correspond to $T_2$ and should come out automatically in the causal
construction. We therefore start from the following general ansatz \cite{Aste1999}
$$T_1(x)=ig\{f^1_{abc}:A_{\mu a}A_{\nu b}\d^\nu A^\mu_c:+
f^2_{abc}:A_{\mu a}A^\mu_b\d^\nu A^\nu_c:$$ 
\be +f^3_{abc}:A_{\mu a}u_b\d^\mu\tilde u_c+f^4_{abc}:(\d^\mu A_{\mu a})
u_b\tilde u_c:+f^5_{abc}: A_{\mu a}(\d^\mu u_b)\tilde u_c:\}.
\label{3.2.1}\ee
Here we have further assumed that $T_1$ is a Lorentz scalar,
and for the sake of simplicity we only consider CP conserving
terms here. $T_1$ being a Lorentz scalar,
we need an odd number of derivatives in each term. We only consider
one derivative because with three the theory is not renormalizable. The
$f^j_{abc}$ are arbitrary constants, but unitarity requires a
skew-adjoint $T_1$
$$T_1^\dagger(x)=-T_1(x),$$
so that the $f$'s and $g$ must be real. This was the reason for the
imaginary $i$ in Eq. (\ref{3.2.1}). Since the Wick monomial in the second term is 
symmetric in $a$ and $b$, we assume
$$f^2_{abc}=f^2_{bac}$$
without loss of generality.
The reader easily convinces himself that there
is no further possibility to contract the Lorentz indices and place
the derivative. All double indices including $a,b,c$ are summed over.

Next we calculate the gauge variation
$$d_QT_1=-\Bigl\{f^1_{abc}(\d_\mu u_aA_{\nu b}\d^\nu A^\mu_c+A_{\mu a}
\d_\nu u_b\d^\nu A^\mu_c+A_{\mu a}A_{\nu b}\d^\nu\d^\mu u_c)$$ 
$$+f^2_{abc}(2\d_\mu u_aA^\mu_b
\d_\nu A^\nu_c+A_{\mu a}A^\mu_b\d_\nu\d^\nu u_c)$$
$$+f^3_{abc}(\d_\mu u_au_b\d^\mu\tilde u_c+A_{\mu a}u_b\d^\mu\d_\nu
A^\nu_c)$$
$$+f^4_{abc}(\d^\mu\d_\mu u_au_b\tilde u_c+(\d^\mu A_{\mu a})u_b\d_\nu
A^\nu_c)$$ 
\be +f^5_{abc}(\d_\mu u_a\d^\mu u_b\tilde u_c+A_{\mu a}\d^\mu u_b\d_\nu
A^\nu_c)\Bigl\}.\label{3.2.4}\ee
The last term in the second and the first one in the fourth line vanish 
due to the wave equation.
To simplify the notation we do not write the double dots for normal
ordering anymore, all products of field operators with the same
argument are normally ordered {\it if nothing else is said.} 

For gauge invariance the expression Eq. (\ref{3.2.4}) must be a divergence
$$=i\d_\mu T^\mu_{1/1}(x).$$
We therefore write down a general ansatz for $T^\mu_{1/1}$ as well:
$$iT^\mu_{1/1}=g\Bigl\{g^1_{abc}\d^\mu u_aA_{\nu b}A^\nu_c+g^2_{abc}
u_aA_{\nu b}\d^\mu A^\nu_c$$
$$+g^3_{abc}\d_\nu u_aA^\nu_bA^\mu_c+g^4_{abc}u_a\d_\nu
A^\nu_bA^\mu_c$$
\be +g^5_{abc}u_aA^\nu_b\d_\nu A^\mu_c+g^6_{abc}u_au_b\d^\mu\tilde u_c
+g^7_{abc}\d^\mu u_au_b\tilde u_c\Bigl\}.\label{3.2.5}\ee
The symmetry in the first and antisymmetry in the sixth term give
 the relations
$$g^1_{abc}=g^1_{acb},\quad g^6_{abc}=-g^6_{bac}.$$
This ansatz for $T^\mu_{1/1}$ can be further restricted using the
nilpotence property
$$id_Q\d_\mu T^\mu_{1/1}=d_Q^2T_1=0.$$
Substituting Eq. (\ref{3.2.5}) and collecting terms with the same field operators
we obtain the following homogeneous relations:  
\be \d^\mu u_a\d_\mu\d_\nu u_bA^\nu_c:\quad 
2g^1_{abc}+g^2_{acb}+g^3_{abc}-g^3_{bac}-g^3_{bca}+g^5_{acb}=0\label{3.2.8}\ee 
\be \d^\mu u_a\d_\mu A^\nu_b\d_\nu u_c:\quad 2g^1_{abc}+g^2_{acb}-g^3
_{cba}-g^5_{cab}=0\label{3.2.9}\ee
\be \d_\mu u_a\d^\mu u_b\d_\nu A_c^\nu:\quad g^3_{abc}-g^3_{bac}+g^4
_{acb}-g^4_{bac}+g^7_{abc}-g^7_{bac}=0\label{3.2.10}\ee 
\be u_a\d^\mu u_b\d_\mu\d_\nu A^\nu_c:\quad g^4_{abc}+g^5_{abc}+g^6
_{abc}-g^6_{bac}-g^7_{bac}=0\label{3.2.11}\ee 
\be u_a\d^\mu\d_\nu u_b\d_\mu A^\nu_c:\quad g^2_{abc}+g^2_{acb}+g^5
_{abc}+g^5_{acb}=0.\label{3.2.12}\ee

First order gauge invariance according to Eq. (\ref{3.2.4}) now implies linear relations
between the $f$'s and $g$'s:
\be \d^\mu u_a\d_\mu A_{\nu b}A^\nu_c:\quad -f^1_{cab}=2g^1_{abc}
+g^2_{acb}\label{3.2.13}\ee
\be \d^\mu u_aA_{\mu b}\d_\nu A^\nu_c:\quad -2f^2_{abc}-f^5_{bac}=
g^3_{abc}+g^4_{acb}\label{3.2.14}\ee
\be u_a\d_\nu A^\nu_b\d_\mu A^\mu_c:\quad
-f^4_{bac}-f^4_{cab}=g^4_{abc}+g^4_{acb}\label{3.2.15}\ee
\be \d_\mu u_au_b\d^\mu\tilde u_c:\quad -f^3_{abc}=2g^6_{abc}+g^7_{abc}
\label{3.2.16}\ee
\be \d_\mu u_a\d^\mu u_b\tilde u_c:\quad -f^5_{abc}+f^5_{bac}=g^7_{abc}
-g^7_{bac}\label{3.2.17}\ee
\be \d_\mu\d_\nu u_aA^\nu_bA^\mu_c:\quad -f^1_{cba}-f^1_{bca}=g^3
_{abc}+g^3_{acb}\label{3.2.18}\ee
\be \d_\mu u_a\d_\nu A^\mu_bA^\nu_c:\quad -f^1_{acb}=g^3_{abc}+g^5
_{acb}\label{3.2.19}\ee
\be u_a\d_\nu A^\nu_b\d_\mu A^\nu_c:\quad g^2_{abc}=-g^2_{acb}
\label{3.2.20}\ee
\be u_a\d_\mu\d_\nu A^\mu_bA^\nu_c:\quad -f^3_{cab}=g^4_{acb}+g^5_{abc}
\label{3.2.21}\ee
\be u_a\d_\nu A^\mu_b\d_\mu A^\nu_c:\quad g^5_{abc}=-g^5_{acb}.
\label{3.2.22}\ee 

All information comes out of this linear system. Since the elimination
process is somewhat tedious, we give all details to save the readers time.
Let us interchange $b$ and $c$ in Eq. (\ref{3.2.19})
\be -f^1_{abc}=g^3_{acb}+g^5_{acb}\label{3.2.23}\ee
and add this to Eq. (\ref{3.2.19})
\be -f^1_{abc}-f^1_{acb}=g^3_{abc}+g^3_{acb}+g^5_{abc}+g^5_{acb}.
\label{3.2.24}\ee 
By Eq. (\ref{3.2.22}) $g^5$ drops out and by Eq. (\ref{3.2.18}) the right side is equal to
\be =-f^1_{cba}-f^1_{bca}.\label{3.2.25}\ee 
This implies
\be f^1_{abc}-f^1_{cba}=f^1_{bca}-f^1_{acb}.\label{3.2.26}\ee 
Let us now decompose $f^1_{abc}$ into symmetric and antisymmetric parts
in the first and third indices:
\be f^1_{abc}=d_{abc}+f_{abc},\quad d_{abc}=d_{cba},\quad f_{abc}=-
f_{cba},\label{3.2.27}\ee 
then Eq. (\ref{3.2.26}) implies
\be f_{abc}=-f_{acb}=-f_{cba}=f_{cab}=f_{bca}=-f_{bac}.\label{3.2.28}\ee 
So we arrive at the important result that $f_{abc}$ is totally
antisymmetric. The Jacobi identity follows from
second order gauge invariance, hence, $f_{abc}$ can be regarded as
structure constants of a $real$ Lie algebra. 

The total antisymmetry of $f_{abc}$ implies the total symmetry of
$d_{abc}$. Next we use the representation Eq. (\ref{3.2.27}) in Eq. (\ref{3.2.13}):
\be -f_{cab}-d_{cab}=2g^1_{abc}+g^2_{abc}.\label{3.2.30}\ee 
Here $g^2_{abc}$ is antisymmetric in $b, c$ according to Eq. (\ref{3.2.20}), so that $g^1_{abc}$
must be symmetric, hence
\be g^1_{abc}=-\eh d_{cab}=-\eh d_{abc}\label{3.2.31}\ee 
\be g^2_{abc}=f_{cab}=f_{abc}.\label{3.2.32}\ee 
Now we can write Eqns. (\ref{3.2.18}) and (\ref{3.2.19}) in the form
\be g^3_{abc}+g^3_{acb}=-2d_{cba}=-2d_{abc}\label{3.2.33}\ee 
\be -f_{acb}-d_{acb}=g^3_{abc}+g^5_{acb}.\label{3.2.34}\ee 
Since $g^5_{abc}$ is antisymmetric in $b, c$ according to Eq. (\ref{3.2.22}), the symmetric
part of this equation agrees with Eq. (\ref{3.2.33}) and the antisymmetric part is
given by
\be -f_{acb}=\eh(g^3_{abc}-g^3_{acb})+g^5_{acb}.\label{3.2.35}\ee 
Hence, we find
\be g^5_{abc}=g^3_{abc}-f_{abc}+d_{abc},\label{3.2.36}\ee
where Eq. (\ref{3.2.33}) has been taken into account.

Now we turn to Eq. (\ref{3.2.11}) and substitute $g^5$ from Eq. (\ref{3.2.36})
\be g^4_{abc}=g^7_{cab}-2g^6_{acb}-(g^3_{acb}-f_{acb}+
d_{abc}).\label{3.2.37}\ee
Using this in Eq. (\ref{3.2.10}) we see that $g^3$ and $g^7$ cancel out so that
finally
\be g^6_{abc}=\eh f_{abc}.\label{3.2.38}\ee 
Then Eq. (\ref{3.2.37}) can be simplified to
\be g^4_{abc}=g^7_{cab}-g^3_{acb}-d_{abc}.\label{3.2.39}\ee 
Substituting this into Eq. (\ref{3.2.14}) gives
\be f^2_{abc}=-\eh(f^5_{bac}+g^7_{bac}-d_{abc}).\label{3.2.40}\ee 
$f^3$ follows from Eq. (\ref{3.2.16}):
\be f^3_{abc}=-2g^6_{abc}-g^7_{abc}=-g^7_{abc}-f_{abc}.\label{3.2.41}\ee
On the other hand, from Eq. (\ref{3.2.21}) we get a different result
\be f^3_{abc}=-g^7_{cba}+f_{abc},\label{3.2.42}\ee
which implies
\be g^7_{abc}=g^7_{cba}-2f_{abc}.\label{3.2.43}\ee

Finally, from Eq. (\ref{3.2.15}) we conclude
\be f^4_{bac}+g^7_{bac}=-f^4_{cab}-g^7_{cab}\label{3.2.44}\ee
and Eq. (\ref{3.2.17}) gives another symmetry relation
\be f^5_{abc}+g^7_{abc}=f^5_{bac}+g^7_{bac}.\label{3.2.45}\ee
It is easily checked that with the results just obtained all
equations are identically satisfied.

Summing up we have obtained the following form of the trilinear coupling
$$ T_1=ig\Bigl\{(f_{abc}+d_{abc})A_{\mu a}A_{\nu b}\d^\nu A^\mu_c-
\eh (f^5_{bac}+g^7_{bac}-d_{abc})
A_{\mu a}A_b^\mu\d_\nu A_c^\nu$$
\be
-(g^7_{abc}+f_{abc})A_{\mu a}u_b\d^\mu\tilde u_c+f^4_{abc}(\d^\mu
A_{\mu a})u_b\tilde u_c
+f^5_{abc}A_{\mu a}(\d^\mu u_b)\tilde u_c\Bigl\}.\label{3.2.46}
\ee  
The terms proportional to $d_{abc}$ give a divergence
\be
d_{abc}(A_{\mu a}A_{\nu b}\d^\nu A_c^\mu+\eh A_{\mu a}A_b^\mu\d_\nu
A_c^\nu)=\eh d_{abc}\d^\nu(A_{\mu a}A_b^\mu A_{\nu c}). \label{divergence_d}
\ee
This can be left out because it does not change the S-matrix. Next
it is important to remember the relation
Eq. (\ref{3.2.44}) which shows the antisymmetry with
respect to $b$ and $c$. Therefore, we have
\be
C_1\=d ig(f^4_{abc}+g^7_{abc})(\d^\mu A_{\mu a})u_b\tilde u_c= 
-\eh g(f^4_{abc}+g^7_{abc})
d_Q(\tilde u_au_b\tilde u_c). \label{C1def}
\ee
Such a term which is $d_Q$ of "something" is called a coboundary 
in cohomology theory \cite{Massey}. Using 
Eq. (\ref{C1def}) in Eq. (\ref{divergence_d}) we have to add the term with $g^7$ which is taken
into account as follows
$$-g^7_{abc}(A_{\mu a}u_b\d^\mu \tilde u_c+\d^\mu A_{\mu a}u_b
\tilde u_c)$$
$$=-g^7_{abc}\d^\mu(A_{\mu a} u_b\tilde u_c)+g^7_{abc}A_{\mu a} 
\d^\mu u_b\tilde u_c).$$
Now $T_1$ assumes the following form
$$T_1=ig\Bigl\{f_{abc}A_{\mu a}A_{\nu b}\d^\nu A^\mu_c-\eh(f^5_{abc} 
+g^7_{abc})A_{\mu a}A_b^\mu\d^\nu A^\nu_c$$
$$-f_{abc}A_{\mu a}u_b\d^\mu\tilde u_c+(f^5_{abc}+g^7_{abc}) 
A_{\mu a}\d^\mu u_b\tilde u_c\Bigl\}$$
\be +{i\over 2}gd_{abc}\d^\nu(A_{\mu a}A_b^\mu A_{\nu c})-ig
g^7_{abc}\d^\mu(A_{\mu a}u_b\tilde u_c)+C_1.\label{3.2.50}\ee
Due to Eq. (\ref{3.2.45}) the second and fourth term together give a second
coboundary 
$$C_2=ig(f^5_{abc}+g^7_{abc})(A_{\mu a}\d^\mu u_b\tilde u_c-\eh 
A_{\mu a}A^\mu_b\d_\nu A^\nu_c)$$
\be
={i\over 2}g(f^5_{abc}+g^7_{abc})d_Q(A_{\mu a}A^\mu_b\tilde u_c).
\ee
The coboundary terms lead to an equivalent S-matrix as well.

Omitting the trivial divergence and coboundary terms we arrive at the
following final result
\be T_1=ig f_{abc}(A_{\mu a}A_{\nu b}\d^\nu A^\mu_c-A_{\mu a}u_b
\d^\mu\tilde u_c).\label{3.2.52}\ee 
This is the well-known Yang-Mills plus ghost coupling to lowest order.
At second order, gauge invariance gives the remaining coupling terms
of pure Yang-Mills theory (see \cite{Scharf2}, sect. 3.4).

The real strength of the method comes out in massive gauge theories.
Since in S-matrix theory the asymptotic free fields are the basic objects,
one has to start with massive gauge fields from the beginning. Then gauge
invariance of first and second order has to work and fixes all couplings.
In particular, a physical scalar field, the Higgs field is necessary to
satisfy second order gauge invariance. But the Brout-Englert-Higgs mechanism
and spontaneous symmetry breaking plays no immediate role in such an approach.
For details we refer to \cite{Scharf2}.

In order to motivate the formal accomplishments constructed so far,
we conclude by giving a rather qualitative comparison of the present formalism
to the texbook literature.
Above, we observed that QED is gauge invariant,
but the true importance of gauge invariance is the fact
that it allows to prove on a formal level
the {\em{unitarity}} of the S-matrix on the physical subspace
\cite{Dutsch}.
The presence of a skew-adjoint operator $A^0$ in the first order interaction
or the presence of unphysical longitudinal and timelike photon states
causes the S-matrix to be non-unitary on the full Fock space, but it is on
the physical subspace.
In QED, ghosts are introduced only as a formal tool, since
they 'blow up' the Fock space unnecessarily, and they do not
interact with the electrons and photons. But in QCD, the situation
is not so trivial, due to the self-coupling of the gauge fields.

The gluon vector potential can be represented by the traceless Hermitian
$3 \times3$ standard Gell-Mann matrices $\lambda^a$, $a=1,...8$
\be
A_\mu=\sum_{a=1}^{8} A_\mu^a \frac{\lambda^a}{2} =: A_\mu^a
\frac{\lambda^a}{2} .
\ee
The $\lambda$'s satisfy the commutation and normalization relations
\be
\Bigl[ \frac{\lambda^a}{2}, \frac{\lambda^b}{2} \Bigr]=
i f_{abc} \frac{\lambda^c}{2}, \quad \mbox{tr} \, (\lambda^a \lambda^b)=
2 \delta_{ab}, \label{goodrelations}
\ee
and the numerical values of the structure constants
$f_{abc}=-f_{bac}=-f_{acb}$ can be found
in numerous QCD textbooks.
Since we are working with a fixed matrix representation, we do not care
whether the color indices are upper or lower indices.

The natural generalization of the QED Lagrangian to the
Lagrangian of purely gluonic QCD is
\be
{\cal{L}}_{gluon} = -\frac{1}{2} \mbox{tr} \, G_{\mu \nu} G^{\mu \nu}
=-\frac{1}{4} G^a_{\mu \nu} G_a^{\mu \nu},
\ee
with
\be
G_{\mu \nu}=\partial_\mu A_\nu-\partial_\nu A_\mu-ig[A_\mu,A_\nu]
\ee
or, using the first relation of Eq. (\ref{goodrelations})
\be
G_{\mu \nu}^a=\partial_\mu A_\nu^a - \partial_\nu A_\mu^a+
g f_{abc} A_\mu^b A_\nu^c .
\ee
It is an important detail that we are working with {\em{interacting}}
classical fields here, therefore the corresponding field strength tensor $G$
contains a term proportional to the coupling constant in contrast to the free
fields
$F_{\mu \nu}^{free}=\partial_\mu A_\nu^{free}-\partial_\nu A_\mu^{free}$
used throughout this paper. ${\cal{L}}_{gluon}$ is invariant under
classical local gauge transformations
\be
A_\mu(x) \rightarrow U(x) A_\mu(x) U^{-1}(x)
+\frac{i}{g} U(x) \partial_\mu U^{-1}(x),
\ee
where $U(x) \in SU(3)$.

We extract now the first order gluon coupling from the Lagrangian.
The Lagrangian
\be
{\cal{L}}_{gluon}=-\frac{1}{4} [\partial_\mu A_\nu^a - \partial_\nu A_\mu^a+
g f_{abc} A_\mu^b A_\nu^c]
[\partial^{\mu} A^{\nu}_{a} - \partial^{\nu} A^{\mu}_{a}+
g f_{ab'c'} A^{\mu}_{b'} A^{\nu}_{c'}],
\ee
contains obviously the free field part (this terminology is not
really correct, since we are dealing with interacting fields
here)
\be
{\cal{L}}_{gluon}^{free}=-\frac{1}{4} [\partial_\mu A_\nu^a - \partial_\nu A_\mu^a][\partial^{\mu} A^{\nu}_{a} - \partial^{\nu} A^{\mu}_{a}]
\ee
and the first order interaction part is given by
\be
{\cal{L}}_{gluon}^{int}=
-\frac{1}{4} [\partial_\mu A_\nu^a - \partial_\nu A_\mu^a]
[g f_{ab'c'} A^{\mu}_{b'} A^{\nu}_{c'}]-\frac{1}{4}
[g f_{abc} A_\mu^b A_\nu^c]
[\partial^{\mu} A^{\nu}_{a} - \partial^{\nu} A^{\mu}_{a}]
\ee
\be
=-\frac{g}{2} f_{abc} A_\mu^b A_\nu^c
[\partial^{\mu} A^{\nu}_{a} - \partial^{\nu} A^{\mu}_{a}]=
-\frac{g}{2} f_{abc} A_\mu^a A_\nu^b
[\partial^{\mu} A^{\nu}_{c} - \partial^{\nu} A^{\mu}_{c}]
\ee
\be
=g f_{abc} A_\mu^a A_\nu^b \partial^{\nu} A^{\mu}_{c}.
\ee
The first interaction terms comes out from classical symmetry considerations
here; in the framework presented in this paper, it is the consequence
of purely quantum mechanical considerations.

Since we are working in Feynman gauge, we add
the corresponding gauge fixing term ${\cal{L}}_{gf}$ to the Lagrangian. Additionally,
we add the ghost term which describes the ghost interaction.
The total Lagrangian then reads
\be
{\cal{L}}_{QCD}={\cal{L}}_{gluon}+{\cal{L}}_{gf}
+{\cal{L}}_{ghost}
\ee
\be
={\cal{L}}_{gluon}-\frac{1}{2}(\partial_\mu A^\mu_a)^2+
\partial^\mu \tilde{u} ( \partial_\mu u_a-g f_{abc} u_b A_{\mu c}).
\ee
The classical ghosts are anticommuting Grassmann numbers, i.e.
$u^2=\tilde{u}^2=0, u \tilde{u}=-\tilde{u} u$.

The BRST transformation is defined by
\be
\delta A_\mu^a= i\lambda (\partial_\mu u_a-g f_{abc} u_b A_{\mu c}),
\ee
\be
\delta \tilde{u}_a=-i \lambda \partial_\mu A^\mu_a,
\ee
\be
\delta u_a= \frac{g}{2} \lambda f_{abc} u_b u_c,
\ee
where $\lambda$ is a space-time independent anticommuting
Grassmann variable.
The special property of the BRST transformation is the fact that
the actions
\be
S_{gluon}=\int d^4 x \, {\cal{L}}_{gluon}, \quad
S_{gf}+S_{ghost}=\int d^4 x \, ({\cal{L}}_{gf}
+{\cal{L}}_{ghost})
\ee
and $S_{total}=S_{gluon}+S_{gf}+S_{ghost}$
are all invariant under the transformation:
\be
\delta S_{gluon}=0, \quad \delta(S_{gf}+S_{ghost})=0.
\ee
The similarity of free quantum gauge transformation introduced
in this paper to the BRST transformation is obvious.
One important difference is the absence of interaction terms
$\sim g$. Furthermore, the free quantum gauge transformation
is a transformation of free quantum fields, whereas the
BRST transformation is a transformation of classical fields,
which enter in path integrals when the theory is
quantized. Finally, the free gauge transformation leaves
the $T_n$'s invariant up to divergences, whereas the
BRST transformation is a symmetry of the full QCD Lagrangian.
How the two symmetries are intertwined perturbatively is explained
in \cite{Hurth1}. A more rigorous axiomatic approach is discussed in
\cite{MasterWard,Dutsch1}.

\subsection{\it Spin 2}
The crucial test of the gauge principle is spin 2 where it should
lead to a quantum theory of gravity.
In this case we supplement the gauge invariance condition
\be [Q,T(x)]=d_Q T(x)=i\d_\al T^\al(x),\label{1.1}\ee
in the following way. Since $d_Q$ and
the space-time derivative $\d_\al$ commute it follows from nilpotency that
\be
\d_\al d_Q T^\al=0.
\ee
If the appropriate form of the Poincar\'e lemma is true, this implies
\be d_QT^\al=[Q,T^\al]=i\d_\beta T^{\al\beta}\label{1.2}\ee
with antisymmetric $T^{\al\beta}$. In the same way we get
\be [Q,T^{\al\beta}]=i\d_\ga T^{\al\beta\ga}\ldots\label{1.3}\ee
with totally antisymmetric $T^{\al\beta\ga}$ and so on. These are the
so-called descent equations (similar to Wess-Zumino consistency conditions).
It is our aim to find a solution of these equations describing the self-coupling
of the symmetric tensor field $h^{\mu\nu}$ considered in sect. 3.1, Eq. (\ref{(1.7.2)}).
We recall the gauge variations Eq. (\ref{(1.7.9)}):
\be d_Qh^{\mu\nu}=-{i\over 2}(\d^\nu u^\mu+\d^\mu u^\nu-\eta^{\mu\nu}\d_\al u^\al)
\label{2.6}\ee 
$$d_Q u^\mu=0$$
\be d_Q\tilde u^\mu=i\d_\nu h^{\mu\nu},\label{2.7}\ee 
where we now denote the Minkowski tensor by $\eta^{\mu\nu}$ to distinguish it from
Einstein's $g^{\mu\nu}$.

The descent procedure starts from $T^{\al\beta\gamma}$ which must contain three
ghost fields $u$ and two derivatives and is totally antisymmetric. To exclude
trivial couplings we require that it does not contain a co-boundary $d_QB$ for
some $B\ne 0$. Then there are the following two possibilities only:
\be
\d^\beta u^\al u^\mu\d_\mu u^\gamma,\quad \d^\al u^\mu\d_\mu u^\beta u^\gamma.
\ee
Therefore we start the descent procedure with the expression
$$ T^{\al\beta\gamma}=a_1(\d^\beta u^\al u^\mu\d_\mu u^\gamma-\d^\al u^\beta u^\mu\d_\mu u^\gamma-\d^\beta u^\gamma u^\mu\d_\mu u^\al-\d^\gamma u^\al u^\mu\d_\mu u^\beta$$
$$ +\d^\al u^\gamma u^\mu\d_\mu u^\beta+\d^\gamma u^\beta u^\mu\d_\mu u^\al)+
a_2(\d^\al u^\mu \d_\mu u^\beta u^\gamma-\d^\beta u^\mu \d_\mu u^\al u^\gamma$$
\be -\d^\gamma u^\mu \d_\mu u^\beta u^\al-\d^\al u^\mu \d_\mu u^\gamma u^\beta
+\d^\gamma u^\mu \d_\mu u^\al u^\beta+\d^\beta u^\mu \d_\mu u^\gamma u^\al).
\label{2.12}\ee

Next we have to compute $\d_\ga T^{\al\beta\ga}$ and this is equal to
$-id_Q T^{\al\beta}$ by Eq. (\ref{1.3}). To determine $T^{\al\beta}$ requires an
"integration" $d_Q^{-1}$. As always in calculus this integration can be
achieved by making a suitable ansatz for $T^{\al\beta}$ and fixing the
free parameters. The following 5 parameter expression will do:
$$T^{\al\beta}=b_1u^\mu\d_\mu u_\nu\d^\beta h^{\al\nu}+b_2u^\mu\d_\nu u^\al
\d_\mu h^{\beta\nu}+b_3u^\al\d_\nu u^\mu\d_\mu h^{\beta\nu}$$
\be +{b_4\over 2}\d_\mu u^\al\d_\nu u^\beta h^{\mu\nu}+b_5\d_\mu u^\mu\d_\nu u^\al
h^{\beta\nu}-(\al\leftrightarrow\beta).\label{2.13}\ee
Substituting this into Eq. (\ref{1.3}) leads to
$$b_1=-2a_1,\> b_2=-2a_2=-2a_1,\> b_3=2a_1,\> b_4=-4a_1,\> b_5=-2a_1.$$  
An overall factor is arbitrary, we take $a_1=-1$ which gives
\be T^{\al\beta}=2(u^\mu\d_\mu u_\nu\d^\beta h^{\al\nu}+u^\mu\d_\nu u^\al
\d_\mu h^{\beta\nu}-u^\al\d_\nu u^\mu\d_\mu h^{\beta\nu}
+\d_\mu u^\al\d_\nu u^\beta h^{\mu\nu}+\d_\nu u^\nu\d_\mu u^\al h^{\beta\mu})
-(\al\leftrightarrow\beta).\label{2.14}\ee

In a similar way we compute $\d_\beta T^{\al\beta}$ and make an ansatz 
for $T^\al$.
The latter now has to contain ghost-antighost couplings also. The
precise form can be taken from the following final result:
$$T^\al=4u^\mu\d_\mu h_{\beta\nu}\d^\beta h^{\al\nu}-2u^\mu\d_\mu h^{\beta\nu}
\d^\al h_{\beta\nu}-2u^\al\d^\beta h^{\mu\nu}\d_\mu h_{\beta\nu}
-4\d_\nu u_\beta\d_\mu h^{\al\beta}h^{\mu\nu}+4\d_\nu u^\nu\d^\mu h^{\al\beta}
h_{\beta\mu}$$
$$+u^\al\d_\beta h_{\mu\nu}\d^\beta h^{\mu\nu} 
-2\d_\nu u^\nu h_{\mu\beta}\d^\al h^{\mu\beta}-{1\over 2}u^\al\d_\mu h\d^\mu h
+\d_\nu u^\nu hh^\al+u^\nu\d_\nu h\d^\al h
-2\d_\nu u^\mu h^{\mu\nu}\d^\al h$$
$$+4\d^\nu u_\mu\d^\al h^{\mu\beta} h_{\beta\nu}
-4\d^\nu u^\mu\d_\mu h^{\al\beta}h_{\beta\nu}
-2u^\mu\d_\mu u^\nu\d^\al\tilde u_\nu+2u^\mu\d_\nu u^\al\d_\mu\tilde u^\nu$$
\be -2u^\al\d_\nu u^\mu\d_\mu\tilde u^\nu+2\d_\nu u^\nu\d_\mu u^\al\tilde u^\mu
+2u^\mu\d_\mu\d_\nu u^\nu\tilde u^\al-2u^\al\d_\mu\d_\nu u^\mu\tilde u^\nu.
\label{2.15}\ee 
The last step calculating $\d_\al T^\al$ and setting it equal to
$-id_Q T$ gives the trilinear coupling of massless gravity
$$T=-h^{\al\beta}\d_\al h\d_\beta h+2h^{\al\beta}\d_\al h_{\mu\nu}\d_\beta h^{\mu\nu}
+4h_{\al\beta}\d_\nu h^{\beta\mu}\d_\mu h^{\al\nu}
+2h_{\al\beta}\d_\mu h^{\al\beta}\d^\mu h-4h_{\al\beta}\d_\nu h^{\al\mu}\d^\nu 
h_\mu^{\beta}$$
\be -4u^\mu\d_\beta\tilde u_\nu\d_\mu h^{\nu\beta}+4\d_\nu u^\beta\d_\mu\tilde u_\beta 
h^{\mu\nu}-4\d_\nu u^\nu\d_\mu\tilde u^\beta h^{\beta\mu}+4\d_\nu u^\mu\d_\mu
\tilde u_\beta h^{\nu\beta}.\label{2.16}\ee 

This coupling should have something to do with general relativity. To see this
we leave quantum field theory aside and take the metric tensor $g_{\mu\nu}$ as the
fundamental classical field. The indices are no longer Lorentz indices, they are
raised and lowered with $g^{\nu\ro}$ itself which is defined as the inverse
$g_{\mu\nu}g^{\nu\ro}=\delta_\mu^\ro.$
One also introduces the determinant
\be
g=\det g_{\mu\nu}.
\ee

Our starting point is the Einstein-Hilbert action given by
\be S_{EH}=-{2\over\kappa^2}\int d^4x\,\sqrt{-g}R,\quad
\kappa^2=32\pi G,\label{5.5.3}\ee 
where $G$ is Newton's constant. $R$ is the scalar
curvature
\be
R=g^{\mu\nu}R_{\mu\nu}
\ee
which follows from the Ricci tensor
\be R_{\mu\nu}=\d_\al\Gamma^\al_{\mu\nu}-\d_\nu\Gamma^\al_{\mu\al}+
\Gamma^\al_{\al\beta}\Gamma^\beta_{\mu\nu}-\Gamma^\al_{\nu\beta}
\Gamma^\beta_{\al\mu},\label{5.5.5}\ee 
where
\be \Gamma_{\beta\gamma}^\al=\eh g^{\al\mu}(g_{\beta\mu},_\gamma
+g_{\mu\gamma},_\beta-g_{\beta\gamma},_\mu)\label{5.5.6}\ee 
are the Christoffel symbols.

The variation of Eq. (\ref{5.5.3}) is given by
\be S_{EH}[g+\eps f]-S_{EH}[g]=\eps\int d^4x\,\Bigl({\d\over\d g^{\mu\nu}}
\sqrt{-g}g^{\al\beta}\Bigl)R_{\al\beta}f^{\mu\nu}(x)$$
$$+\int d^4x\,\sqrt{-g}g^{\al\beta}\Bigl(R_{\al\beta}[g+\eps f]-
R_{\al\beta}[g]\Bigl)+O(\eps^2).\label{5.5.7}\ee 
By calculating in geodesic coordinates one finds that the last term
vanishes. Since
$${\d\over\d g^{\mu\nu}}\sqrt{-g}g^{\al\beta}={1\over 2\sqrt{-g}}g
g_{\mu\nu}g^{\al\beta}+\sqrt{-g}\delta_\mu^\al\delta_\nu^\beta$$
\be
=\sqrt{-g}\Bigl(-{1\over 2} g_{\mu\nu}g^{\al\beta}+\delta_\mu^\al
\delta_\nu^\beta\Bigl),
\ee
we finally obtain
\be S_{EH}[g+\eps f]-S_{EH}[g]=\eps\int d^4x\,
\sqrt{-g}\Bigl(-\eh g_{\al\beta}R
+R_{\al\beta}\Bigl)f^{\al\beta}(x)+O(\eps^2).\label{5.5.8}\ee 
This implies Einstein's field equations in vacuum
\be
R_{\al\beta}-\eh g_{\al\beta}R=0.
\ee
For this reason the Lagrangian
\be L_{EH}=-{2\over\kappa^2}\sqrt{-g}R \label{5.5.10}\ee 
can be taken as starting point of the classical theory.

A glance at Eq. (\ref{5.5.5}) and Eq. (\ref{5.5.6}) shows that the first two terms 
in Eq. (\ref{5.5.5}) contain
second derivatives of the fundamental tensor field $g_{\mu\nu}$. This
defect can be removed by splitting off a divergence. We rewrite the
first term in Eq. (\ref{5.5.5}) as 
\be \sqrt{-g}g^{\mu\nu}\Gamma^\al_{\mu\nu},_\al=(
\sqrt{-g}g^{\mu\nu}\Gamma^\al_{\mu\nu}),_\al-
\Gamma^\al_{\mu\nu}(\sqrt{-g}g^{\mu\nu}),_\al\label{5.5.11}\ee 
and calculate the last derivative with the help of
\be
g^{\mu\nu},_\al=-\Gamma^\mu_{\beta\al}g^{\beta\nu}
-\Gamma^\nu_{\al\beta}g^{\beta\mu}.
\ee
Proceeding with the second term in the same way we find
\be \sqrt{-g}R=\sqrt{-g}G-\Bigl(\sqrt{-g}g^{\mu\nu}\Gamma^\al_{\mu\nu}-
\sqrt{-g}g^{\mu\al}\Gamma^\nu_{\mu\nu}\Bigl),_\al\label{5.5.12}\ee 
where
\be G=g^{\mu\nu}\Bigl(\Gamma^\al_{\nu\beta}\Gamma^\beta_{\mu\al}-
\Gamma^\al_{\mu\nu}\Gamma^\beta_{\al\beta}\Bigl).\label{5.5.13}\ee 
Since the divergence in Eq. (\ref{5.5.12}) does not matter in the variational 
principle, we can go on with the Lagrangian
\be L=-{2\over\kappa^2}\sqrt{-g}g^{\mu\nu}\Bigl(\Gamma^\al_{\nu\beta} 
\Gamma^\beta_{\mu\al}-
\Gamma^\al_{\mu\nu}\Gamma^\beta_{\al\beta}\Bigl),\label{5.5.14}\ee 
which contains first derivatives of $g$ only.

For the following it is convenient to remove the square root $\sqrt{-g}$
by introducing the so-called  Goldberg variables
\be \ti g^{\mu\nu}=\sqrt{-g}g^{\mu\nu},\quad
\ti g_{\mu\nu}=(-g)^{-1/2}g_{\mu\nu}.\label{5.5.15}\ee 
Using
$$\d_\ro g=g\ti g_{\al\beta}\ti g^{\al\beta},_\ro$$
$$\d_\ro g^{\mu\nu}=(-g)^{-1/2}\Bigl(\ti g^{\mu\nu},_\ro-\eh
\ti g^{\mu\nu}\ti g_{\al\beta}\ti g^{\al\beta},_\ro\Bigl),$$
\be \d_\ro g_{\mu\nu}=\sqrt{-g}\Bigl(\eh\ti g_{\mu\nu}\ti g_{\al\beta} 
\ti g^{\al\beta},_\ro-\ti g_{\mu\al}\ti g_{\nu\beta}\ti g^{\al\beta},_\ro 
\Bigl)\label{5.5.16}\ee
in Eq. (\ref{5.5.6}) we obtain
$$\Gamma^\al_{\beta\gamma}=\eh\Bigl(\eh\delta_\beta^\al\ti g_{\mu\nu}
\ti g^{\mu\nu},_\gamma+\eh\delta_\gamma^\al\ti g_{\mu\nu}
\ti g^{\mu\nu},_\beta-\ti g_{\beta\mu}\ti g^{\al\mu},_\gamma$$
\be -\ti g_{\gamma\mu}\ti g^{\al\mu},_\beta+\ti g^{\al\ro}\ti g_{\gamma\mu}
\ti g_{\beta\nu}\ti g^{\mu\nu},_\ro-\eh\ti g^{\al\ro}\ti g_{\beta\gamma}
\ti g_{\mu\nu}\ti g^{\mu\nu},_\ro\Bigl).\label{5.5.17}\ee 
This enables us to express the Lagrangian $L$ in Eq. (\ref{5.5.14}) by $\ti
g_{\mu\nu}$. It is simple to compute the second term
\be \ti g^{\mu\nu}\Gamma^\al_{\mu\nu}\Gamma^\beta_{\al\beta}=-\eh\ti 
g^{\mu\al},_\mu\ti g_{\nu\beta}\ti g^{\nu\beta},_\al.\label{5.5.18}\ee
But the first term in Eq. (\ref{5.5.14}) requires the collection of many terms,
until one arrives at the simple result
$$\ti g^{\mu\nu}\Gamma^\al_{\nu\beta}\Gamma^\beta_{\mu\al}={1\over 4} 
\Bigl(-2\ti g_{\mu\nu}\ti g^{\mu\nu},_\al\ti g^{\al\beta},_\beta+
2\ti g_{\al\beta}\ti g^{\al\mu},_\nu\ti g^{\beta\nu},_\mu$$
\be -\ti g_{\al\ro}\ti g_{\beta\si}\ti g^{\mu\nu}\ti g^{\ro\beta},_\mu 
\ti g^{\si\al},_\nu+\eh\ti g^{\al\beta}\ti g_{\mu\nu}\ti g^{\mu\nu},_\al
\ti g_{\ro\si}\ti g^{\ro\si},_\beta\Bigl).\label{5.5.19}\ee 
Then the total Lagrangian is given by
$$L={1\over\kappa^2}\Bigl(-\ti g_{\al\beta}\ti g^{\al\mu},_\nu\ti
g^{\beta\nu},_\mu+ \eh\ti g_{\al\ro}\ti g_{\beta\si}\ti g^{\ro\beta},_\mu
\ti g^{\al\si},_\nu\ti g^{\mu\nu}$$
\be -{1\over 4}\ti g_{\mu\nu}\ti g^{\mu\nu},_\al\ti g_{\ro\si}
\ti g^{\ro\si},_\beta\ti g^{\al\beta}\Bigl).\label{5.5.20}\ee 

To make contact with quantum field theory on Minkowski space we
consider the situation in scattering theory where at large distances in
space and time the geometry is flat and given by the Minkowski metric
$\eta^{\mu\nu}$. Then we write the metric tensor as a sum
\be \ti g^{\mu\nu}(x)=\eta^{\mu\nu}+\kappa h^{\mu\nu}(x).\label{5.5.21}\ee 
We do not assume that the new dynamical field $h^{\mu\nu}(x)$ is small
in some sense, it only goes to zero at large distances because of the
asymptotically flat situation. The indices of $h^{\mu\nu}$ are ordinary
Lorentz indices which can be raised and lowered with the Minkowski
metric. Then the inverse of Eq. (\ref{5.5.21}) is given by
\be \ti g_{\mu\nu}(x)=\eta_{\mu\nu}-\kappa h_{\mu\nu}(x)+\kappa^2 
h_{\mu\al}h^\al\,_\nu-\ldots.\label{5.5.22}\ee 
Substituting these expressions into Eq. (\ref{5.5.20}), the Lagrangian $L$ becomes
an infinite sum
\be L=\sum_{n=0}^\infty\kappa^n L^{(n)}.\label{5.5.23}\ee
Here is the proliferation of couplings which is typical for gravity.
It can be traced back to the infinite series Eq. (\ref{5.5.22}). The three 
terms in Eq. (\ref{5.5.20}) give the contributions
\be L_1=\Bigl(-\eta_{\al\beta}+\kappa h_{\al\beta}-\kappa^2h_{\al\ro}
h^\ro_\beta+\ldots\Bigl)h^{\al\mu},_\nu h^{\beta\nu},_\mu\label{5.5.24}\ee 
$$ L_2=\eh\Bigl(\eta_{\al\ro}-\kappa h_{\al\ro}+\kappa^2h_{\al\al'}
h^{\al'}_\ro-\ldots\Bigl)\Bigl(\eta_{\beta\si}-\kappa h_{\beta\si} 
+\kappa^2h_{\beta\beta'}h^{\beta'}_\si-\ldots\Bigl)$$
\be \times\,(\eta^{\mu\nu}+\kappa h^{\mu\nu})h^{\ro\beta},_\mu 
h^{\al\si},_\nu\label{5.5.25}\ee 
$$ L_3={1\over 4}\Bigl(-\eta_{\mu\nu}+\kappa h_{\mu\nu}-\kappa^2h_{\mu\al'}
h^{\al'}_\nu-\ldots\Bigl)\Bigl(\eta_{\ro\si}-\kappa h_{\ro\si} 
+\kappa^2h_{\ro\beta'}h^{\beta'}_\si-\ldots\Bigl)$$
\be \times\,(\eta^{\al\beta}+\kappa h^{\al\beta})h^{\mu\nu},_\al 
h^{\ro\si},_\beta.\label{5.5.26}\ee 

The lowest order
\be L^{(0)}={1\over 2}h^{\al\beta},_\mu h_{\al\beta}^{,\mu}-
h^{\al\beta},_\mu h_\al^\mu,_\beta-{1\over 4}h,_\al h^{,\al},
\label{5.5.27}\ee 
where $h=h^\mu\,_\mu$, defines the free theory. Indeed, the
corresponding Euler-Lagrange equations reads
\be \sq h^{\al\beta}-\eh\eta^{\al\beta}\sq h-h^{\al\mu,\beta},_\mu
-h^{\beta\mu,\al},_\mu=0.\label{5.5.28}\ee 
Both Eq. (\ref{5.5.27}) and Eq. (\ref{5.5.28}) are invariant under the classical gauge
transformation
\be h^{\al\beta}\rightarrow \ti h^{\al\beta}=h^{\al\beta}+f^{\al,
\beta}+f^{\beta,\al}-\eta^{\al\beta}f^\mu,_\mu.\label{5.5.29}\ee 
The gauge can be specified by the Hilbert condition
\be \ti h^{\al\beta},_\beta=0.\label{5.5.30}\ee 
This can be achieved by choosing the solution of the inhomogeneous
wave equation
$$\sq f^\al=-h^{\al\beta},_\beta$$
as gauge function in Eq. (\ref{5.5.29}). In the Hilbert gauge the equation of 
motion Eq. (\ref{5.5.28}) gets simplified
\be
\sq h^{\al\beta}-\eh\eta^{\al\beta}\sq h=0.
\ee
Taking the trace we conclude
\be
\sq h=0,\quad \sq h^{\al\beta}=0,
\ee
so that we precisely arrive at the free tensor field as it was
assumed in the QFT.

The first order coupling $O(\kappa)$ in Eqns. (\ref{5.5.24})-(\ref{5.5.26}) 
can easily be computed
$$L^{(1)}=-{1\over 4}h^{\al\beta}h,_\al h,_\beta+{1\over 2}h^{\mu\nu}
h^{\al\beta},_\mu h_{\al\beta},_\nu+h_{\al\beta}h^{\al\mu},_\nu
h^{\beta\nu},_\mu$$
\be
+{1\over 2}h_{\mu\nu}h^{\mu\nu},_\al h^{\prime\al}-h_{\mu\nu}h^{\al\mu},_\ro
h^{\nu,\ro}_\al.
\ee
The first three terms herein agree precisely with the first three terms
in Eq. (\ref{2.16}). The last two terms and the forth and fifth terms in 
Eq. (\ref{2.16}) are 
divergences. This is due to Lorentz contraction of the two derivatives.
Indeed, if $f_1$, $f_2$, $f_3$ are massless fields satisfying the wave
equation, then the following identity is true
\be
2\d_\al f_1 \d^\al f_2f_3=\d^\al(\d_\al f_1f_2f_3+f_1\d_\al f_2f_3
-f_1f_2\d_\al f_3).
\ee
Since divergence couplings do not change the physics, the coupling
Eq. (\ref{2.16}) derived from spin-2 quantum gauge theory agrees with general
relativity in lowest order. It agrees at higher orders, too (see \cite{Scharf2},
sect. 5.7). The gauge principle even works in massive gravity \cite{Scharf3}.
The cohomological nature of gauge invariance was analyzed in \cite{Grigore_cohom}.

The approach presented above is perturbative in nature and lives on a trivial
background. Presently, no fully satisfactory theory of quantum gravity exists,
and other ambitious approaches like, e.g., loop quantum gravity aim at a
formally background independent description of quantum gravity, and
they are expected to give rise to spacetime itself at distances
which are large compared to the Planck length. How Einstein's classical
geometric view on spacetime is related to such a theory is another story.
Here, we content ourselves with the observation that we have found a
gauge principle which uses the cohomological formulation of gauge invariance in
Eq. (\ref{1.1}) etc for the time-ordered products, having the character of a universal principle.
Consequently it must be respected in any conventional regularization method.

\section{Conclusion}
Causality is a fundamental guiding element for the construction of perturbative
quantum field theories. Using causality in conjunction with a proper mathematical handling of
distribution theory enables one to avoid ultraviolet divergences in perturbative quantum
field theory from the start. Whereas standard methods like dimensional regularization
have calculational advantages compared to the causal method, the causal method provides
a mathematically well-defined construction scheme of the perturbative $S$-matrix.

In this review, a condensed introduction and overview of the causal approach to regularization
theory has been given, which goes back to a classic paper by Henry Epstein and
Vladimir Jurko Glaser \cite{Epstein}. The causal approach was taken up by Michael D\"utsch
and G\"unter Scharf in 1985. During the last two decades, several important aspects
of the theory have been worked out, which constitute the basis of this review.
It should be mentioned that several topics which are not part of this work have been
treated in the recent literature, like e.g. interacting fields \cite{Interacting},
a complete discussion of perturbative QCD to all orders was worked out \cite{Dutsch},
and gauge theories like the full standard model (including phenomena like spontaneous
symmetry breaking) were studied in \cite{Electroweak,Gracia}. Theories in dimensions
other than four were also considered \cite{QED3} and specific analytic calculations
of multi-loop diagrams were carried out \cite{Multiloop,Twoloop}.
Supersymmetric theories \cite{SUSY} have been investigated, and the causal method was
generalized to field theories on curved space-times \cite{Fredenhagen} and studied in
the framework of light cone quantum field theory \cite{Werner}.

As mentioned before, there are severe conceptual differences between the causal method
and other regularization methods, which make it difficult to compare the different
approaches in a reasonable way. Therefore, specific examples have been used in this
work in order to demonstrate the differences and connections between the causal and
dimensional regularization. On the one hand, dimensional regularization has many attractive
features concerning the preservation of gauge invariance and in actual calculations
due to its well established methods. On the other hand, the causal method is a
strictly mathematical approach without any "intuitive" aspects like continuous
spacetime dimensions. Furthermore, the formulation of quantum gauge invariance
found during the study of gauge theories in the causal framework has a cogent
structure when compared to the standard BRST approach \cite{BRS,Tyutin}. In this sense,
the causal method constitutes an independent framework in its own right with many
attractive features. Critical issues like, e.g. axial anomalies can be discussed in a
unambiguous manner,  and the strong mathematical background of the method permits to apply
it to problems on curved spacetime and to quantum gravity, as we have illustrated
in the last section.

The mere observation that ultraviolet divergences can be avoided by a proper
mathematical construction of Feynman diagrams certainly puts some arguments in the literature
concerning the short-range behavior of quantum field theories in connection with ultraviolet
divergences into perspective.
Several approaches to QFT have been developed so far, and it is obvious that all considerations
presented in this review are based on perturbation theory.
Even if perturbation theory is well-defined order by order, it is far from being clear
that the perturbation expansion can be summed up for physically relevant theories,
even Borel summability is most probably not fulfilled due to Landau ghosts and renormalons.
Despite these problems,  perturbative QFT is a very successful and promising approach, since theoretical
predictions of physical quantities made by using renormalized Feynman graph
calculations match experimental results with a vertiginous high precision. In all these
calculations, one should not think that it is impossible to avoid ill-defined integrals,
as the causal approach proves. However, it should be mentioned that formal infinities
are admissible if they are treated within a rigorous mathematical framework. Recent developments
by Alain Connes and Dirk Kreimer based on Hopf algebras \cite{KreimerConnes} have
lead to some profound understanding how to "absorb" ultraviolet divergences in a consistent manner
by a redefinition of the parameters defining the QFT.
The Hopf algebra approach has also been applied to the causal Epstein-Glaser approach in
\cite{Graciabondia}, in order to overcome the separation between the causal method and
mainstream QFT.

Richard Feynman in his Nobel lecture remarked: "I think that the renormalization theory
is simply a way to sweep the difficulties of the divergencies of electrodynamics
under the rug." This problem has been solved by the causal method, at least
on a perturbative level.

\appendix
\section{Special Distributions in 3+1-Dimensional Spacetime}
In this appendix, we give a condensed account of the most important properties of the causal
commutators and propagators used in the present review. The distributions
used in the causal framework typically differ from the most common definitions
found in the literature and by a sign or a normalization factor, since we use the
"mathematical" symmetric definition of the (inverse) Fourier transform. 
Accordingly, the distributions used in the present text are related to the
distributions below by the simple redefinitions
\be
(2 \pi)^2 \hat{D}_{F,ret}^m (k) = - \hat{\Delta}^{F,ret}_m (k),
\quad
(2 \pi)^2 \hat{D}^{(\pm)}_m (k) = - \hat{\Delta}^\pm_m (k)
\ee
in momentum space and by ${D}_{F,ret}^m (x)= -{\Delta}^{F,ret}_m (x)$,
${D}^{(\pm)}_m (x) = -{\Delta}^\pm_m (x)$ in real space, omitting potential
mass indices.

The free (non-interacting) neutral scalar quantum field $\varphi(x)$
for particles with a given mass $m$ is given by
($k x= k_\mu x^\mu$, $k^0=\sqrt (\vec{k}^2+m^2)$)
\be
\varphi(x)=\varphi^{-}(x)+\varphi^{+}(x)=
\frac{1}{(2 \pi)^{3/2}} \int \frac{d^3 k}{\sqrt{2 k^0}}
[a(\vec{k}) e^{-ikx}+a^\dagger(\vec{k}) e^{+ikx}],
\ee
where $\varphi^{-}(x)$ and $\varphi^{+}(x)$ refer to the corresponding frequency parts,
respectively, whereas the charged field $\varphi_c(x)$ has the form
\be
\varphi_c(x)=\varphi_c^{-}(x)+\varphi_c^{+}(x)=
\frac{1}{(2 \pi)^{3/2}} \int \frac{d^3 k}{\sqrt{2 k^0}}
[a(\vec{k}) e^{-ikx}+b^\dagger(\vec{k}) e^{+ikx}].
\ee
The commutators of the operator-valued distributions $a(\vec{k})$, $b(\vec{k})$
("annihilation operators") and $a^\dagger(\vec{k})$, $b^\dagger(\vec{k})$
("creation operators") are
\be
[a(\vec{k}),a^\dagger(\vec{k'})]=[b(\vec{k}),b^\dagger(\vec{k'})]=
\delta^{(3)} (\vec{k}-\vec{k'}),
\ee
\be
[a(\vec{k}),a(\vec{k'})]=[b(\vec{k}),b(\vec{k'})]=
[a^\dagger(\vec{k}),a^\dagger(\vec{k'})]=[b^\dagger(\vec{k}),b^\dagger(\vec{k'})]=0
\quad \forall \, \vec{k}, \vec{k}',
\ee
and the annihilation operators destroy the unique perturbative vacuum $|0 \rangle$ according to
$a(\vec{k}) |0 \rangle=b(\vec{k}) |0 \rangle=0$ $\forall \,  \vec{k}$.

The commutation relations of the scalar fields lead to the so-called
positive and negative frequency Jordan-Pauli distributions
\be
\Delta^\pm_m(x)=-i [\varphi^{\mp}(x) , \, \varphi^{\pm}(0)] =
-i \langle 0 |  [\varphi^{\mp}(x) , \, \varphi^{\pm}(0)] | 0 \rangle \, ,
\label{commutator}
\ee
with the distributional Fourier transforms
\be
\hat{\Delta}^{\pm}_m(k)= \int d^4x \, \Delta^{\pm}_m(x) e^{ikx}=
\mp (2 \pi i) \,  \Theta(\pm k^0) \delta(k^2-m^2) . \label{dfourier}
\ee
$\delta$ is the one-dimensional Dirac distribution depending
on $k^2=k_\mu k^\mu=(k^0)^2-(k^1)^2-(k^2)^2-(k^3)^2=k_0^2-{\vec{k}}^2$.
The fact that the commutator
\be
[\varphi(x),\varphi(0)]=i\Delta^+_m(x)+i\Delta^-_m(x) =: i \Delta_m(x)
\ee
vanishes for spacelike arguments $x^2 < 0$
due to the requirement of microcausality, leads to
the important property that the Jordan-Pauli distribution
$\Delta_m$ has {\emph{causal support}}, i.e. it vanishes outside the closed
forward and backward light-cone such that
\be
\mbox{supp} \, \Delta_m(x) \subseteq \overline{V}^- \cup \overline{V}^+  \, , \quad
\overline{V}^\pm=\{x \, | \, x^2 \ge 0, \, \pm x^0 \ge 0 \}
\ee
in the sense of distributions.

The retarded propagator $\Delta^{ret}_m(x)$ is defined in configuration space by
\be
\Delta^{ret}_m(x)=\Theta(x^0) \Delta_m(x) ,
\ee
leading to the Fourier transformed expression
\be
\hat{\Delta}^{ret}_m(k)=\frac{1}{k^2-m^2+ik^0 0}.
\ee

The Feynman propagator is given in configuration space by the vacuum expectation value
\be
\Delta^F_m(x)=-i \langle 0 | T (\varphi_c (x) \varphi_c^\dagger(0) | 0 \rangle=
-i \langle 0 | T (\varphi_c^\dagger (x) \varphi_c(0) | 0 \rangle
-i \langle 0 | T (\varphi(x) \varphi(0) | 0 \rangle,
\ee
the well-known distributional Fourier transform reads
\be
\hat{\Delta}^F_m(k)=\frac{1}{k^2-m^2+i0}.
\ee
In the massless case, one has
\be
\Delta^F_0(x)=\int \frac{d^4 k}{(2 \pi)^4} \frac{e^{-ikx}}{k^2+i0}
=\frac{i}{4 \pi^2} \frac{1}{x^2-i0} = \frac{i}{4 \pi^2} P \frac{1}{x^2}-
\frac{1}{4 \pi} \delta (x^2),
\ee
where $T$ is the time-ordering operator, $P$ denotes principal value
regularization.

It is straightforward to show that the distributions introduced above fulfill the
distributional differential equations displayed below.
From the wave equation $(\Box+m^2) \varphi^{(\pm)}(x) = (\Box+m^2) \varphi_c^{\pm} (x)=0$ follows
\be
(\Box+m^2) \Delta^{\pm}_m(x)=(\Box+m^2) \Delta_m(x)=0.
\ee
Furthermore, one has
\be
(\Box+m^2) \Delta^F_m(x)=(\partial_\mu \partial^\mu+m^2) \Delta^F_m(x)=- \delta^{(4)} (x),
\ee
and
\be
(\Box+m^2) \Delta^{ret}_m(x)=- \delta^{(4)} (x).
\ee
The Feynman propagator and the retarded propagator are related via
\be
\Delta^{ret}_m=\Delta^F_m+\Delta^{-}_m.
\ee


\begin{thebibliography}{99}
\itemsep -2pt

\bibitem{Schwartz1966}
L. Schwartz, \emph{Th\'eorie des distributions}, 
Hermann, Paris, 1966.

\bibitem{Streater}
R.F. Streater, A.S. Wightman,
\emph{PCT, spin and statistics, and all that}, Benjamin Cummings, New York, 1964,
Princeton University Press, Princeton, 2000.

\bibitem{Constantinescu}
F. Constantinescu,
\emph{Distributions and their applications in physics},
Pergamon Press, Oxford, New York, 1980.

\bibitem{Senata}
E. Senata, \emph{Regularly Varying Functions},
Lecture Notes in Mathematics 508, Springer, Berlin, 1976.

\bibitem{Bogol}
N.N. Bogoliubov, D.V. Shirkov, \emph{Introduction to the theory of quantized fields},
Wiley-Interscience, New York, 1959.

\bibitem{Scharf2}
G. Scharf, \emph{Quantum Gauge Theories - a True Ghost Story}, John Wiley,
New York, 2001.

\bibitem{Scharf}
G. Scharf, \emph{Finite quantum electrodynamics: The causal approach}, 2nd ed.,
Springer-Verlag, New York, Berlin, 1995.

\bibitem{ADS}
A. Aste, G. Scharf, M. Duetsch,
\emph{Gauge independence of the S-matrix in the causal approach},
J. Phys. A31 (1563-1579) 1998.

\bibitem{DeWitt}
B. DeWitt, \emph{Quantum theory of gravity. I. The canonical theory},
\Journal{\PREV}{160} {1113-1148} {1967}.

\bibitem{Epstein}
H. Epstein \& V. Glaser, \emph{The role of locality in perturbation theory},
Annales de l'institut Henri Poincar\'e (A), Physique th\'eorique, 19 (211-295) 1973.

\bibitem{Narison}
S. Narison, \emph{Techniques of dimensional regularization and the two-point functions
of QCD and QED}, \Journal{\PREP}{84} {263-399}{1982}.

\bibitem{Jegerlehner}
F. Jegerlehner,
\emph{Renormalizing the standard model},
Boulder TASI 90:0476-590 (QCD161:T45:1990).

\bibitem{Hooft1972}
G. 't Hooft, M. Veltman,
\emph{Regularization and renormalization of gauge fields},
\Journal{\NPB}{44} {189-213} {1972}.

\bibitem{Bollini}
C. G. Bollini, J. J. Gambiagi, Nuovo Cimento 12A (1972) 20.

\bibitem{Schwinger1962} J. Schwinger, \emph{Gauge invariance and mass II},
\Journal{\PREV}{128} {2425-2429} {1962}.

\bibitem{Walther1998}
A. Aste, G. Scharf, and U. Walther, 
\emph{Power counting degree versus singular order in the Schwinger model},
Nuovo Cim. A 111 (1998) 323-327.

\bibitem{Casher1973}
A. Casher, J. Kogut, and L. Susskind, \emph{Vacuum polarization and the quark-parton puzzle},
\Journal{\PRL}{31} {792-795} {1973}.

\bibitem{Casher1974}
A. Casher, J. Kogut, and L. Susskind, \emph{Vacuum polarization and the absence of free quarks},
\Journal{\PRD}{10} {732-745} {1974}.

\bibitem{Strocchi}
F. Strocchi, \emph{Selected topics on the general properties of quantum field theory},
Lecture Notes in Physics 51, World Scientific, Singapore, New Jersey, London, Hong Kong, 1993.

\bibitem{Weinberg1960}
S. Weinberg, \emph{High-energy behavior in quantum field theory},
\Journal{\PREV}{128} {838-849} {1960}.

\bibitem{Vladimirov}
V.S. Vladimirov, Y.N. Drozhzhinov, B.I. Zavialov, 
\emph{Tauberian theorems for generalized functions}, Kluwer Acadademic Publishers,
Maine, 1988.

\bibitem{Adam1992}
C. Adam, R.A. Bertlmann, and P. Hofer,
\emph{Dispersion relation approach to the anomaly in 2 dimensions},
Z. Phys. C 56 (123-127) 1992.

\bibitem{Rohrlich}
F. Rohrlich, \emph{Quantum electrodynamics of charged particles without spin},
\Journal{\PREV}{80} {666-687} {1950}.

\bibitem{Duetsch_sQED}
M. D\"utsch, F. Krahe, G. Scharf, \emph{Scalar QED revisited},
Nuovo Cim. 106 (1993) 277-308.

\bibitem{Axial}
M. D\"utsch, F. Krahe, G. Scharf,
\emph{Axial anomalies in massless finite sQED},
Phys. Lett. B258 (457-460) 1991.

\bibitem{Schroer1}
B. Schroer,
\emph{A note on infraparticles and unparticles},
arXiv:0804.3563v5 [hep-th].

\bibitem{Schroer2}
B. Schroer,
\emph{Infrateilchen in der Quantenfeldtheorie
(Infraparticles in quantum field theory)},
Fortsch. Phys. 11 (1-31) 1963.

\bibitem{Aste2003}
A. Aste, D. Trautmann,
\emph{Finite calculation of divergent selfenergy diagrams},
Can. J. Phys. 81 (1433-1455) 2003.

\bibitem{Twoloop}
A. Aste, \emph{The two loop master diagram in the causal approach},
Annals Phys. 257 (158-204) 1997.

\bibitem{Epstein_adiabatic}
H. Epstein, V. Glaser,
\emph{Adiabatic limit in perturbation theory},
CERN-TH-1344 (1975), Erice 1975 Proceedings, Renormalization Theory,
Dordrecht (193-254) 1976.

\bibitem{Aste1999}
A. Aste, G. Scharf, \emph{Non-abelian gauge theories as a consequence of perturbative
quantum gauge invariance}, Int. J. Mod. Phys. A14 (3421-3434) 1999.

\bibitem{Massey}
W.S. Massey, \emph{Homology and cohomology theory},
Dekker, New York, 1978.

\bibitem{BRS}
C. Becchi, A. Rouet, R. Stora,
\emph{Renormalization of gauge theories},
\Journal{\ANNP}{98} {287-321} {1976}.

\bibitem{Tyutin}
I. V. Tyutin, Lebedev preprint FIAN 39 (1975), unpublished.

\bibitem{Hurth1}
T. Hurth, K. Skenderis, \emph{Quantum Noether method},
\Journal{\NPB} {541} {566-614} {1999}.

\bibitem{MasterWard}
M. D\"utsch, K. Fredenhagen,
\emph{The master Ward identity and generalized Schwinger-Dyson equation in
classical field theory}, Comm. Math. Phys. 243 (2003) 275-314.

\bibitem{Dutsch1}
M. D\"utsch, F.M. Boas,
\emph{The master Ward identity},
Rev. Math. Phys. 14 (977-1049) 2002.

\bibitem{Scharf3}
D.R. Grigore, G. Scharf, \emph{Massive gravity from descent equations},
Class. Quantum Grav. 25 (2008) 225008.

\bibitem{Grigore_cohom}
D. R. Grigore,
\emph{Perturbative gravity in the causal approach},
arXiv:0805.3438v2 [hep-th].

\bibitem{Interacting}
M. D\"utsch, F. Krahe, G. Scharf, \emph{Interacting fields in finite sQED},
Nuovo Cim. A103 (871-901) 1990.

\bibitem{Dutsch}
M. D\"utsch, T. Hurth, G. Scharf, \emph{Causal construction of Yang-Mills
theories. 4. Unitarity},  Nuovo Cim. A108 (737-774) 1995.

\bibitem{Electroweak}
A. Aste, G. Scharf, M. D\"utsch, \emph{Perturbative gauge invariance:
Electroweak theory. II}, Annalen Phys. 8 (389-404) 1999.

\bibitem{Gracia}
J.M. Gracia-Bondia, \emph{On the causal gauge principle}, [hep-th/0809.0160].

\bibitem{QED3}
G. Scharf, W.F. Wreszinski, B.M. Pimentel, J.L. Tomazelli,
\emph{Causal approach to (2+1)-dimensional QED},
Annals Phys. 231 (185-208) 1994.

\bibitem{Multiloop}
A. Aste, \emph{Dispersive calculation of the massless multi-loop sunrise diagram},
Lett. Math. Phys. 77 (209-218) 2006.

\bibitem{SUSY}
D.R. Grigore, G. Scharf,
\emph{The Quantum supersymmetric vector multiplet and some problems in
nonfileian supergauge theory},
Annalen Phys. 12 (643-683) 2003.

\bibitem{Fredenhagen}
R. Brunetti, K. Fredenhagen,
\emph{Quantum field theory on curved backgrounds}, to appear in the proceedings of
\emph{Quantum field theory on curved spacetimes}, Potsdam, Germany, October 8-12, 2007.

\bibitem{Werner}
P. Grange and E. Werner, \emph{UV and IR behaviour for QFT and LCQFT with fields as
operator valued distributions: Epstein and Glaser revisited},
Nucl. Phys. Proc. Suppl. 161 (75-80) 2006.

\bibitem{KreimerConnes}
A. Connes, D. Kreimer,
\emph{Renormalization in quantum field theory and the Riemann-Hilbert problem 1:
The Hopf algebra structure of graphs and the main theorem},
Commun. Math. Phys. 210 (249-273) 2000.

\bibitem{Graciabondia}
J.M. Gracia-Bondia,
\emph{Improved Epstein-Glaser renormalization in coordinate space I. Euclidean framework},
Math. Phys. Anal. Geom. 6 (59-88) 2003.

\end{thebibliography}
\end{document}